\documentclass[%
reprint,
superscriptaddress,
showpacs,
nofootinbib,
nobibnotes,
 amsmath,amssymb,
 aps,
prc,
floatfix,
]{revtex4-1}

\usepackage{graphicx}% Include figure files
\usepackage{dcolumn}% Align table columns on decimal point
\newcolumntype{.}{D{.}{.}{-1}}
\usepackage{bm}% bold math
\usepackage{float,array,booktabs,amsmath,amssymb}
\usepackage{csvsimple,longtable}
\usepackage{siunitx}
\usepackage{dcolumn}
\newcolumntype{.}{D{.}{.}{-1}} 

\begin{document}
\preprint{APS/123-QED}

%\preprint{}

\title{\boldmath Spectroscopy of states in $^{136}\rm{Ba}$ using the $^{138}\rm{Ba}(p,t)$ reaction}

\author{B.\,M.~Rebeiro}
\email{b.rebeiro@gmail.com \\ Present address:  Univ Lyon, Univ Claude Bernard Lyon 1,CNRS/IN2P3, IP2I Lyon, UMR 5822, F-69622, Villeurbanne, France}
\affiliation{Department of Physics and Astronomy, University of the Western Cape, P/B X17, Bellville 7535, South Africa.}
\author{S.~Triambak}
\affiliation{Department of Physics and Astronomy, University of the Western Cape, P/B X17, Bellville 7535, South Africa.}
\author{P.\,E.~Garrett}
\affiliation{Department of Physics, University of Guelph, Guelph, Ontario N1G 2W1, Canada.}%
\affiliation{Department of Physics and Astronomy, University of the Western Cape, P/B X17, Bellville 7535, South Africa.}
\author{B.\,A.~Brown}
\affiliation{Department of Physics and Astronomy and National Superconducting Cyclotron Laboratory, Michigan State University, East Lansing, Michigan 48824-1321, USA}
\author{G.\,C.~Ball}
\affiliation{TRIUMF, 4004 Wesbrook Mall, Vancouver, British Columbia V6T 2A3, Canada.}
\author{R.~Lindsay}
\affiliation{Department of Physics and Astronomy, University of the Western Cape, P/B X17, Bellville 7535, South Africa.}
\author{P.~Adsley}
\affiliation{School of Physics, University of the Witwatersrand, Johannesburg 2050, South Africa}%
\affiliation{iThemba LABS, P.O. Box 722, Somerset West 7129, South Africa}%
\author{V.~Bildstein}
\affiliation{Department of Physics, University of Guelph, Guelph, Ontario N1G 2W1, Canada.}%
\author{C.~Burbadge}
\affiliation{Department of Physics, University of Guelph, Guelph, Ontario N1G 2W1, Canada.}%
\author{A.~Diaz-Varela}
\affiliation{Department of Physics, University of Guelph, Guelph, Ontario N1G 2W1, Canada.}%
\author{T.~Faestermann }
\affiliation{Physik Department, Technische Universit\"{a}t M\"{u}nchen, D-85748 Garching, Germany.}%
\author{R.~Hertenberger}
\affiliation{Fakult\"{a}t f\"{u}r Physik, Ludwig-Maximilians-Universit\"{a}t M\"{u}nchen, D-85748 Garching, Germany.}%
\author{B.~Jigmeddorj}
\affiliation{Department of Physics, University of Guelph, Guelph, Ontario N1G 2W1, Canada.}%
\author{M.~Kamil}
\affiliation{Department of Physics and Astronomy, University of the Western Cape, P/B X17, Bellville 7535, South Africa.}
\author{K.\,G.~Leach}
\affiliation{Department of Physics, Colorado School of Mines, Golden, Colorado 80401, USA}
\author{P.\,Z.~Mabika}
\affiliation{Department of Physics and Astronomy, University of the Western Cape, P/B X17, Bellville 7535, South Africa.}
\affiliation{Department of Physics and Engineering, University of Zululand, Private Bag X1001, KwaDlangezwa 3886, South Africa.}
\author{J.\,C.~Nzobadila Ondze}
\affiliation{Department of Physics and Astronomy, University of the Western Cape, P/B X17, Bellville 7535, South Africa.}
\author{J.\,N.~Orce}
\affiliation{Department of Physics and Astronomy, University of the Western Cape, P/B X17, Bellville 7535, South Africa.}
\author{A.~Radich}
\affiliation{Department of Physics, University of Guelph, Guelph, Ontario N1G 2W1, Canada.}%
% \author{E.~Rand}
% \affiliation{Department of Physics, University of Guelph, Guelph, Ontario N1G 2W1, Canada.}%
\author{H.\,-F.~Wirth}
\affiliation{Fakult\"{a}t f\"{u}r Physik, Ludwig-Maximilians-Universit\"{a}t M\"{u}nchen, D-85748 Garching, Germany.}%

\date{\today}

\begin{abstract}
\begin{description}
 \item[Background] The $^{136}$Ba isotope is the daughter nucleus in $^{136}$Xe $\beta\beta$ decay. It also lies in a shape transitional region of the nuclear chart, making it a suitable candidate to test a variety of nuclear models.
% %
\item[Purpose] To obtain spectroscopic information on states in $^{136}$Ba, which will allow a better understanding of its low-lying structure. These data may prove useful to constrain future $^{136}$Xe $\to$ $^{136}$Ba neutrinoless $\beta\beta$ decay matrix element calculations.  
\item[Methods] A $^{138}\mathrm{Ba}(p,t)$ reaction was used to populate states in $^{136}$Ba up to approximately 4.6~MeV in excitation energy. The tritons were detected using a high-resolution Q3D magnetic spectrograph. A distorted wave Born approximation (DWBA) analysis was performed for the measured triton angular distributions.   
% %
\item[Results] One hundred and two excited states in $^{136}$Ba were observed, out of which fifty two are reported for the first time. Definite spin-parity assignments are made for twenty six newly observed states, while previously ambiguous assignments for twelve other states are resolved.
%\item[Conclusion] 
\end{description}

\end{abstract}

\pacs{}
\keywords{}

\maketitle

\section{\label{sec:intro}Introduction} There has been significant interest in studying the structure of nuclei in Xe-Ba-Ce region of the Segr\`e chart~\cite{bucher1,bucher2,Dewald,Donau,Kaya,nomura1,nomura2,Vanhoy}, with particular emphasis on the $A \sim 130$ shape-transitional isotopes~\cite{Peters,li:2010,Pascu_132_134Ba,Kusakari1980,ukal1,ukal2,jespere}. In this context, $^{136}$Ba (with neutron number $N = 80$) is an interesting case. Its low-lying excitations have been variously described in terms of two-quasiparticle configurations~\cite{Griffioen1975}, vibrational two-phonon~\cite{meyer} as well as multiphonon and mixed symmetry states~\cite{sharmishta}, and the coupling of two neutron holes with a quadrupole vibrational $N = 82$ core~\cite{Levy}. Furthermore, $^{136}$Ba is the daughter nucleus for $^{136}$Xe $\beta\beta$ decay, an attractive candidate to search for neutrinoless double beta ($0\nu \beta\beta$) decays. An important issue concerning $0\nu \beta\beta$ decay measurements is addressing the observed model dependence in calculated decay matrix elements~\cite{Engel_2017,Ejiri}. This is particularly relevant for $^{136}$Xe decay, which has several next-generation experiments that aim to go online in the near future~\cite{nexo,lux,darwin,next}. The maximum discrepancy between calculated matrix elements for $^{136}$Xe $0\nu\beta\beta$ decay, from using different many-body techniques is around a factor of four~\cite{Rebeiro}.  
In this regard, spectroscopic studies of both parent and daughter nuclei play an important role~\cite{Freeman:2012,Rebeiro,Schiffer,ge76} in benchmarking the $0\nu\beta\beta$ decay matrix element calculations. Recently we performed a $^{138}{\rm Ba}(p,t)$ study~\cite{Rebeiro} that focused  on pair-correlated $0^+$ states in $^{136}$Ba, to extract the $J=0$ component of the Gamow-Teller matrix element for $^{136}$Xe $0\nu\beta\beta$ decay more accurately. This paper follows up on that work, with a comprehensive report on all states in $^{136}$Ba that were observed in the experiment.

 \section{\label{sec:expt_details}Experimental Details}
The %$^{138}$Ba$(p,t)$ 
experiment was performed at the Maier-Leibnitz Laboratorium (MLL) in Garching (Germany).  A 1.5~$\mu$A, 23 MeV proton beam from MLL tandem accelerator impinged on a $99.8 \%$ isotopically enriched 40~$\mu$g/cm$^2$ thick $^{138}$BaO target, which was evaporated onto a 30~$\mu$g/cm$^2$ carbon backing. 
%  The reaction products then pass through the high resolution Q3D magnetic spectrometer~\cite{LOFFLER19731} where they are separated based on their magnetic rigidity and detected onto the 1-m long focal plane detector \cite{FP,HERTENBERGER1987201}.
The beam was stopped by a Faraday cup placed at $0^\circ$, downstream of the target. A Brookhaven Instruments Corporation (BIC) current integrator recorded the integrated charge from the beam on a run-by-run basis.  
The light reaction ejectiles were momentum analyzed using a high resolution Q3D magnetic spectrograph~\cite{LOFFLER19731}. They were detected at the focal plane of the spectrograph which consisted of two gas proportional counters, with the downstream counter coupled to a cathode strip readout that gave high-resolution position information. After passing through the proportional counters, the ejectiles were stopped in a 7-mm-thick plastic scintillator. 
Energy loss information from the two proportional counters and the residual energy deposited in the scintillator were used for particle identification.
%and detected on the 1-m long focal plane detector \cite{FP,HERTENBERGER1987201}.
In different stages of the experiment we acquired data at ten laboratory angles, from $\theta_{\rm{lab}} = 5^\circ$ to $50^\circ$, with the solid angle acceptance of the spectrograph ranging from $2.3-14.6$~msr. We used four magnetic field settings for the spectrograph, that covered up to $\sim$ 4.6 MeV in excitation energy for $^{136}$Ba. %These measurements are essential to determine the best proton optical model parameter set as well as to determine the number of $^{138}$Ba nuclei in $^{138}$BaO target.

 \section{\label{sec:data_analysis}Data Analysis}
  \subsection{\label{subsect:Energy_Calib} Energy Calibration}

%   For calibrating the focal plane we used well-known states in $^{136}$Ba available in the nuclear data sheets \cite{Sonzogni_A136} upto 2.5 MeV. Beyond that as the errors on the extracted excitation energies became large and peak identification unreliable from the self calibration, states in $^{134}$Ba populated via $^{136}$Ba$(p,t)$ reaction at the same magnetic setting as the $^{138}$Ba$(p,t)$ were used. An excitation energy spectrum of $^{136}$Ba obtained at $25^\circ$ from this method is illustrated in Fig. \ref{fig:136Ba_calib2}. 
\begin{figure*}[t]
 \centering
 \includegraphics[scale=0.58]{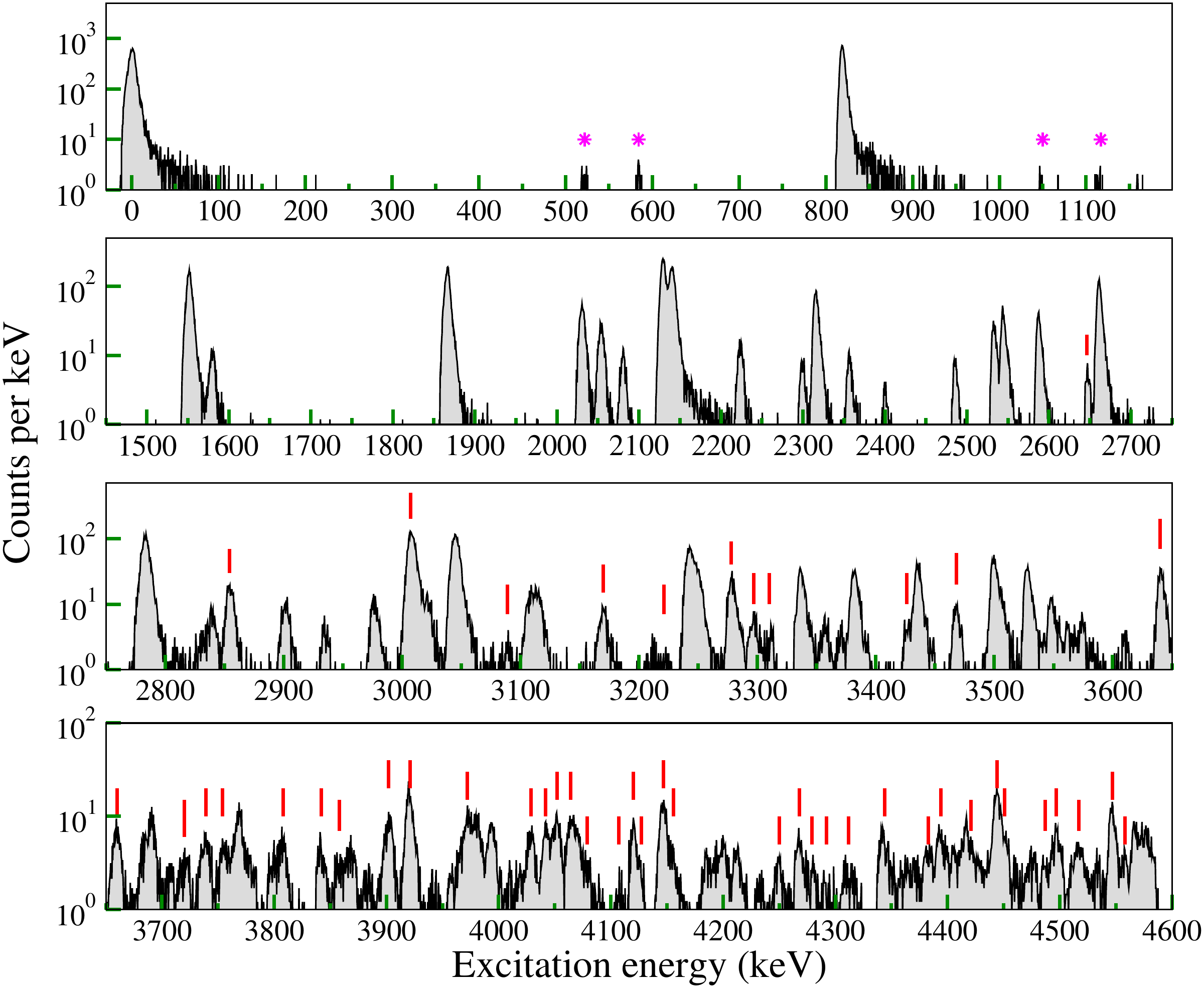}
   \caption{\label{fig:136Ba_calib2}{Excitation energy spectrum for $^{136}$Ba, obtained from $^{138}\mathrm{Ba}(p,t)$ at $\theta_{{\rm{lab}}}=15^\circ$. The red lines indicate new states observed in this work. Since our target was enriched to 99.8$\%$, significant contributions from isotopic impurities are not expected. This is observed in the spectra, where the small triton peaks from the dominant $^{137}$Ba$(p,t)$ contaminant reaction are marked with asterisks. All other peaks correspond to previously known states in $^{136}$Ba}}
  \end{figure*}
% 
%     
  %\noindent The large negative $Q$ value of the $^{138}$Ba$(p,t)$ reaction restricts the observable excitation energy range on the focal plane to  $\sim 1$~MeV    per momentum setting.  Hence we measured the $^{136}$Ba excitation spectra at 4 different momentum settings as shown in 
  Fig.~\ref{fig:136Ba_calib2} shows calibrated triton spectra obtained with the four magnetic field settings. The triton peaks had full widths at half maximum (FWHM) of $\lesssim$ 10~keV. Since the energies of many of the states in $^{136}$Ba below $\sim$ 3.5~MeV are well known~\cite{Mccutchan2018}, we used a quadratic fit~\cite{Mukwevho2018} to the triton momenta for an intrinsic calibration of states corresponding to an energy range $1.5~{\rm MeV} \lesssim E_x \lesssim  3.5$~MeV. The higher energy states were calibrated separately, using the $^{136}$Ba$(p,t)$ reaction at $\theta_\mathrm{lab} = 25^\circ$, after taking into consideration differences in reaction kinematics, energy losses, etc. %(at the same energy and magnetic field settings) 
  %For this part of the calibration, it was important to make corrections for differences in both reaction kinematics as well as energy-losses in the two targets. 
  Details describing the energy calibration procedure can be found elsewhere~\cite{BRThesis, Mukwevho2018}. 
  
 Table~\ref{tab:136Ba_Ex_Jpi} lists excitation energies for all $^{136}$Ba states observed in this experiment. Both systematic and statistical uncertainties were added in quadrature to obtain the final uncertainties in the energy values. The former include contributions from known ground state masses, $\theta_\mathrm{lab}$, %($\pm 1.5^\circ$),  
  the uncertainty in beam energy, the stopping powers (for both protons and tritons), the uncertainties in target thicknesses and the location of the reaction~\cite{Mukwevho2018} in the target.%\footnote{It was assumed that all reactions took place at the center of the targets.} 
  %of the $^{138}$Ba0 and $^{136}$Ba0 targets. 
%We also conservatively included an uncertainty contribution from the reaction location~\cite{Mukwevho2018} within the target. 
  %An estimate of this uncertainty is based on the assumption that the thickness of the target is uniform. In Table~\ref{tab_Ex_errors} we give an estimate of the relative systematic uncertainties for the 4558~keV state.
  
  \begingroup
  \renewcommand{\arraystretch}{1} % Default value: 1
   \LTcapwidth=0.82\textwidth %initially 0.47
  \begin{longtable*}{.c.c..}
  \caption{ {\label{tab:136Ba_Ex_Jpi}} Excited states in $^{136}$Ba observed in this work. Spin-parity determinations were made from the angular distribution analyses described in the following sections. The last column lists $\epsilon_i$, the relative $(p,t)$ strengths for each $L$ value, after correcting for differences in $Q$-values similar to Refs.~\cite{Rebeiro, jespere}. }  \\
   \hline
   \hline
   \multicolumn{2}{c}{Nuclear Data Sheets~\cite{Mccutchan2018}} & & & \multicolumn{1}{c}{This work }  & \cr
   % \cline{1-2}  
   \multicolumn{1}{c}{$E_x$} &  \multicolumn{1}{c}{$J^{\pi}$} & \multicolumn{1}{c}{$E_x$} & \multicolumn{1}{c}{$J^\pi$} & \multicolumn{1}{c}{$\left({d\sigma}/{d\Omega}\right)_{5^\circ}$} & \epsilon_i \cr% \\
    \multicolumn{1}{c}{[keV]} &   & \multicolumn{1}{c}{[keV]} &  & \multicolumn{1}{c}{[mb/sr]} & \multicolumn{1}{c}{[$\%$]}\cr% \\
   \hline
   \endfirsthead
   \multicolumn{6}{c}%
   {{\bfseries \tablename\ \thetable{} -- continued}} \\
   \hline  
%     \multicolumn{1}{c}{$E_x$ (keV)} &  \multicolumn{1}{c}{$J^{\pi}$} & \multicolumn{1}{c}{$E_x$ (keV)} & \multicolumn{1}{c}{$J^\pi$} & \multicolumn{1}{l}{$\left({d\sigma}/{d\Omega}\right)$}(mb/sr) \\
    \multicolumn{1}{c}{$E_x$} &  \multicolumn{1}{c}{$J^{\pi}$} & \multicolumn{1}{c}{$E_x$} & \multicolumn{1}{c}{$J^\pi$} & \multicolumn{1}{c}{$\left({d\sigma}/{d\Omega}\right)_{5^\circ}$}& \epsilon_i \cr% \\
    \multicolumn{1}{c}{[keV]} &   & \multicolumn{1}{c}{[keV]} &  & \multicolumn{1}{c}{[mb/sr]} & \multicolumn{1}{c}{[$\%$]}\cr% \\
   \hline  
   \endhead
   %   \hline 
   %   \multicolumn{4}{|r|}{{ Continued on next page}} \\ 
   %   \hline
   \endfoot
   %   \hline 
   \endlastfoot
   %   \hline
   0.0          & $0^+$                 & 0.0               & $0^+$        & 2.2(1)             & 100.0    \\ %\hline
   818.522(10)  & $2^+$                 & 818.5(6)          & $2^+$        & 0.093(6)           & 100.0    \\ %\hline
   1550.987(13) & $2^+$                 & 1551.4(6)         & $2^+$        & 0.034(2)           & 35.0(20) \\ %\hline
   1578.969(13) & $0^+$                 & 1579.7(6)         & $0^+$        & 0.071(4)           & 5.1(7)   \\ %\hline
   1866.611(18) & $4^+$                 & 1866.1(6)         & $4^+$        & 0.061(4)           & 100.0    \\ %\hline
   2030.535(18) & $7^-$                 & 2030.3(6)         & $7^-$        & 0.019(1)           & 100.0    \\ %\hline
   2053.892(18) & $4^+$                 & 2053.6(6)         & $4^+$        & 0.0119(9)          & 0.067(2) \\ %\hline
   2080.13(3)   & $2^+$                 & 2080.3(6)         & $2^+$        & 0.0043(5)          & 1.7(1)   \\ %\hline
   2128.869(25) & $2^+$                 & 2129.3(6)         & $2^+$        & 0.120(7)           & 73.0(30) \\ %\hline
   2140.237(18) & $5^-$                 & 2140.2(6)         & $5^-$        & 0.037(3)           & 100.0    \\ %\hline
   2222.709(19) & $(1,2)^+$             & 2223.4(6)         & $2^+$        & 0.0073(7)          & 2.9(1)   \\ %\hline
   2298.69(4)   & $(6^-)$               & 2299.0(6)         & $6^+$        & 0.0027(4)          & 100.0    \\ %\hline
   2315.26(7)   & $0^+$                 & 2315.5(6)         & $0^+$        & 0.170(9)           & 15.2(19) \\ %\hline
   2356.497(22) & $4^+$                 & 2356.3(7)^a       & $(5^-)$      & 0.0018(4)          & ...      \\ %\hline
   2399.94(5)   & $(1)^+$               & 2399.8(7)         & $(1,2^+)$    & 0.0057(7)          & ...      \\ %\hline
   2485.13(5)   & $2^+$                 & 2485.3(7)         & $2^+$        & 0.0068(7)          & 2.9(1)   \\ %\hline
   2532.653(23) & $3^-$                 & 2532.4(6)         & $3^-$        & 0.012(1)           & 100.0    \\ %\hline
   2544.481(24) & $4^+$                 & 2543.8(6)^a       & $(5^-, 6^+)$ & 0.0050(7)          & ...      \\ %\hline
   % 2587.09 (5)& $3-5$                 & 2587.2(6)         & $4^+$        & 0.0073(7)          & ...      \\ %\hline
   %            &                       & 2647.2(7)         & $7^-$        & 0.0009(3)          & ...      \\ %\hline
   %            &                       & 2661.0(6)         & $2^+$        & 0.022(2)           & ...      \\ %\hline
   2587.08(3)   & $(5)^+$               & 2587.6(7)         & $4^+$        & 0.0077(7)          & 19.6(7)  \\ %\hline
                &                       & 2646.4(8)         & $7^-$        & 0.0008(3)          & 5.6(3)   \\ %\hline
   2661.48(5)   & $1, 2^+$              & 2660.4(7)^a       & $2^+$        & 0.026(2)           & 13.4(6)  \\ %\hline
   2784.44(13)  & $0^+$                 & 2783.4(7)         & $0^+$        & 0.148(8)           & 14.6(17) \\ %\hline
                &                       & 2829.9(8)         & $(6^+,7^-)$  & 0.0006(3)          & ...      \\ %\hline
   2840.74(10)  & $(4^+)$               & 2839.1(7)         & $4^+$        & 0.0029(5)          & 1.6(1)   \\ %\hline
                &                       & 2854.3(7)         & $5^-$        & 0.0098(8)          & 13.8(5)  \\ %\hline
   2905.0(5)    &                       & 2902.0(7)         & $(4^+,5^-)$  & 0.0019(5)          & ...      \\ %\hline
   2935.1(9)    &$(1,2^+)$              & 2935.1(7)         & $(1,2^+)$    & 0.0044(5)          & ...      \\ %\hline
   2977.67(18)  &                       & 2977.1(7)         & $0^+$        & 0.0046(6)          & 0.65(9)  \\ %\hline
                &                       & 3007.2(8)         & $5^-$        & 0.036(2)           & 101.0(40)\\ %\hline
   3022.10(8)   &$(1,2^+)$              & 3021.0(10)        & $2^+$        & 0.0052(8)          & 3.6(2)   \\ %\hline
   3044.54(5)   &$1^{(-)}$              & 3044.5(7)^a       & $(0^+,2^+)$  & 0.088(5)           & ...      \\ %\hline
                &                       & 3089.0(10)        & $(4^+, 5^-)$ & 0.0005(2)^b        & ...      \\ %\hline
   3109.59(9)   & $2^+$                 & 3108.7(8)^a         & $(2^+)$      & 0.018(2)           & ...      \\ %\hline
   3116.08(6)   & $2^+$                 & 3115.3(9)         & $2^+$        & 0.005(1)           & 3.1(2)   \\ %\hline
                &                       & 3170.0(7)         & $6^+$        & 0.0044(4)^c        & 176.0(80)\\ %\hline
   3212.0(5)    & $0^{(+)},1,2,3^{(+)}$ & 3210.0(10)        & $(2^+,3^-)$  & 0.0005(1)^c        & ...      \\ %\hline
                &                       & 3221.0(20)        & $(2^+)$      & 0.0014(2)^c        & 0.61(5)  \\ %\hline
   3241.89(17)  &                       & 3244.7(7)         & $2^+$        & 0.040(2)           & 26.0(10) \\ %\hline
                &                       & 3278.6(7)         & $0^+$        & 0.040(2)           & 3.3(3)   \\ %\hline
                &                       & 3297.1(8)         & $5^-$        & 0.0008(3)          & 4.0(2)   \\ %\hline
                &                       & 3310.0(10)        & $(1,2^+)$    & 0.0012(3)          & ...      \\ %\hline
   3335.6(3)    &                       & 3336.2(7)         & $2^+$        & 0.024(2)           & 12.1(5)  \\ %\hline
   3354.5(3)    &                       & 3356.7(8)         & ...          & 0.0016(3)          & ...      \\ %\hline
   3370.07(21)  &    1                  & 3369.0(10)        & $(1,2^+)$    & 0.0026(4)          & ...      \\ %\hline
   3378.0(5)    &                       & 3381.0(10)        & $2^+$        & 0.015(1)           & 8.2(4)   \\ %\hline
                &                       & 3426.7(8)         & $0^+$        & 0.0082(8)          & 1.1(1)   \\ %\hline
   3435.0(1)    & $1^-$                 & 3435.1(7)^a       & $(1,2^+)$    & 0.017(1)           & ...      \\ %\hline
                &                       & 3468.2(9)         & ...          & 0.0031(5)          & ...      \\ %\hline
%  3505.5(9)    & $0^{(+)},1,2,3^+$     & 3498.7(8)         & $(4^+,5^-)$  & 0.015(1)           & ...      \\ %\hline
                &                       & 3498.7(8)         & $(4^+,5^-)$  & 0.015(1)           & ...      \\ %\hline
   3526.7(4)    & $2^+$                 & 3527.6(7)         & $2^+$        & 0.017(1)           & 9.5(4)   \\ %\hline
%  3550.70(20)  &                       & 3547.9(7)         & $(4^+)$      & 0.0011(5)          & ...      \\ %\hline
                &                       & 3547.9(7)         & $(4^+)$      & 0.0011(5)          & ...      \\ %\hline
%               &                       & 3639.9(7)         & $4^+$        & 0.009(1)           & ...      \\ %\hline
%               &                       & 3659.8(7)         & $2^+$        & 0.0066(8)          & ...      \\ %\hline
%               &                       & 3682.8(8)         & ...          & 0.0012(2)$^{{b}}$  & ...      \\ %\hline
                &                       & 3640.0(10)        & $4^+$        & 0.0052(6)          & 27.0(10) \\ %\hline
                &                       & 3660.0(10)        & $2^+$        & 0.0047(6)          & 4.2(2)   \\ %\hline
%               &                       & 3684.0(10)        & ...          & 0.0008(2)$^{{c}}$  & ...      \\ %\hline
%  3691.92(13)  & $(1-3)$               & 3691.0(10)        & $5^-$        & 0.0022(4)          & 7.6(4)   \\ %\hline
                &                       & 3691.0(10)^a      & $5^-$        & 0.0022(4)          & 7.6(4)   \\ %\hline
   3706.1(6)    &$(1,2^+)$              & 3708.0(20)        & $(1,2^+)$    & 0.0016(4)          & ...      \\ %\hline
                &                       & 3720.0(10)        & $(J>5)$      & 0.0012(3)^c        & ...      \\ %\hline
                &                       & 3739.0(10)        & $(1,2^+)$    & 0.014(2)           & ...      \\ %\hline
                &                       & 3754.0(10)        & $(4^+,5^-)$  & 0.0023(5)          & ...      \\ %\hline
   3768.9(3)    & $1^{(-)},2,3^+$       & 3768.0(10)^a      & $(3^-)$      & 0.0068(8)          & 76.0(30) \\ %\hline
   3795.34(15)  &$(1,2^+)$              & 3799.0(20)        & $(1, 2^+)$   & 0.0015(3)          & ...      \\ %\hline
                &                       & 3808.0(20)        & $(2^+, 3^-)$ & 0.0031(5)          & ...      \\ %\hline
                &                       & 3842.0(20)        & $2^+$        & 0.0072(7)          & 2.7(1)   \\ %\hline
                &                       & 3858.0(30)        & $(5^-,6^+)$  & 0.002(3)           & ...      \\ %\hline
%  3863.47(23)  &$(1,2^+)$              & 3868.0(20)        & $(2^+,6^+)$  & 0.004(2)           & ...      \\ %\hline
   3863.47(23)  &$(1,2^+)$              & 3868.0(20)^a      & ...          & 0.004(2)           & ...      \\ %\hline
   3881.17(10)  &$(1,2^+)$              & 3883.0(30)^a      & $(7^-,8^+)$  & 0.0017(6)^b        & ...      \\ %\hline
                &                       & 3902.0(20)        & $2^+$        & 0.0076(7)          & 4.0(2)   \\ %\hline
                &                       & 3921.0(20)        & $0^+$        & 0.0096(8)          & 2.2(3)   \\ %\hline
% 3962.9 (8)    &                       & 3961.0 (2)        & ...          & 0.0016(3)$^{{c}}$  & ...      \\ %\hline
                &                       & 3972.0(20)        & $2^+$        & 0.0083(8)          & 4.4(2)   \\ %\hline
   3979.76(20)  & $(1)$                 & 3980.0(20)^a      & $(4^+)$      & 0.0050(6)          & 10.9(5)  \\ %\hline
   3992.56(19)  & $0^{(+)},1,2,3^+$     & 3994.0(20)^a      & $(3^-)$      & 0.0037(4)^c        & 52.0(30) \\ %\hline
%  4008.6(3)    & $1,2^+$               & 4011.0(30)        & $(2^+, 3^-)$ & 0.0007(2)^b        & ...      \\ %\hline
                &                       & 4029.0(20)        & $(1,2^+)$    & 0.0032(5)          & ...      \\ %\hline
                &                       & 4042.0(20)        & ...          & 0.0018(3)          & ...      \\ %\hline
                &                       & 4052.0(20)        & $2^+$        & 0.0102(8)          & 5.0(2)   \\ %\hline
                &                       & 4064.0(20)        & $(5^-,6^+)$  & 0.0037(5)          & ...      \\ %\hline
   4075.0(100)  &                       & 4070.0(30)        & $(2^+,3^-)$  & 0.0059(6)^c        & ...      \\ %\hline
                &                       & 4079.0(30)        & ...          & 0.0011(3)^c        & ...      \\ %\hline
                &                       & 4107.0(30)        & ...          & 0.0011(2)          & ...      \\ %\hline
                &                       & 4120.0(20)        & $4^+$        & 0.0034(5)^c        & 8.9(5)   \\ %\hline
                &                       & 4127.0(30)        & $2^+$        & 0.0037(5)          & 1.0(1)   \\ %\hline
                &                       & 4147.0(20)        & $0^+$        & 0.018(1)           & 5.4(7)   \\ %\hline
                &                       & 4156.0(20)        & $5^-$        & 0.008(1)           & 4.0(2)   \\ %\hline
%               &                       & 4185.0(20)        & ...          &                    & ...      \\ %\hline
%               &                       & 4193.0(20)        & ...          &                    & ...      \\ %\hline
%               &                       & 4201.0(20)        & ...          & 0.0012(3)          & ...      \\ %\hline
   4214.9       &                       & 4213.0(20)        & ...          & 0.0013(3)          & ...      \\ %\hline
   4231.17(20)  & 1                     & 4233.0(30)        & $(1,2^+)$    & 0.0018(5)          & ...      \\ %\hline
                &                       & 4250.0(20)        & $2^+$        & 0.0057(6)          & 2.4(2)   \\ %\hline
                &                       & 4268.0(20)        & $(2^+,3^-)$  & 0.0050(6)          & ...      \\ %\hline
                &                       & 4279.0(30)        & $(1,2^+)$    & 0.0027(4)^c        & ...      \\ %\hline
                &                       & 4292.0(30)        & $(2^+,3^-)$  & 0.002(1)           & ...      \\ %\hline
                &                       & 4312.0(30)        & $3^-$        & 0.0019(3)          & 18.0(10) \\ %\hline
                &                       & 4344.0(20)        & $0^+$        & 0.0055(5)          & 1.8(3)   \\ %\hline
                &                       & 4383.0(30)        & $(4^+,5^-)$  & 0.0009(3)          & ...      \\ %\hline
                &                       & 4394.0(30)        & $2^+$        & 0.0042(6)          & 2.3(1)   \\ %\hline
%               &                       & 4406.0(20)        & ...          & 0.001(1)           & ...      \\ %\hline
   4413.28(10)  & (1)                   & 4416.0(20)        & $(1,2^+)$    & 0.0012(3)          & ...      \\ %\hline
                &                       & 4421.0(30)        & $(1,2^+)$    & 0.0004(2)          & ...      \\ %\hline
                &                       & 4444.0(20)        & $0^+$        & 0.0075(7)          & 3.2(4)   \\ %\hline
                &                       & 4451.0(30)        & $(3^-)$      & 0.001(1)^c         & ...      \\ %\hline
   4475.18(10)  & (1)                   & 4475.0(30)        & $(1,2^+)$    & 0.0015(3)^c        & ...      \\ %\hline
                &                       & 4487.0(30)        & $2^+$        & 0.0031(4)^c        & 1.4(1)   \\ %\hline
                &                       & 4497.0(20)        & $2^+$        & 0.0044(5)^c        & 3.4(2)   \\ %\hline
                &                       & 4517.0(30)        & ...          & 0.0006(3)^b        & ...      \\ %\hline
   4536.4(3)    & 1                     & 4534.0(30)^a      & $(0^+)$      & 0.0006(3)^b        & 0.6(3)   \\ %\hline
                &                       & 4547.0(20)        & $2^+$        & 0.0054(5)^b        & 4.0(3)   \\ %\hline
                &                       & 4558.0(30)        & ...          & 0.0012(2)^b        & ...      \\ %\hline
   \hline     
   \hline
   \multicolumn{6}{l}{$^{{a}}$ Potential unresolved doublet. }    \\
   \multicolumn{6}{l}{$^{{b}}$ This value is $(d\sigma/d\Omega)_{15^\circ}$ due to the unavailability of $\theta_{\rm lab} = 5^\circ$ and $10^\circ$ data.}    \\
   \multicolumn{6}{l}{$^{{c}}$ This value is $(d\sigma/d\Omega)_{10^\circ}$ due to the unavailability of $\theta_{\rm lab} = 5^\circ$ datum. }    \\
   
  \end{longtable*}
 
  \endgroup
% % 
%    \begin{table}[h!]
%   \caption{\label{tab_Ex_errors}Relative systematic uncertainties on the excitation energies of the 4558~keV state in $^{136}$Ba. For $E_\chi > 3.6$~MeV there is an additional contribution due the $^{136}$Ba$(p,t)$ reaction used for calibration.}
%   \begin{ruledtabular}
%   \begin{tabular}{lc}
%        Source      & $\Delta E_\chi/E_\chi (\%)$  \cr 
%   \hline
% 	Ground state masses & 0.0002\cr
% 	Q3D Angle & 0.002 \cr
% 	Stopping powers (SRIM)&  0.002 \cr
% 	$^{138}$BaO target thickness & 0.002 \cr
% 	Ejectile energy loss & 0.03\cr
% 	Target reaction location & 0.03 \cr
% 	\hline
% 	Total & \cr
%   \end{tabular}
%   \end{ruledtabular}
%   \end{table}
% % 
% 
  \subsection{\label{subsect:Cross_Sect_Calc}Differential Scattering Cross Sections}
  %\noindent To identify the spin-parity ($J^\pi$) of  different states observed in this experiment we compared experimental angular distributions with Distorted Wave Born Approximation (DWBA) predictions. used 
  In the next step of the analysis, we determined $^{138}\mathrm{Ba}(p,t)$ differential scattering cross sections. These were used for a Distorted Wave Born Approximation (DWBA) analysis of the data. The laboratory differential scattering cross section for each observed level in $^{136}$Ba was obtained using the formula
%   \\ For calculating the absolute cross sections, the number of protons incident on the $^{138}$Ba target was extracted from the total integrated current on a Faraday cup place at $0^\circ$ inside the scattering chamber with respect to the beam direction.  \\
 %the expression
%  is proportional to the ratio of the total counts under the peak $N_c$, with respect to the number of reaction centers $N_t$ in the target and the integrated beam current $N_b$ for each run,
  \begin{equation}
  \left(\frac{d\sigma}{d\Omega}\right)_{\theta_{\rm lab}} = \frac{N_c(\theta_{\rm lab})}{N_b(LT)~N_t~d\Omega_{\rm lab}}.
  \label{eqn:cross_section}
  \end{equation}
  In the above, for each experimental run, $N_c$ is the total number of counts under a triton peak, $N_b$ is the number of integrated beam particles recorded during the combined live times $(LT)$ of both the detector and the data acquisition system, $N_t$ is the areal density of the $^{138}$Ba target atoms and $d\Omega_\mathrm{lab}$ is the solid angle acceptance of the spectrograph. We determined $N_t$ from a measured proton elastic scattering cross section at $\theta_{\rm{lab}} = 15^\circ$ (an angle at which the cross section is dominated by Rutherford scattering) together with DWBA calculations of the $^{138}\mathrm{Ba}(p,p)$ angular distribution, described in Section~\ref{Sect:Elastic_Scattering}. 
   
 The final step of data reduction required a transformation to the center-of-mass frame, so that
 \begin{equation}
 \left(\frac{d\sigma}{d\Omega}\right)_{\theta_{\rm c.m.}} = \left(\frac{d\sigma}{d\Omega}\right)_{\theta_{\rm lab}} \left( \frac{1+\gamma~\cos(\theta_{\rm{c.m.}})}{ (1+2\gamma~\cos(\theta_{\rm{c.m.}})+\gamma^2)^{3/2}}  \right),
 \label{eqn:lab_to_com_cs}
 \end{equation}
 with
 \begin{equation}
 \theta_{\rm{c.m.}} = {\sin}^{-1}(\gamma~{\sin(\theta_{\rm{lab}}})) + \theta_{\rm{lab}}
 \label{eqn:lab_to_com_theta}
 \end{equation}
and
 \begin{equation}
 \gamma \approx \sqrt{ \frac{m_1~m_3}{M_2~M_4} \left( \frac{1}{1+ \left(1+\frac{m_1}{M_2} \right) \frac{Q}{E_p} }\right) }.
 \label{eqn:gamma}
 \end{equation}
In the above $m_1,m_3,M_2,M_4$ are the masses of the projectile, ejectile, target and recoil nucleus, $Q$ is the reaction $Q$-value to the relevant excited state, and $E_p$ is the  incident proton energy in the laboratory frame.  
\subsection{Elastic Scattering Cross Sections}\label{Sect:Elastic_Scattering}
\begin{figure}[t]
 \includegraphics[width=8.6 cm]{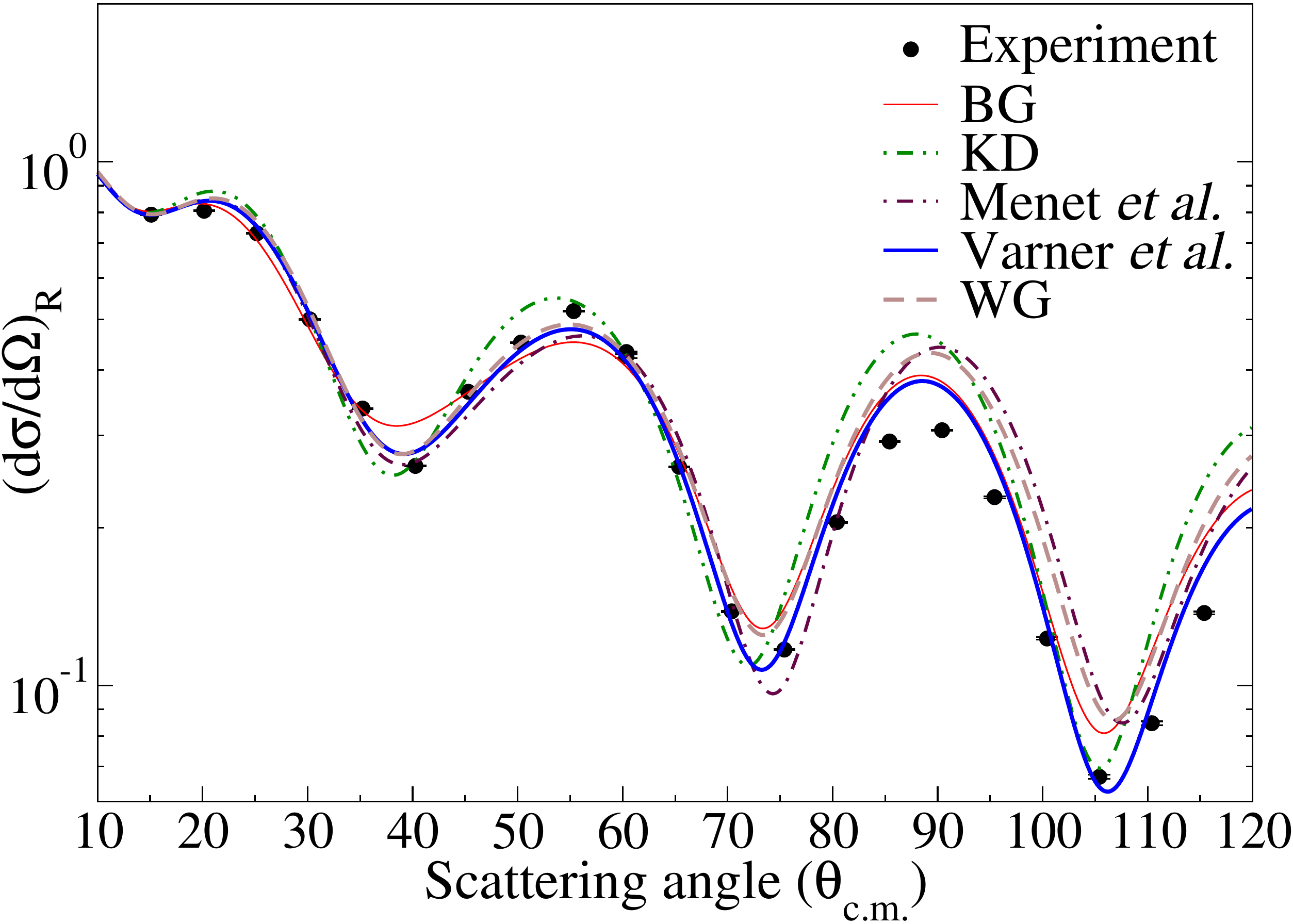}
 \caption{\label{Fig:138Ba_elastic}{Experimental $^{138}\mathrm{Ba}(p,p)$ angular distribution from this work, compared with various DWBA predictions. For the latter we used global proton optical model parameters recommended by Becchetti and Greenlees (BG)~\cite{Becchetti1969}, Koning and Delaroche (KD)~\cite{KONING2003}, Varner {\it{et al.}}~\cite{VARNER199157}, Menet {\it{et al.}}~\cite{Menet1971} and Walter and Guss (WG)~\cite{WalterGuss1986}. 
  %The proton OMP by Varner {\it{et al.}}~\cite{VARNER199157} 
   %that best reproduces the elastic scattering data is the one by . This OMP is further used in our DWBA analysis.
   }}
  \end{figure}

As mentioned previously,  the $(p,p)$ scattering cross section is dominated by Rutherford scattering at small angles. Consequently the DWBA predictions for elastic scattering at these angles (and hence the target thickness determination from small-angle elastic scattering) are largely independent of the choice of optical model potential (OMP) parameters used in the analysis. However, for a DWBA analysis of the $^{138}\mathrm{Ba}(p,t)$ data, an appropriate choice of OMP parameters is critical. To guide us along these lines, we measured $^{138}\mathrm{Ba}(p,p)$ elastic scattering cross sections over an angular range $15^\circ \le  \theta_{\rm{lab}} \le 115^\circ$, in $5^\circ$ steps. The measured angular distribution is shown in Fig.~\ref{Fig:138Ba_elastic}, where the ordinate is expressed as a ratio to the Rutherford differential scattering cross section at that angle.  
To choose the optimal OMP parameters for the incoming $p + ^{138}$Ba channel, we compared these data to DWBA predictions, obtained using the DWUCK4 code~\cite{DWUCK4}. Five available global OMPs~\cite{Becchetti1969, VARNER199157, KONING2003, Menet1971, WalterGuss1986} were used to make the comparison. As evident in Fig.~\ref{Fig:138Ba_elastic}, the DWBA distribution obtained using the parameters recommended by Varner~\emph{et al.}~\cite{VARNER199157} showed best agreement with our experimental data.

 \subsection{DWBA Calculations}\label{Sect:DWBA_Calculations}
  
\begin{figure}[t]
  \centering 
  \includegraphics[width=8.6cm]{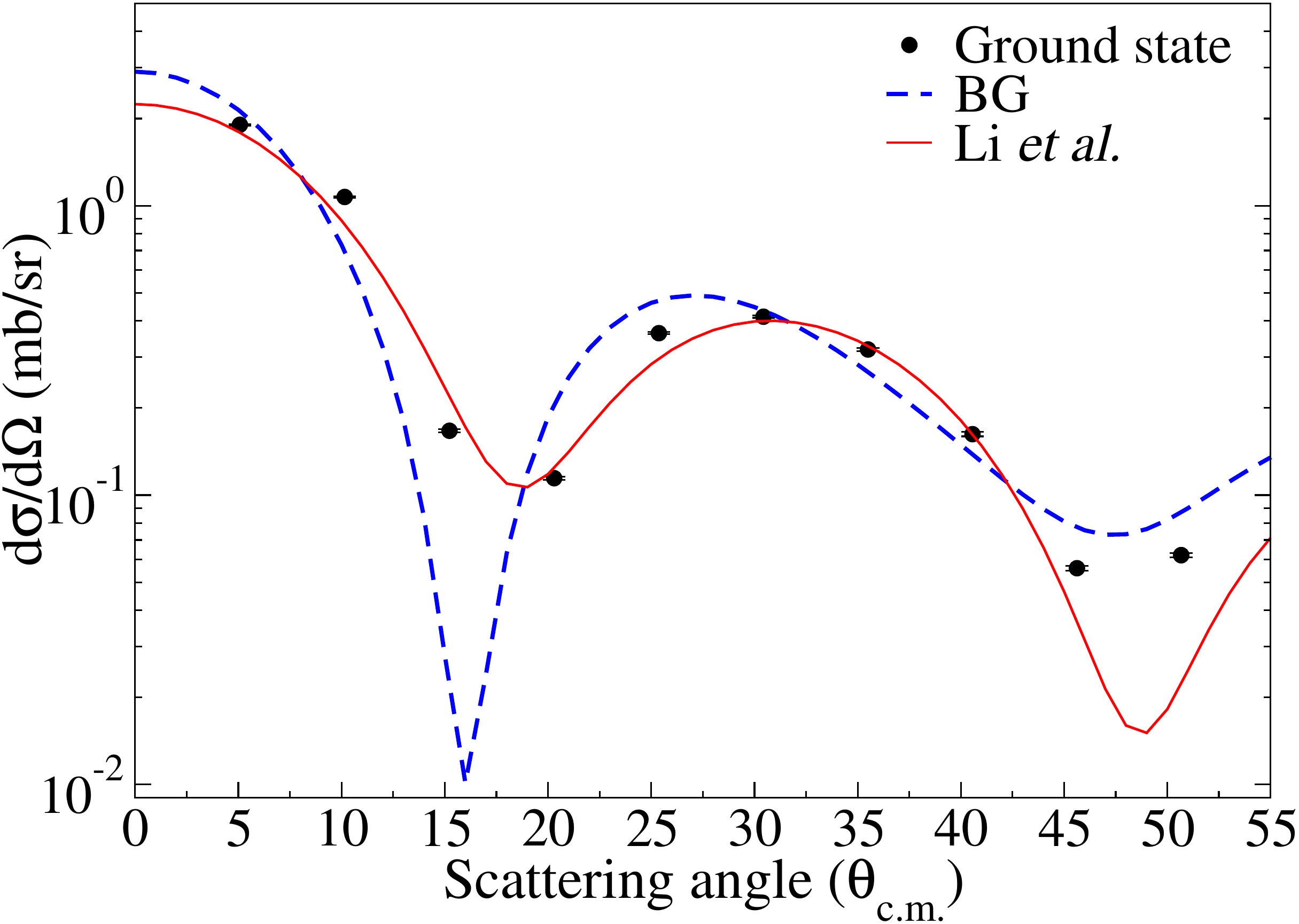}
  \caption{{\label{Fig:GS_OMP_compare}}Measured $^{138}\mathrm{Ba}(p,t)$ angular distribution for the ground state in $^{136}$Ba. The data are compared with normalized DWBA predictions obtained using the proton OMPs by Varner {\it et al.}~\cite{VARNER199157} and the triton OMP parameters by Becchetti and Greenlees (BG)~\cite{ripl1,Becchetti_triton} and Li~{\it et al.}~\cite{LI2007103}. 
  %The OMP from Li {\it{et al.}} (LLC)~\cite{LI2007103} agrees better with our data compared to the one recommended by Becchetti and Greenlees (BG)~\cite{Becchetti_triton}.
  }
 \end{figure}

 %To calculate the $^{138}$Ba$(p,t)$ reaction cross section in the Distorted Wave Born Approximation (DWBA) formalism we used the DWUCK4 code~\cite{DWUCK4}. 
 The DWUCK4 DWBA calculations described here used Woods-Saxon potentials. The $^{138}\mathrm{Ba}(p,t)$ reaction was modeled assuming the zero-range approximation, as a single-step transfer of a di-neutron in a $S=0$ (singlet) state.  
 As mentioned previously, we chose to use the proton global OMPs by Varner~\emph{et al.}~\cite{VARNER199157} for the entrance (proton) channel of the reaction. For the exit $t+^{136}\mathrm{Ba}$ channel, we compared the measured $^{136}$Ba ground state angular distribution with normalized DWBA predictions using available global triton OMP parameters~\cite{LI2007103,ripl1,Becchetti_triton}. As Fig.~\ref{Fig:GS_OMP_compare} shows,   the angular distribution calculated with the OMP parameters recommended by Li~{\it et al.}~\cite{LI2007103} showed better agreement with our data. We therefore chose to use these parameters for our analysis. 
 %
%  %   
%   \begin{table}[b]
%   \caption{\label{tab:OMP_params}Wood-Saxon optical model parameters used for the proton, triton and the transferred di-neutron used in our DWBA analysis.}
%   \begin{ruledtabular}
%   \begin{tabular}{cccc}
%         & proton      & triton\footnote{$E_t$ is the kinetic energy of the outgoing triton calculated for each excited state.}  & neutron\footnote{The well depth was adjusted to reproduce the neutron binding energy in Eq.~\eqref{Eq:neutronBE}.} \\
%   \hline
%   $V_r$ & 52.938   & $0.436E_t^2 - 0.1456E_t +186.8304 $ & --  \\
%   $r_r$ & 1.206    & 1.094    & 1.170\\
%   $a_r$ & 0.690    & 0.795    & 0.75 \\
%   $W_v$ & 1.226    & $-0.0097E_t^2 + 0.5025E_t +7.383$    & -\\
%   $r_v$ & 1.249    & 1.2898    & -\\
%   $a_v$ & 0.690    & 1.2307    & -\\
%   $W_s$ & 9.105    & $-0.6451E_t+27.8117$    & -\\
%   $r_s$ & 1.249    & 1.1718    & -\\
%   $a_s$ & 0.69     & 0.8791    & -\\
%   $V_{so}$ & 5.90  & 1.9029    & -\\
%   $r_{so}$ & 1.108 & 0.4921    & -\\
%   $a_{so}$ & 0.63  & 0.0497    & -\\
%   $r_c$    & 1.260 & 1.4219    & -\\
%   $\lambda$    &  &     & 25\\
%   \end{tabular}
%   \end{ruledtabular}
%   \end{table}
  
 We calculated the two-neutron transfer form factors using the neutron OMPs provided by Ref.~\cite{Becchetti1969}. The form factor for each state was determined by varying the depth of the real volume term of the potential, so that it matched the binding energy of each transferred neutron 
  \begin{equation}
   BE = \frac{S_{2n}(^{138}{\rm{Ba}}) +E_x (^{136}\rm{Ba})}{2}, 
   \label{Eq:neutronBE}
  \end{equation}
  %In the above, $E_x(^{136}\rm{Ba})$ is the excitation energy of the state and 
  where $S_{2n}$ is the two-neutron separation energy of $^{138}$Ba.
%   We summarize all OMP parameters used for our data analysis in Table~\ref{tab:OMP_params}. % 
%   

  The two-neutron transfer amplitudes for different orbital angular momentum $(L)$ values were determined using various configurations in the DWUCK4 calculations. The selectivity of a single-step $(p,t)$ reaction demands that only natural parity states, with total angular momentum $J = L$ and parity $\pi = (-)^L$ are strongly populated. For the even  $L$ transfers (i.e. final states with spin-parity $J^\pi=$ $0^+$, $2^+$ etc.), we chose the $(0h_{11/2})^2$ configuration~\cite{Szwec2016} for the form factor. For the $L=1$  and $L=3$ transfers we used the $(2s_{1/2})(1p_{1/2})$ and $(0h_{11/2})(0g_{7/2})$ orbitals respectively, while for $L=5$ and $L=7$ we used the $(0h_{11/2}) (1d_{3/2})$ orbitals~\cite{Gelletly1969}.%The $L=4$ and $L=6$ transitions were calculated using the $(0g_{7/2})^2$ configuration~\cite{Gelletly1969}.
  \footnote{The shapes of angular distributions in single-step $(p,t)$ reactions are nearly independent of the orbitals from which the neutrons are picked up.}

\section{\label{sec:results}Results} 
 As listed in Table~\ref{tab:136Ba_Ex_Jpi}, we identify a total of one hundred and two excited states in $^{136}$Ba below 4.6 MeV. Fifty two  of these states are reported for the first time. %In this work, we could assign definitive spin-parity to $\_\_$ 
 In the following sections we present angular distribution results for most of the states in $^{136}$Ba that we observe in this experiment. We mainly limit our discussions to those states that are observed for the first time and others for which we disagree with previous work or could make only tentative assignments for their spins and parities.

   \subsection{\label{Sect:0+_states} $\mathbf{J^\pi=0^+}$ states}
\begin{figure}[htpb]
    \includegraphics[width=8.4 cm]{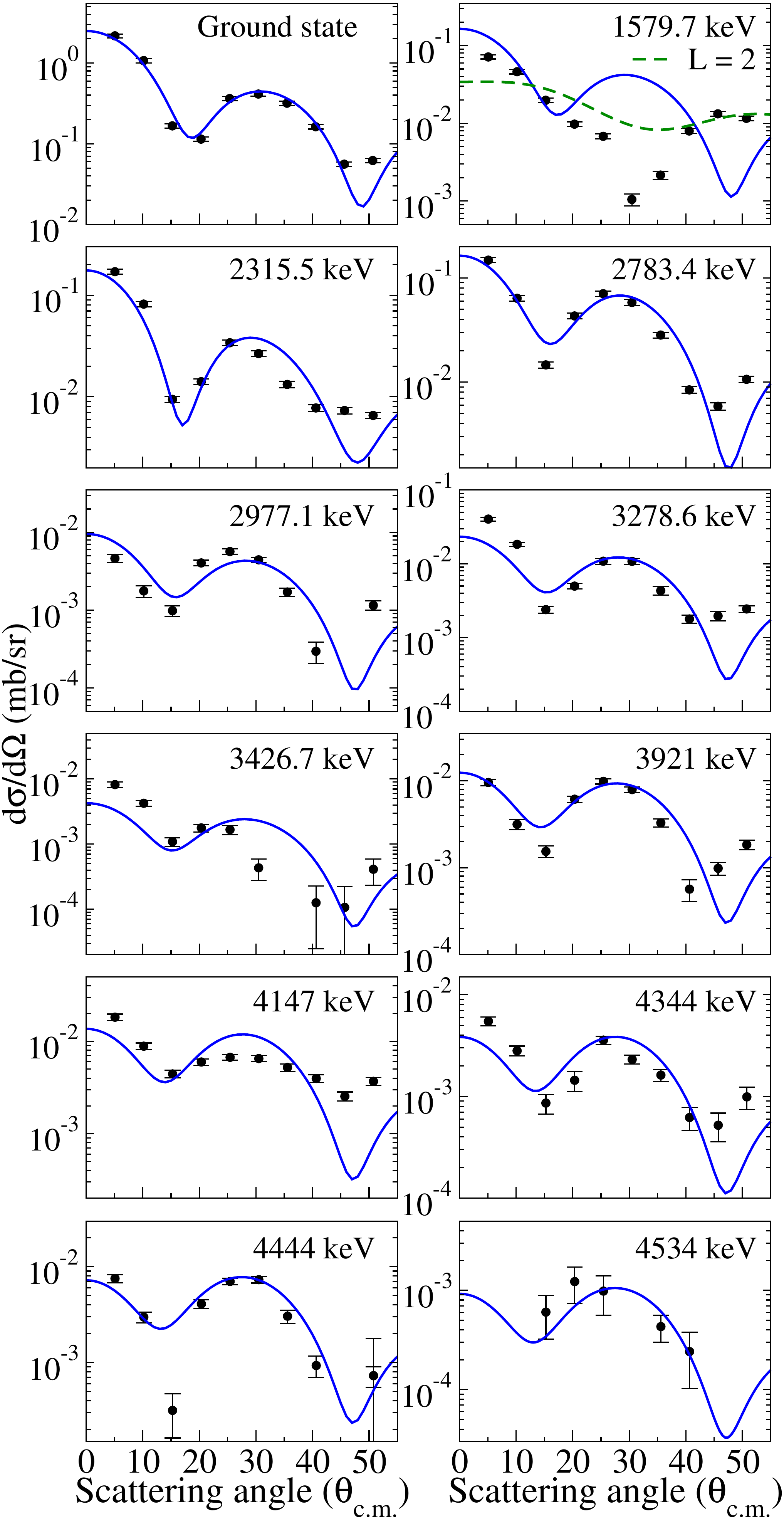}
    \caption{\label{Fig:138Ba_pt_0+}{Experimental angular distributions for all observed $0^+$ states in $^{136}$Ba.
  The solid blue lines represent normalized DWUCK4 cross sections for $L=0$ transfer. The dashed green line, corresponding to $L=2$ transfer is shown for comparison.
  }}
  \end{figure}

   $0^+$ states produced via $(p,t)$ reactions on even-even target nuclei can be easily identified from their angular distributions, that are characteristic of $L = 0$ transfer, with large forward angle cross sections that drop rapidly around $\theta_{\rm c.m.}= 15^\circ$. This is evident in Fig. \ref{Fig:138Ba_pt_0+}, which compares our experimental data with normalized DWBA results. Although we have already published~\cite{Rebeiro} a comprehensive study of $0^+$ states in $^{136}$Ba from this work, we emphasize here some salient features of our observations for completeness. Our analysis showed at least eleven $0^+$ states in $^{136}\rm{Ba}$, of which six are reported for the first time.  We also resolved a discrepancy in the spin-parity assignment for one of these states. One $0^+$ level at 2141.38(3)~keV, listed in the $A = 136$ Nuclear Data Sheets (NDS)~\cite{Mccutchan2018} overlaps with a well-known $5^-$ state at the same energy. As shown in Section~\ref{Sect:5-_states}, the triton angular distribution corresponding to this excitation energy is consistent with an $L=5$ orbital angular momentum transfer, which implies that the $0^+$ state is weakly populated. Based on the measured cross section for this level at the forward angle of $5^\circ$, we place an upper limit of $\leq 3\%$ for the $L=0$ strength to the 2141.38~keV state. We briefly summarize other aspects of our results below. 
 \\
 \\
%   \begin{itemize}
%    \item 
   {\bf $\mathbf{E_x=1579.7}$ keV:} This state is listed as $J^\pi = 0^+$ in the NDS~\cite{Mccutchan2018}, which is consistent with $\gamma$-ray analysis from both $^{136}$La $\beta$ decay~\cite{meyer} and $^{135}$Ba$(n,\gamma)$~\cite{Becvar1969} data. However, our measured triton angular distribution for this level has a peculiar feature. Its shape differs from its DWBA prediction and other $0^+$ states with 
   its first minimum at around $\theta_\mathrm{c.m.} = 30^\circ$.  We investigated this matter in detail and ruled out possible  contaminant peaks that could lead to such an unusual distribution. It is highly likely that the observed discrepancy is due to sequential and multi-step contributions to the cross section.   
 \\ \\
 %   
%  \item 
 {\bf $\mathbf{{\rm{\bf E_x}}=2977.1}$~keV:} The $A=136$ NDS~\cite{Mccutchan2018} reports this level with an unassigned  $J^\pi$, presumably because of discordant measurements. In this work, the angular distribution for this state is consistent with an $L=0$ transition. We therefore confirm this to be a $0^+$ state.
 \\ \\
%   \item 
  $\mathbf{E_x}$ {\bf = 3278.6, 3426.7, 3921, 4147 4344 and 4444~keV:} There are no reported states in the NDS~\cite{Mccutchan2018} at these excitation energies. The angular distributions for these states are consistent with $L=0$ transfer, allowing an assignment of $J^\pi=0^+$.
  \\ \\  
  {\bf $\mathbf{{{\bf E_x}}=4534}$~keV:} The peak corresponding to this state is plagued by lack of statistics and the presence of kinematically broadened light-ion contamination at forward angles. Its measured angular distribution shows reasonable agreement with $L=0$ transfer. However, a level at 4536.4(3)~keV was observed in $^{136}{\rm Ba}(\gamma,\gamma')$~\cite{Massarczyk2012} with a spin assignment of $J = 1$. Therefore, it is quite likely that the two states are not the same.
  \\ \\   

  \subsection{\label{Sect:2+_states} $\mathbf{J^\pi=2^+}$ states}
\begin{figure}
   \includegraphics[width=8.6 cm]{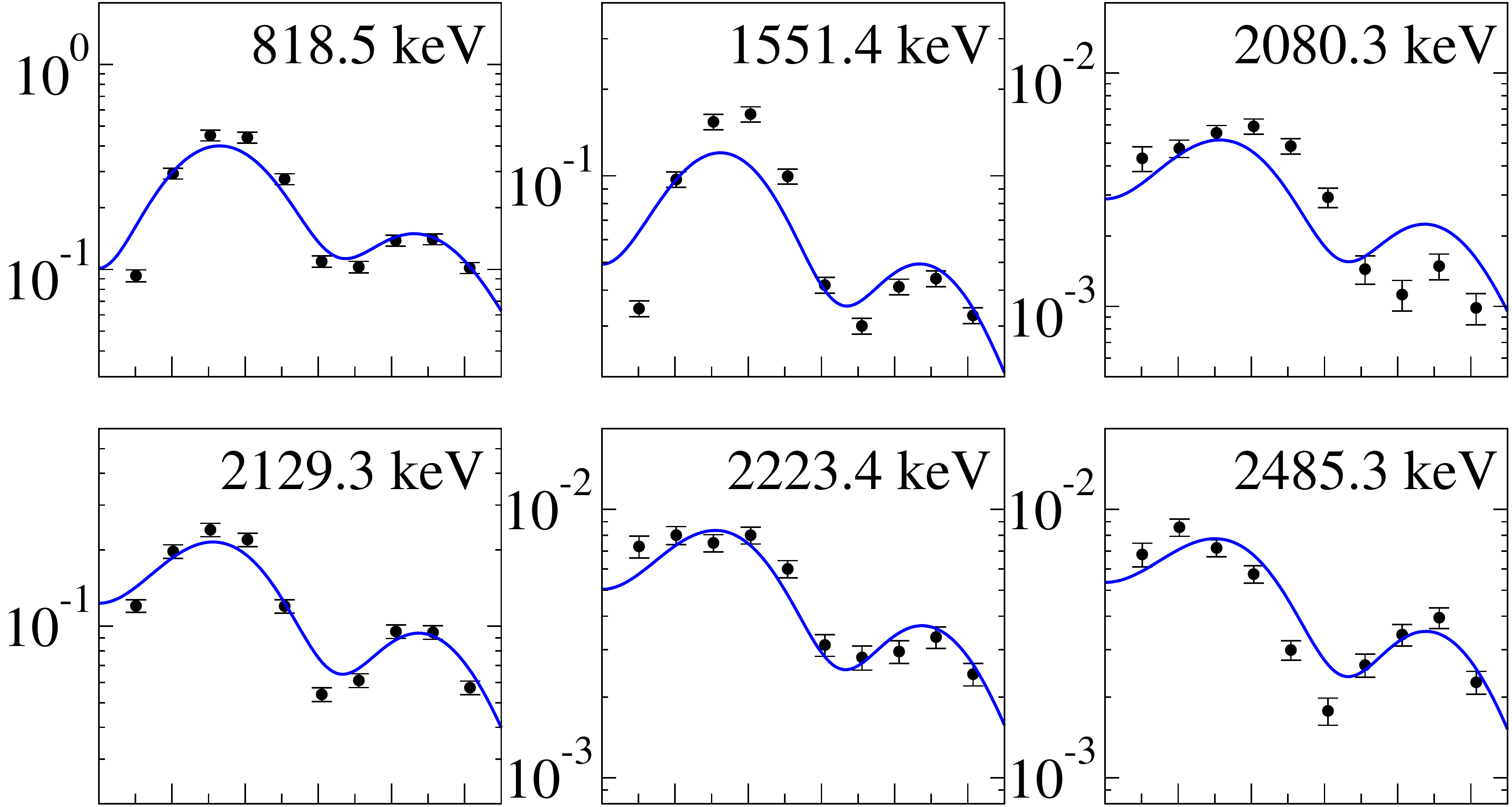} 
   \includegraphics[width=8.6 cm]{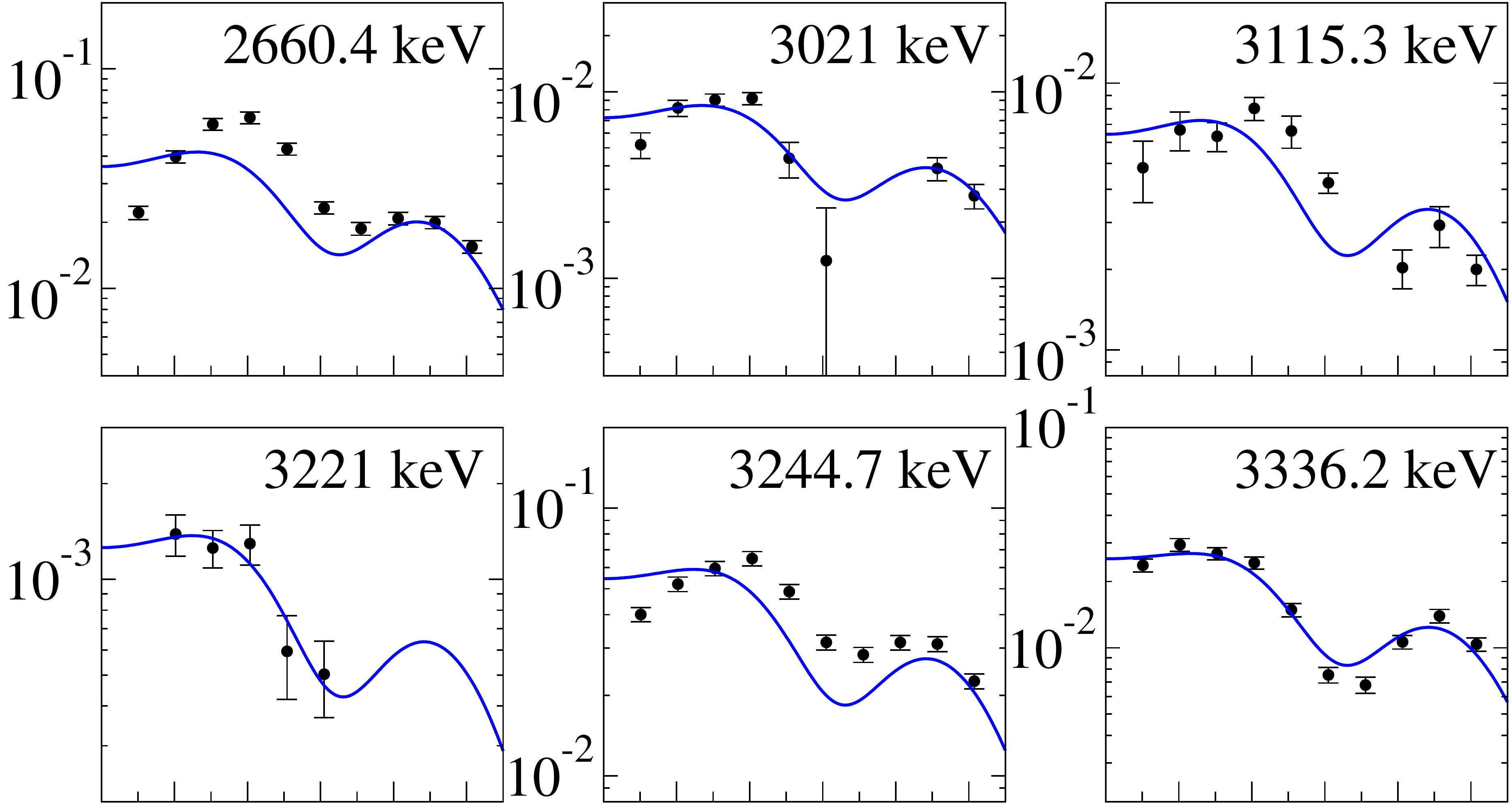}
   \hspace*{-0.3 cm}
   \includegraphics[width=8.85 cm]{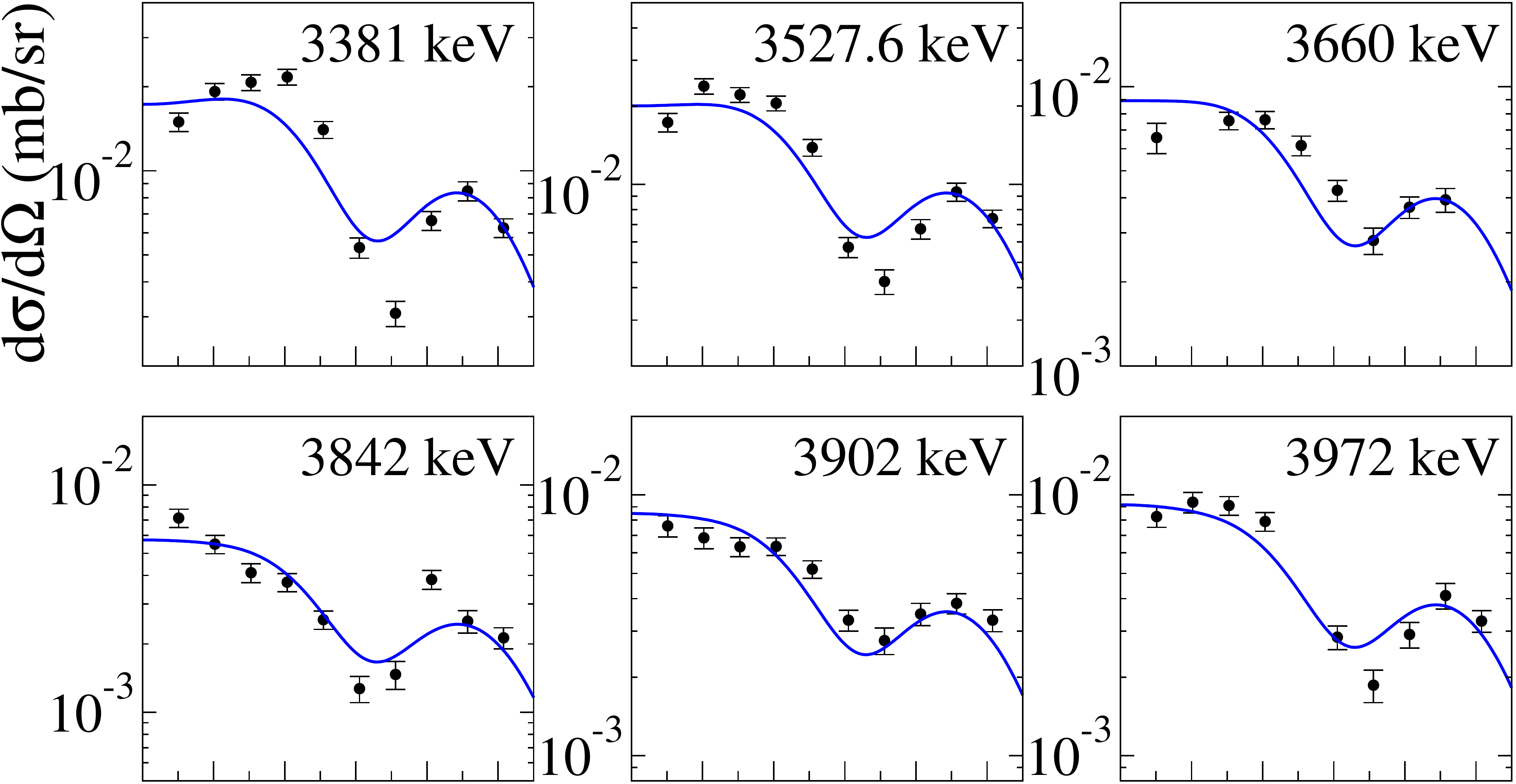}
   \includegraphics[width=8.6 cm]{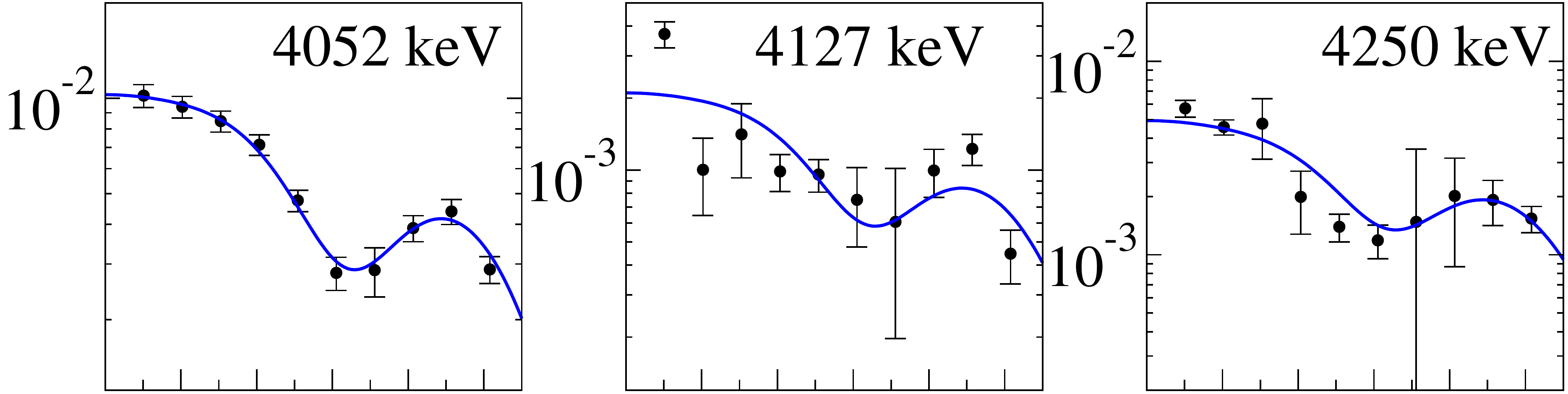}
   \includegraphics[width=8.9 cm]{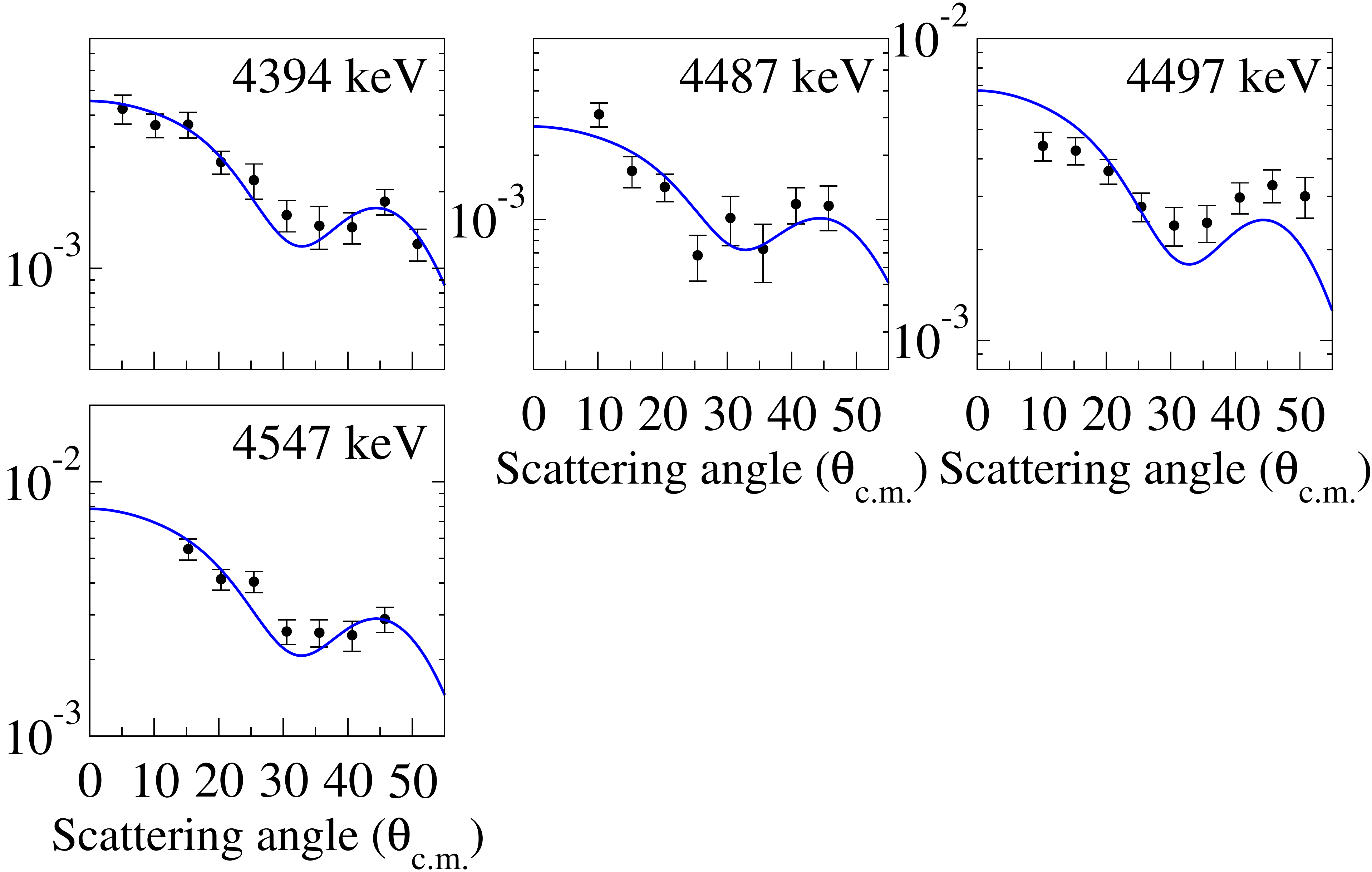}

   \caption{\label{Fig:138Ba_pt_2+}{Experimental angular distribution for all the $2^+$ states observed in this work compared with normalized $L = 2$ DWBA predictions, shown in solid blue lines. States with tentative $2^+$ assignments are presented in Section~\ref{Sect:Tentative}.}}
  \end{figure}

  In addition to the well-known $2^+$ states~\cite{Mccutchan2018}, we use our angular distribution data to identify eleven more (previously unreported) $2^+$ states and resolve six other ambiguous cases. These data are plotted in Fig.~\ref{Fig:138Ba_pt_2+}. We discuss some examples below.
%   
%   \begin{figure}[h!]
% %    \includegraphics[width=0.50\textwidth] {Images_pt_long_Sept2020/all_2+_states_Sep2020_1_mod}\\
%    \includegraphics[width=8.6 cm] {Images_pt_long_Sept2020/all_2+_states_Sep2020_1_new}\\
%    \includegraphics[width=8.6 cm]{Images_pt_long_Sept2020/all_2+_states_Sep2020_2_new}\\
%    \includegraphics[width=8.6 cm]{Images_pt_long_Sept2020/all_2+_states_Sep2020_3_new}\\
%    \includegraphics[width=8.6 cm]{Images_pt_long_Sept2020/all_2+_states_Sep2020_4_new}\\
%    \caption{\label{Fig:138Ba_pt_2+}{Experimental angular distribution for all the $2^+$ states observed in this work. States with tentative $2^+$ assignments are presented in Section~\ref{Sect:Tentative}.}}
%   \end{figure}
  % 
  \\ \\ 
{\bf $\mathbf{{{\bf E_x}}=2223.4}$~keV:} The $A = 136$ NDS  lists an excited state at 2222.709(19)~keV with a tentative $J^\pi$ assignment of $(1,2)^+$. A recent $\gamma$-ray angular distribution measurement from $^{136}{\rm Ba}(n,n'\gamma)$~\cite{Mukhopadhyay2008} assigned this state a spin-parity of $2^+$. Our angular distribution agrees with $L=2$ transfer, consistent with their spin-parity assignment for the state.  
    \\ \\
  {\bf $\mathbf{{{\bf E_x}}=2660.4}$~keV:} The $A = 136$ NDS lists two closely spaced levels at 2659.65(5)~keV with $J^\pi=(3,4,5)^+$  and 2661.48(5)~keV with $J^\pi=~1,2^+$. In comparison, the recent $(n,n'\gamma)$ experiment by Mukhopadhyay {\it et al.}~\cite{Mukhopadhyay2008} determined the spins and parities of these states to be  $J^\pi=5^{(-)}$ and $(2^+,4^+)$ respectively. Our data are limited by experimental resolution to disentangle these two states. However the measured angular distribution for the observed triton peak is consistent with  $L=2$ transfer. This indicates a strong population of the higher energy state, which presumably has $J^\pi = 2^+$.  
  \\ \\ 
  {\bf $\mathbf{{{\bf E_x}}=3021}$~keV:} The recent $^{136}$Ba$(n,n'\gamma)$ study by Mukhopadhyay~{\it et al.}~\cite{Mukhopadhyay2008} determined this state to have $J^\pi=(1,2^+)$, which was adopted by Ref.~\cite{Mccutchan2018}. Our measured $(p,t)$ angular distributions for this level is well reproduced assuming a transferred $L=2$ in the DWBA calculations.% We thus assign it $J^\pi = 2^+$. 
  \\ \\  
  {\bf $\mathbf{{{\bf E_x}}=3221}$~keV:} This state is reported for the first time in this work. Although the data agree well with an $L = 2$ DWBA curve, we make a tentative assignment of $J^\pi = (2^+)$ due to the lack of data points at higher angles.% 
  % Based on the agreement between DWBA and experimental angular distribution for $\theta < 35^\circ$, we tentatively assign this state as $J^\pi=(2^+)$. This state was not observed in any previous work and is being reported for the first time~\cite{Mccutchan2018}.
  \\ \\    
  {\bf $\mathbf{{{\bf E_x}}=3244.7}$~keV:} A level at 3241.89(17)~keV with undetermined spin-parity is reported in the NDS~\cite{Mccutchan2018}. Our measured angular distribution for this level is consistent with a DWBA calculation for $L=2$. Thus we assign $J^\pi=2^+$ for this state.
\\ \\    
  $\mathbf{E_x} = $ {\bf 3336.2 and~3381~keV:} Two levels at $E_x=3335.6$(3) and 3378.0(5)~keV are listed in the $A=136$ NDS~\cite{Mccutchan2018} with no spin-parity assignments. Our work shows that the angular distributions for both these states are well described by $L=2$ transfer. We disagree with the excitation energy of the higher lying state by $\sim$~3~keV.  %Thus we assign $J^\pi=2^+$ for these states.
% \\ \\    
%    {\bf $\mathbf{{{\bf E_x}}=3972}$~keV} The only excited states reported in the NDS~\cite{Mccutchan2018} in the vicinity of this state are 3965.51~keV with $J^\pi=(1,2^+)$ and 3979.76~keV as $J^\pi=(1)$. The later was adopted from previous evaluation~\cite{Peker1979} based on $^{135}$Ba$(n,\gamma)$ reaction~\cite{Chrien1974} and later treated as a possible contaminant~\cite{Islam1990}. We rule out the possibility of a contaminant resulting from the $(p,t)$ reaction on a different isotope of barium. 
%    From our analysis this state is a $2^+$ and is possibly being reported for the first time.  
\\ \\
   {\bf $\mathbf{{{\bf E_x}}}=$ 3660, 3842, 3902, 3972, 4052, 4127, 4250, 4394, 4487, 4497 and 4547~keV:} All these levels are being reported for the first time. The angular distributions for these states exhibit typical $L=2$ behavior. %We therefore assign $J^\pi=2^+$ for all these excited states.
  \subsection{\label{Sect:3-_states} $\mathbf{J^\pi=3^-}$ states}
  %\vspace*{-0.6 cm}
  \begin{figure}[t]
   \includegraphics[width=8.6 cm]{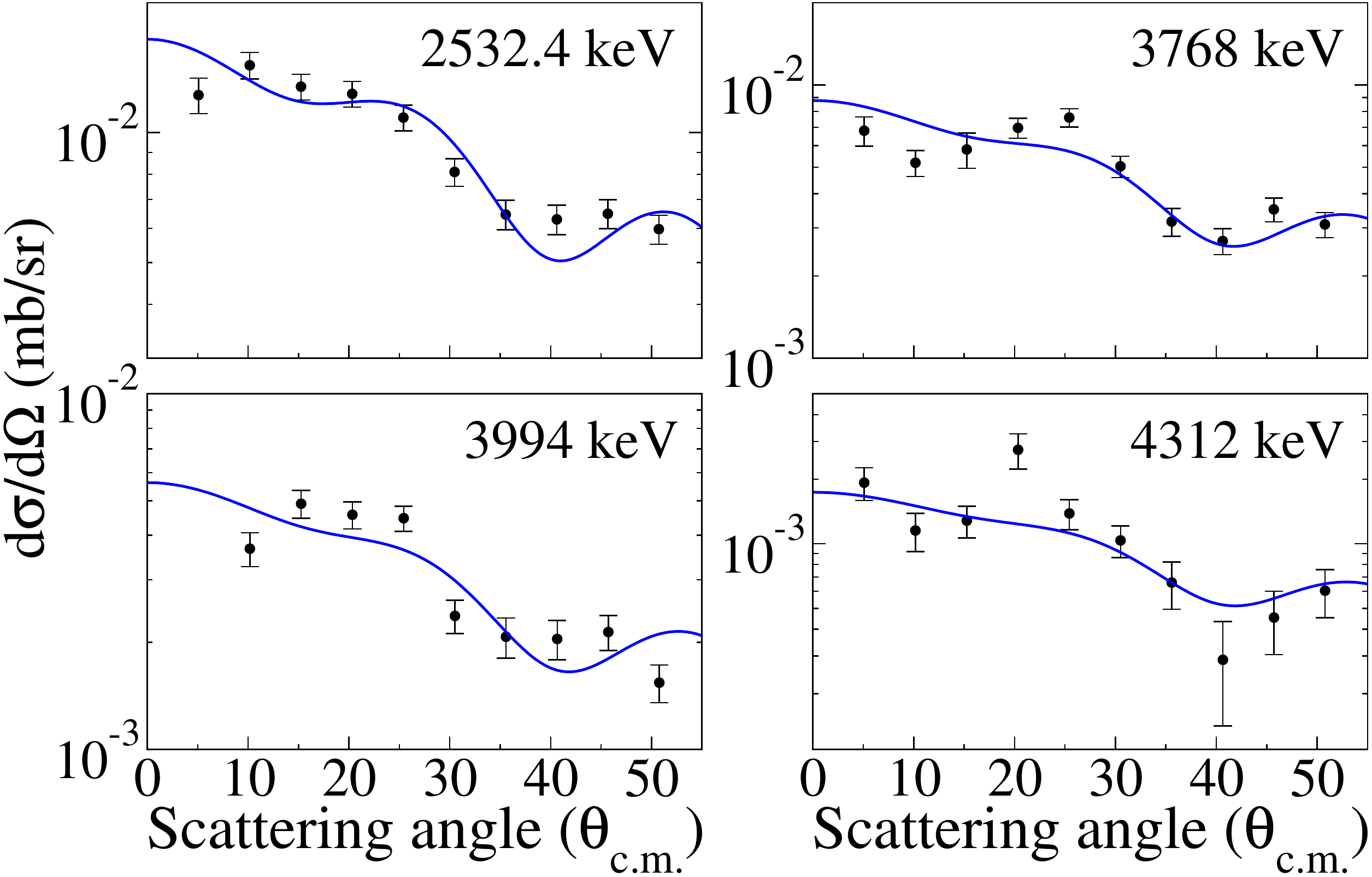}
   \caption{\label{Fig:138Ba_pt_3-}{Experimental angular distribution for all definite $3^-$ assignments from this work. The solid blue lines represent normalized DWBA cross sections for $L=3$ transitions.}}
  \end{figure}
  We do not observe the reported $3_1^-$ state at 2390.17(22)~keV~\cite{Mccutchan2018}. In comparison, the known $3^-$ level at 2532.653(23)~keV~\cite{Mccutchan2018} shows a significant population. This is not surprising, as the latter is known to be an octupole collective state with a $B(E3;0^+ \to 3_1^-)$, transition strength of 0.155(18)~e$^2$b$^3$ or 20.2(23)~W.u.~\cite{Burnett432}.
%  as observed in an $(\alpha,\alpha')$ measurement~\cite{Burnett}.
  Such collective states are expected to have higher cross sections due to a coherent sum of scattering amplitudes. %In comparison, the 2390~keV state most likely has a two-quasiparticle configuration, for which a similar enhancement is not expected.  
  We also identify one new $3^-$ state at 4312~keV as described below.  
% 
% This could be possible indication that the state is proton-dominated. However, we identify two new $3^-$ states in this experiment, in addition to the well-established 2532.653(23)~keV level~\cite{Mccutchan2018}. 
  \\ \\
  \noindent  {\bf $\mathbf{{{\bf E_x}}=3768~{\rm{and}}~3994}$~keV:} The $A = 136$ NDS~\cite{Mccutchan2018} list two levels at 3768.9(3) and 3992.56(19)~keV with $J^\pi=1^{(-)},2,3^+$ and $0^{(+)},1,2,3^+$ respectively. Both were reported from $^{135}\mathrm{Ba}(n,\gamma)$ experiments~\cite{Gelletly1969, Islam1990, Chrien1974}, that also observed $\gamma$-ray transitions from higher-lying resonant $1^+$ states to these levels. Our measured angular distributions for these two states are well reproduced assuming $L=3$ transfer in the DWBA. Therefore, it unlikely that these are the same levels. We thus assign the observed states $J^\pi=(3^-)$. 
\\ \\
%   \noindent  {\bf $\mathbf{{{\bf E_x}}=3768}$~keV:} The $A = 136$ NDS~\cite{Mccutchan2018} identifies this level to have $J^\pi=1^{(-)},2,3^+$, based on prior $^{135}\mathrm{Ba}(n,\gamma)$ work that observed a transition to the state by a higher lying $1^+$ level at 9107~keV~\cite{Gelletly1969, Islam1990, Chrien1974}. Our measured angular distribution for this state is well reproduced assuming $L=3$ transfer in the DWBA. It's unlikely that both these levels are the same. We therefore assign our observed state to have $J^\pi=(3^-)$. 
% \\ \\
% {\bf $\mathbf{{{\bf E_x}}=3994}$~keV:} The $A = 136$ NDS lists a state  at 3992.56(19)~keV~\cite{Mccutchan2018} that is consistent with our determined excitation energy, with $J^\pi =0^{(+)},1,2,3^+$. Our measured angular distribution and DWBA calculations indicate an $L = 3$ transition. However, as in the 3768~keV state, this level is also observed to be fed by a higher$ 1^+$ state~\cite{Chrien1974}, making 
% %This is consistent with a previously observed $(E1)$ $\gamma$ transition from this level to the $2_2^+$ state at 1551~keV~\cite{Islam1990}.   
\\ \\    
 $\mathbf{E_x}=$ {\bf 4312~keV:} This state is reported for the first time in this work. Fig.~\ref{Fig:138Ba_pt_3-} shows that its measured angular distributions are consistent with $L=3$, so that $J^\pi = 3^-$. %Our assignment for these two levels is $J^\pi=3^-$.
\subsection{\label{Sect:4+_states} $\mathbf{J^\pi=4^+}$ states}
\begin{figure}[h!]

 \includegraphics[width=0.153\textwidth, height=0.09\textheight]{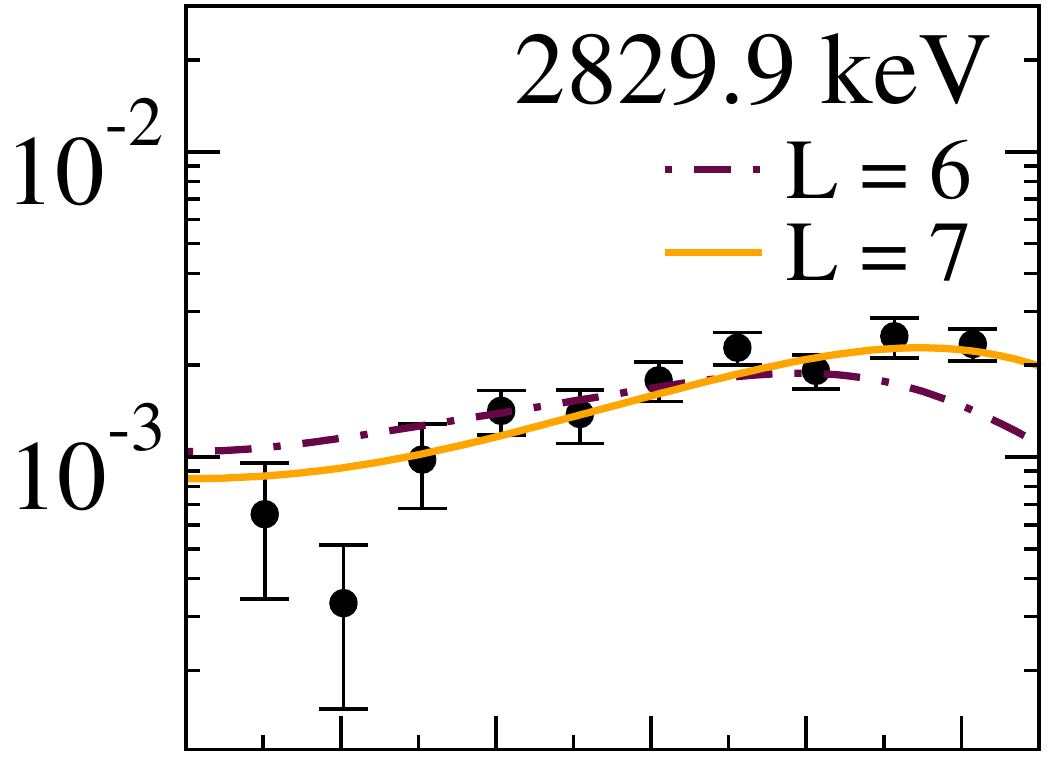}
 \includegraphics[width=0.153\textwidth, height=0.09\textheight]{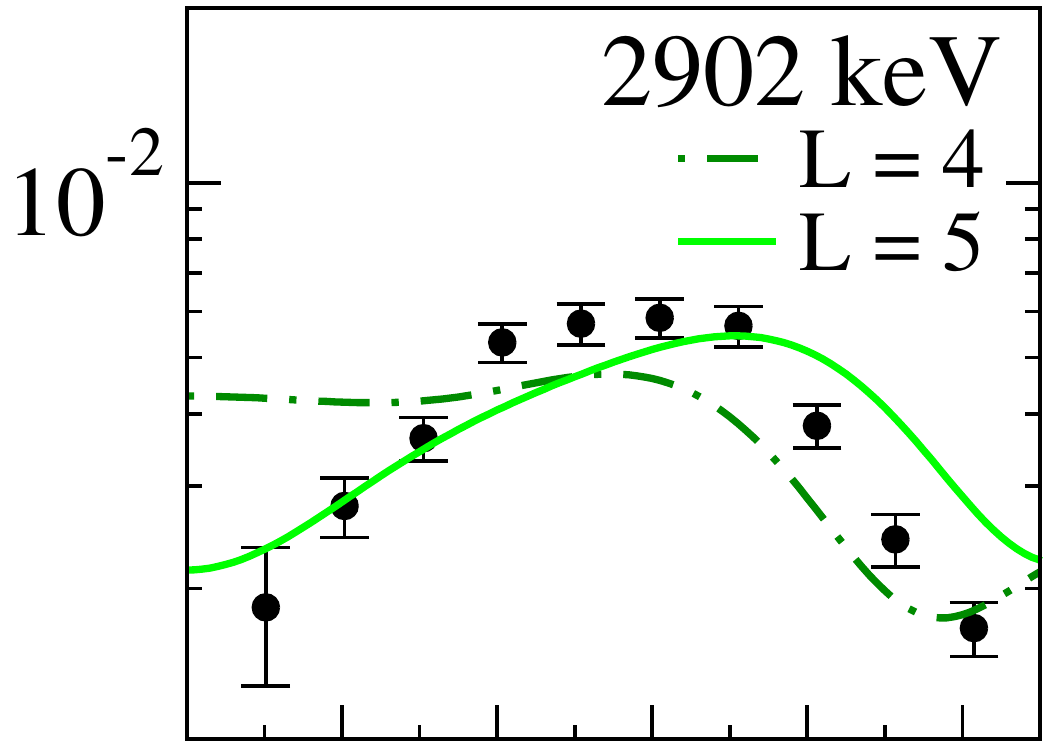}
 \includegraphics[width=0.153\textwidth, height=0.09\textheight]{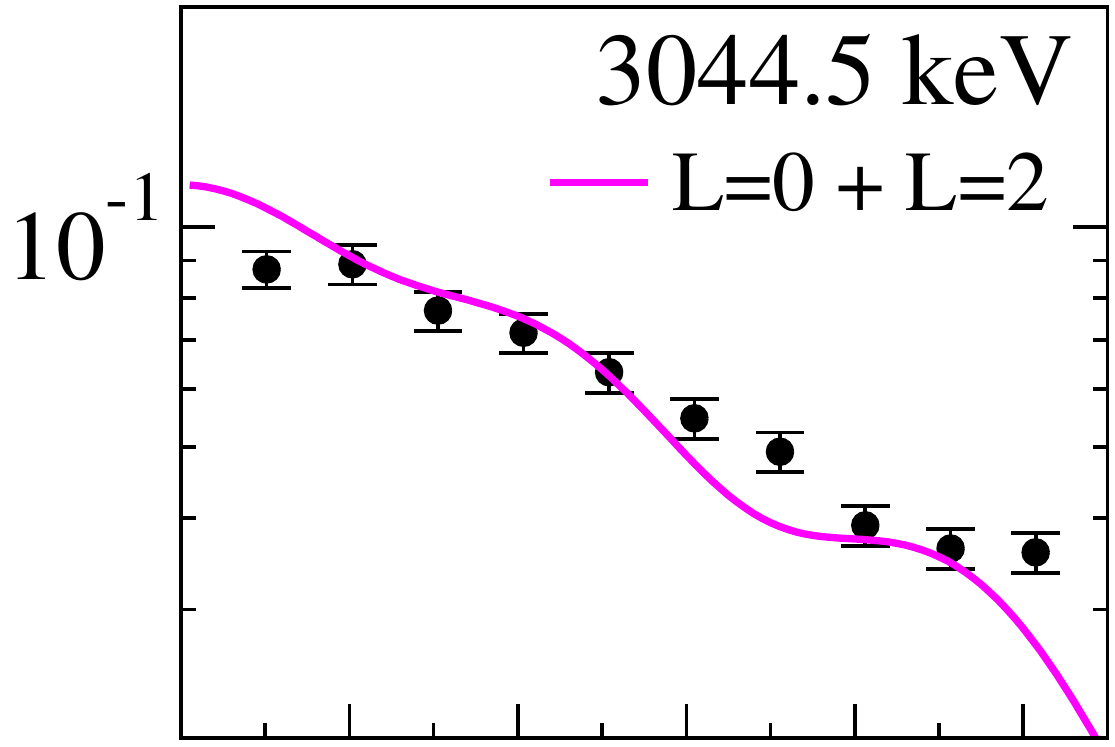}\\
 
 \includegraphics[width=0.153\textwidth, height=0.09\textheight]{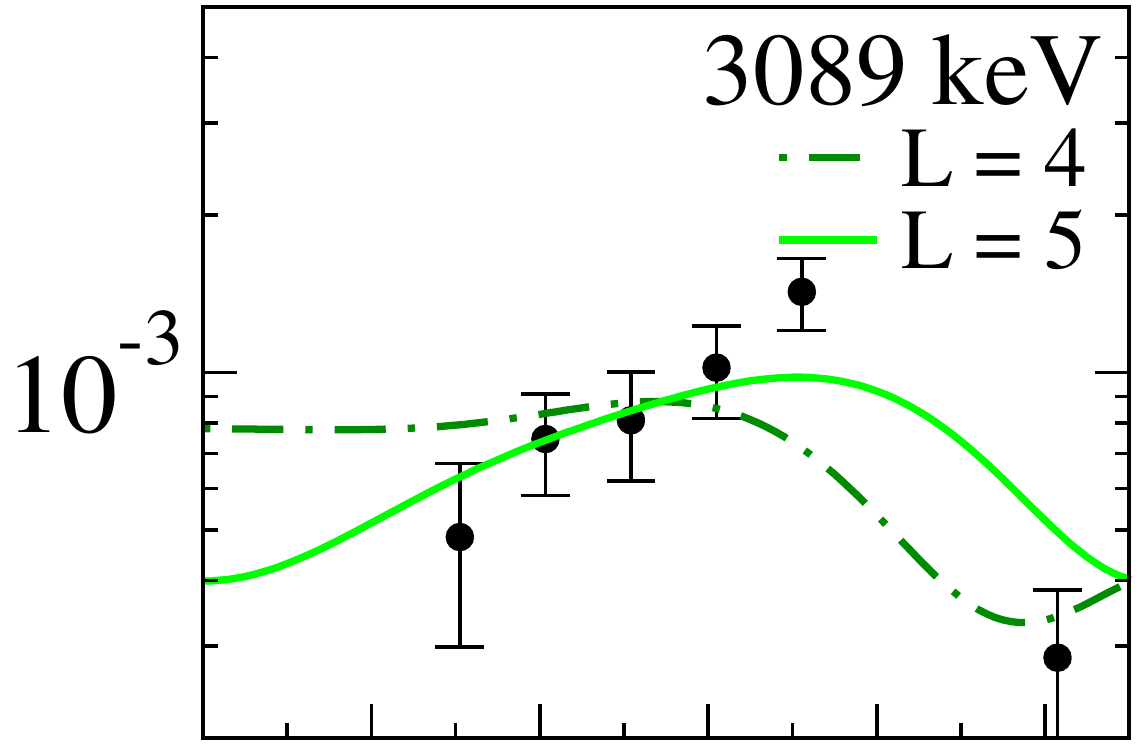}
 \includegraphics[width=0.153\textwidth, height=0.09\textheight]{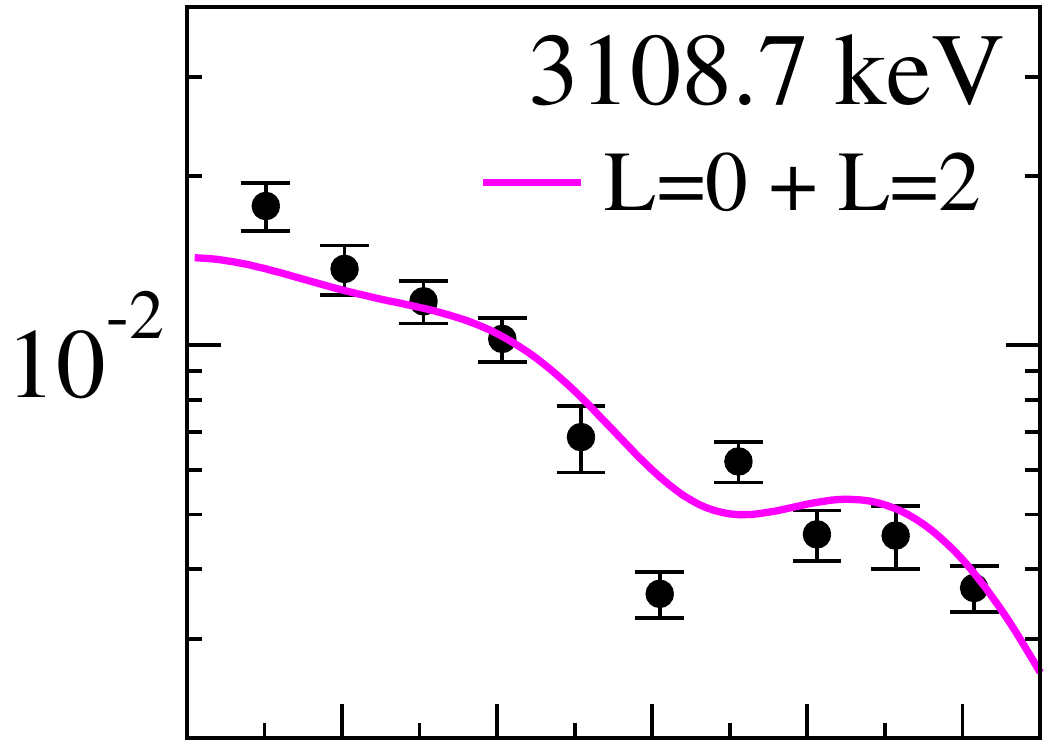}
 \includegraphics[width=0.153\textwidth, height=0.09\textheight]{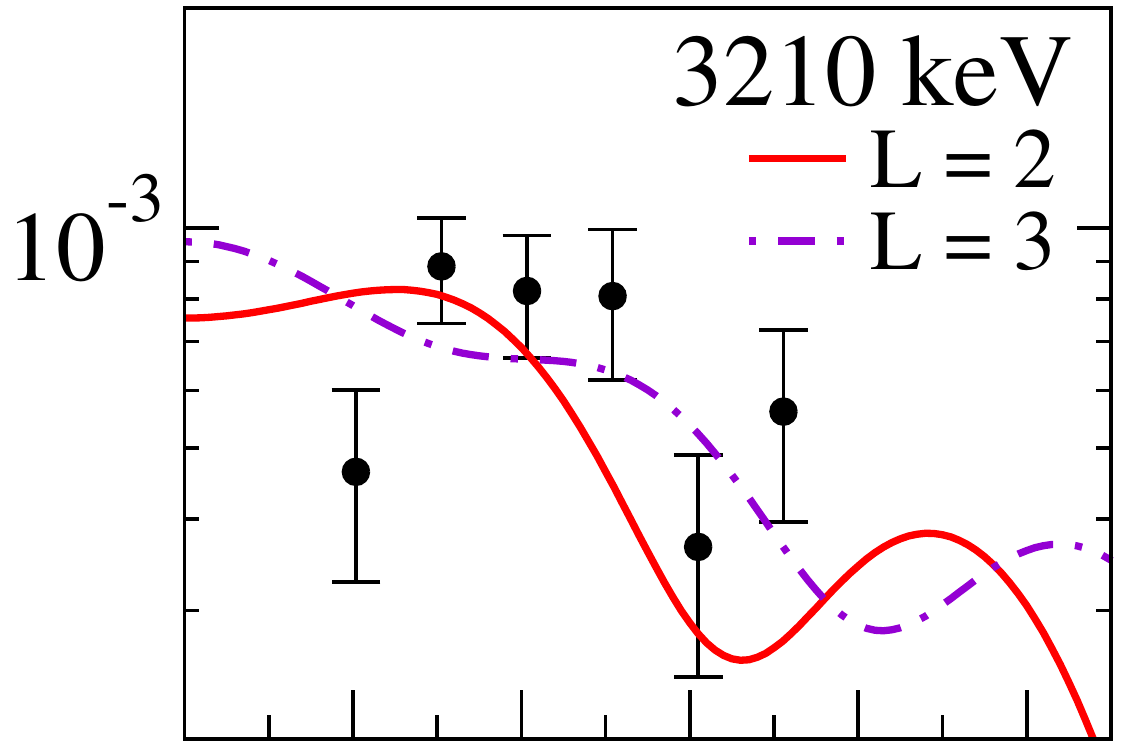}\\

 \includegraphics[width=0.153\textwidth, height=0.09\textheight]{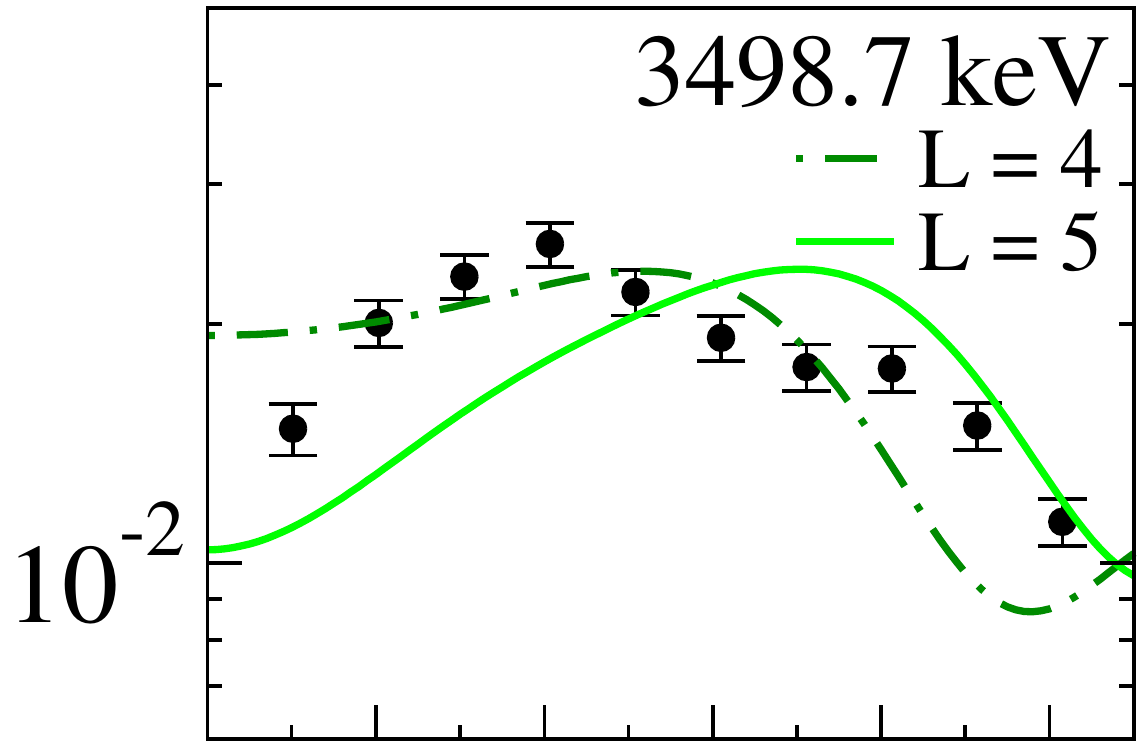}
 \includegraphics[width=0.153\textwidth, height=0.09\textheight]{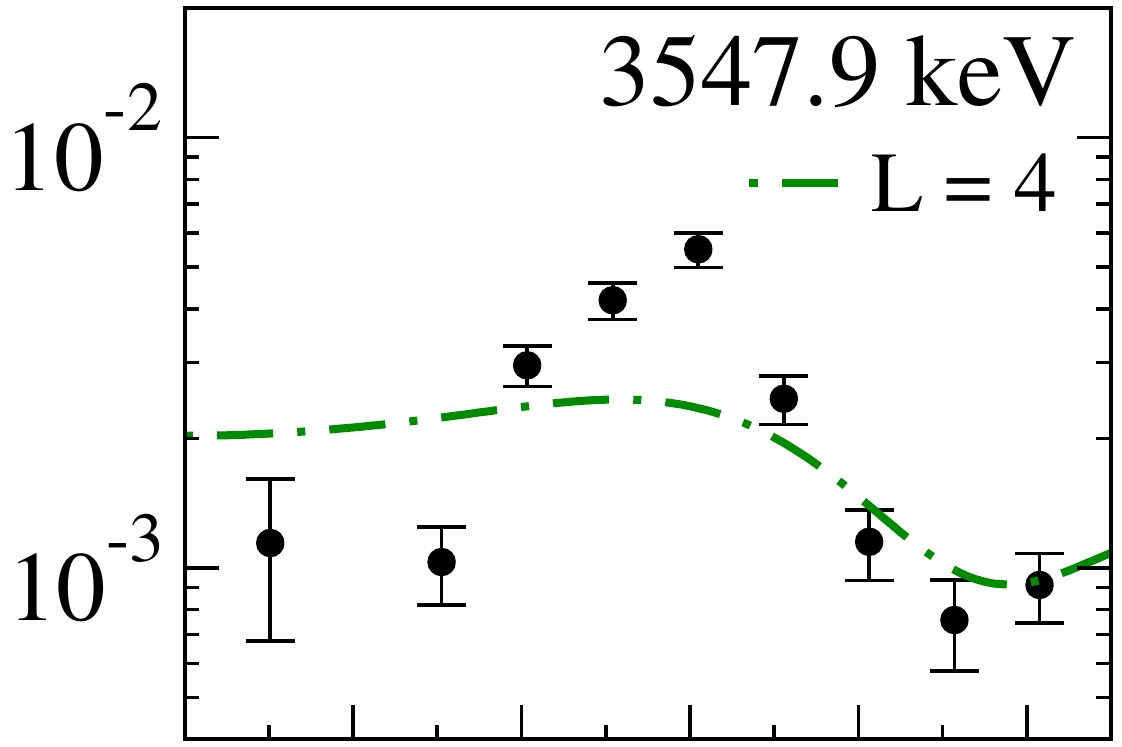}
 \includegraphics[width=0.153\textwidth, height=0.09\textheight]{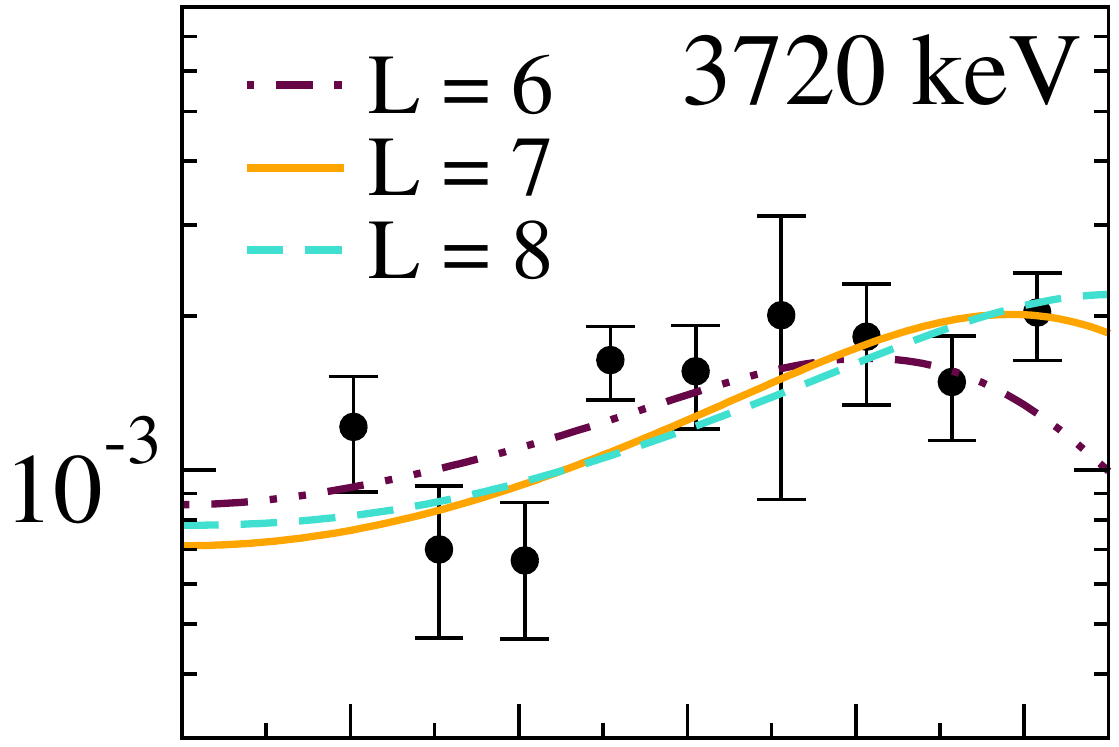}\\
 \hspace*{-0.3 cm}
 \includegraphics[width=0.163\textwidth, height=0.09\textheight]{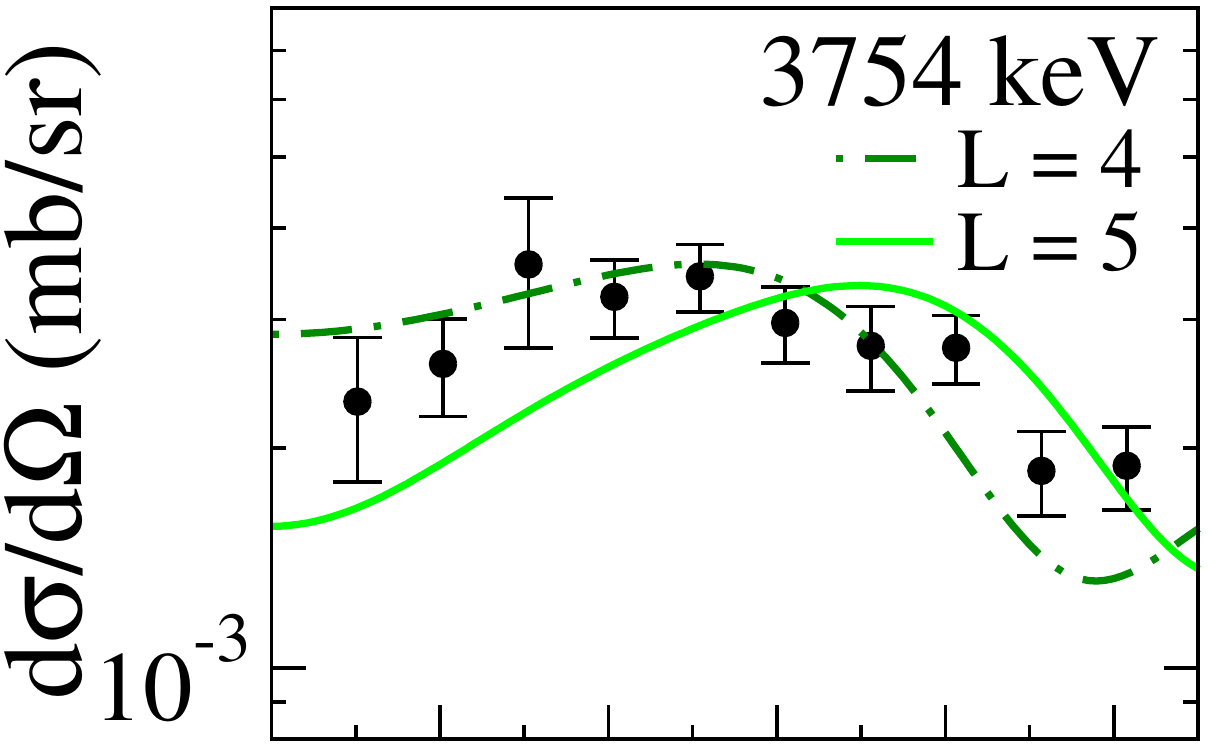} 
 \includegraphics[width=0.153\textwidth, height=0.09\textheight]{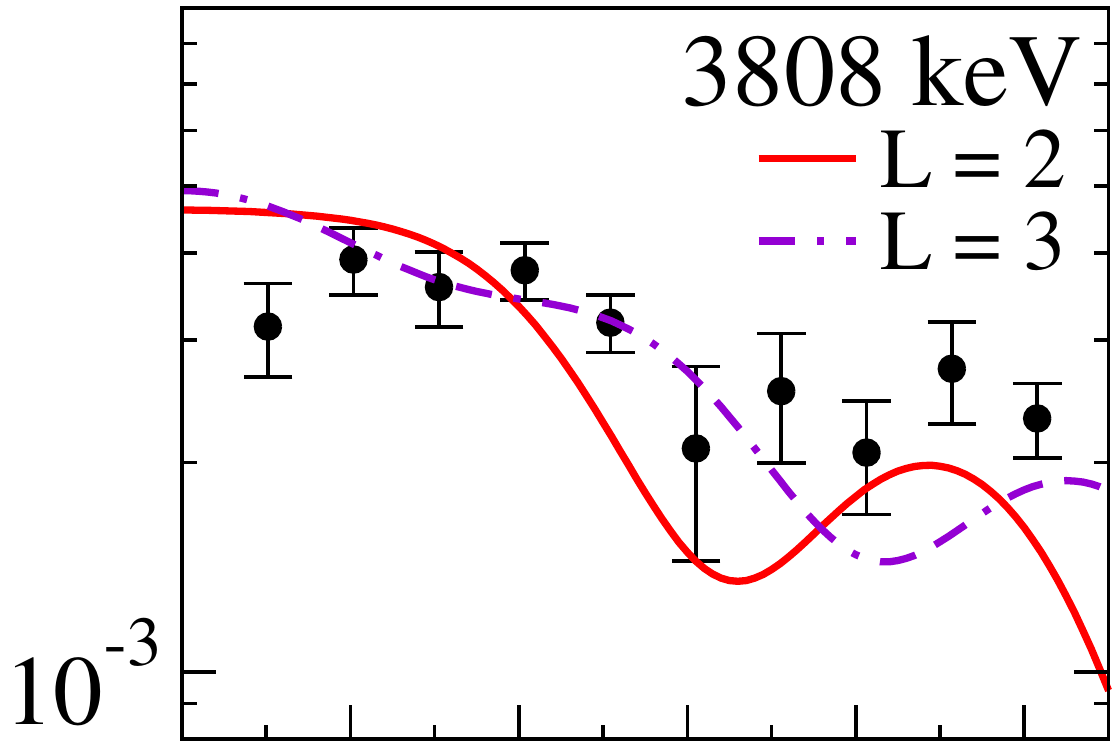}
 \includegraphics[width=0.153\textwidth, height=0.09\textheight]{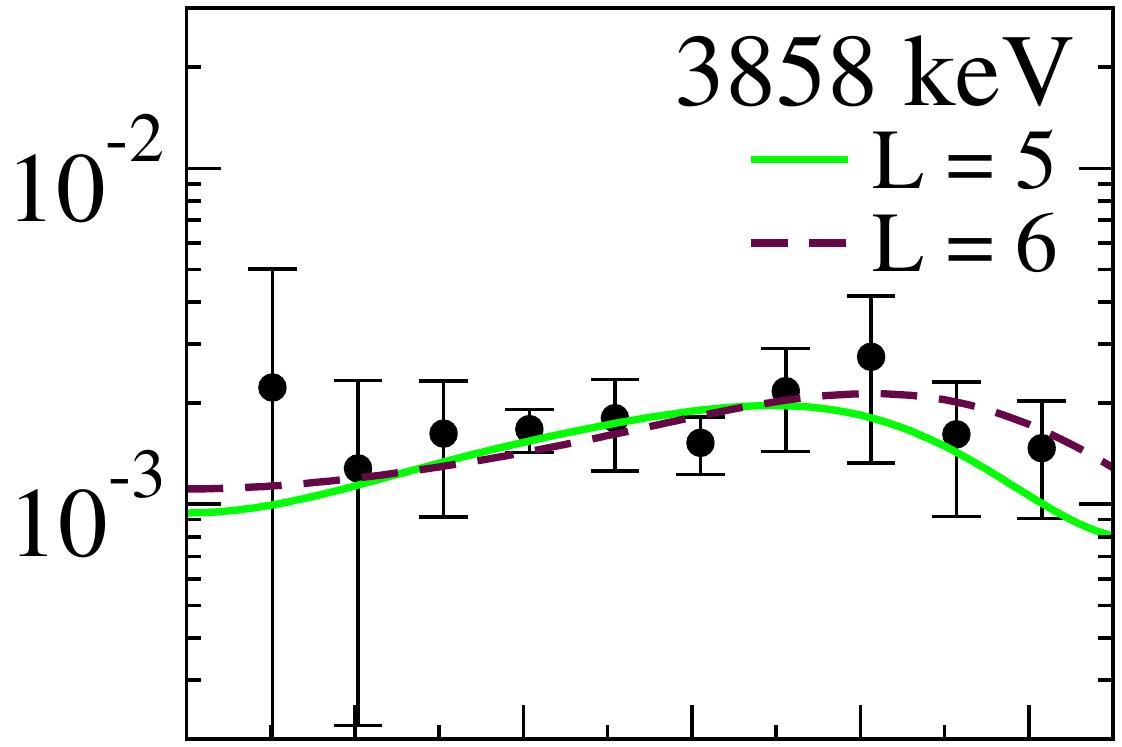}\\
 
 \includegraphics[width=0.153\textwidth, height=0.09\textheight]{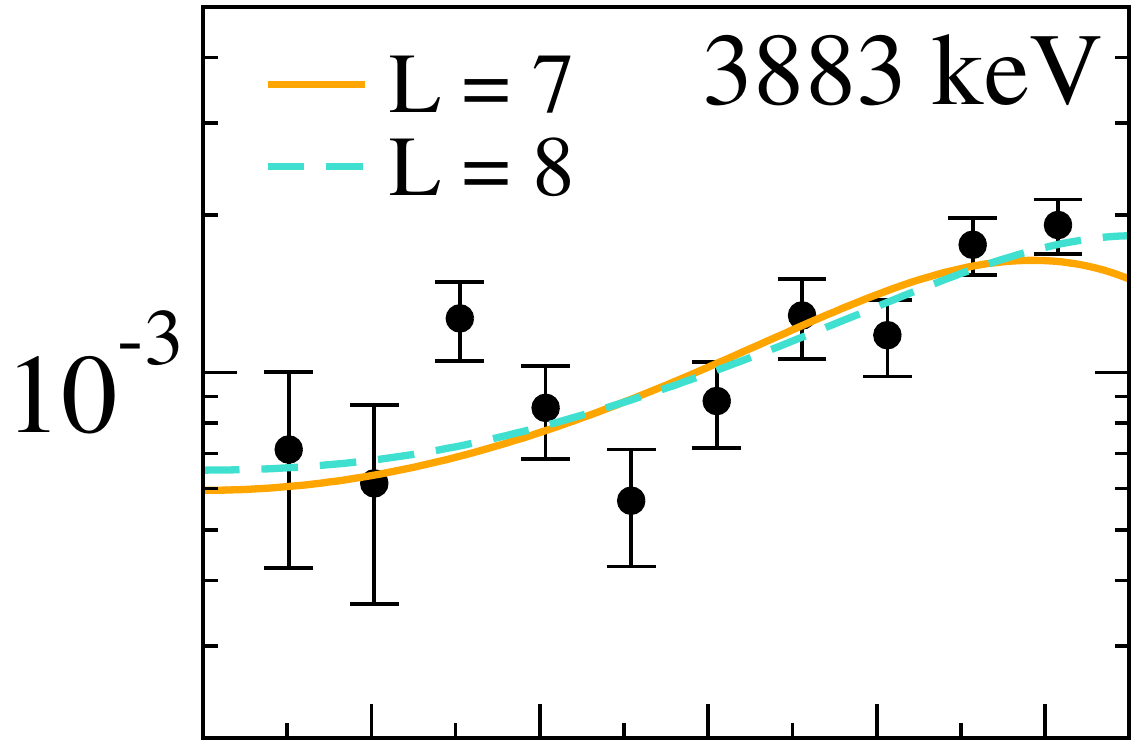}
 \includegraphics[width=0.153\textwidth, height=0.09\textheight]{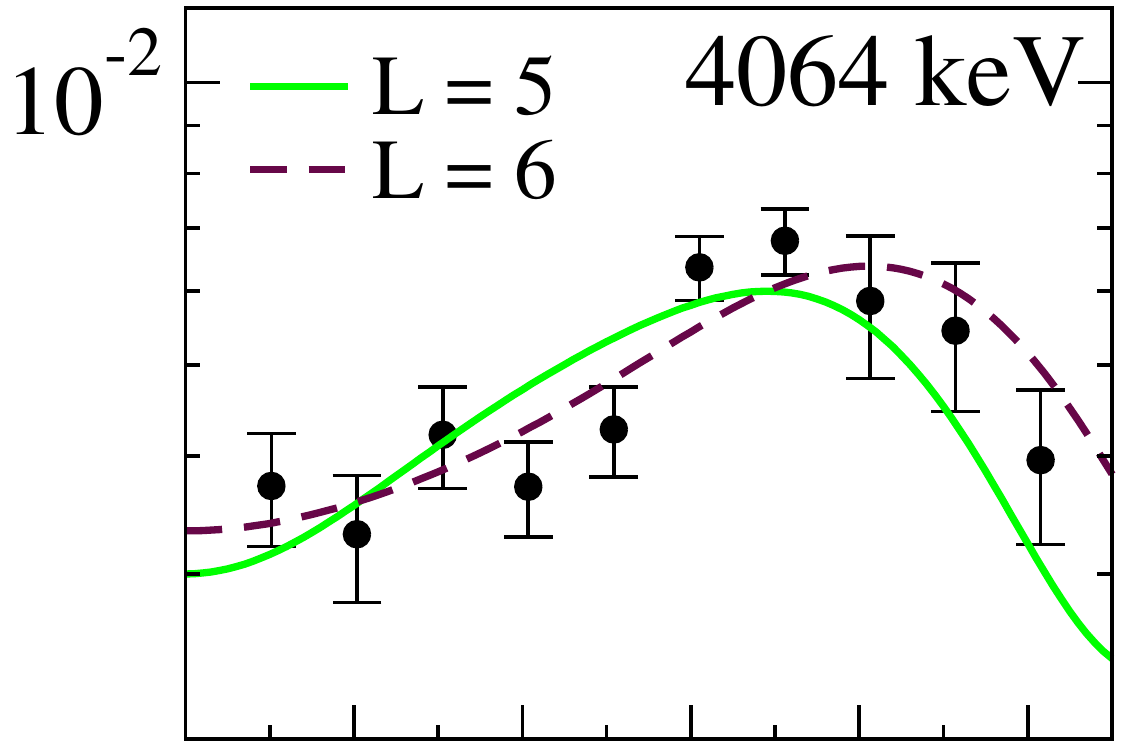} 
 \includegraphics[width=0.153\textwidth, height=0.09\textheight]{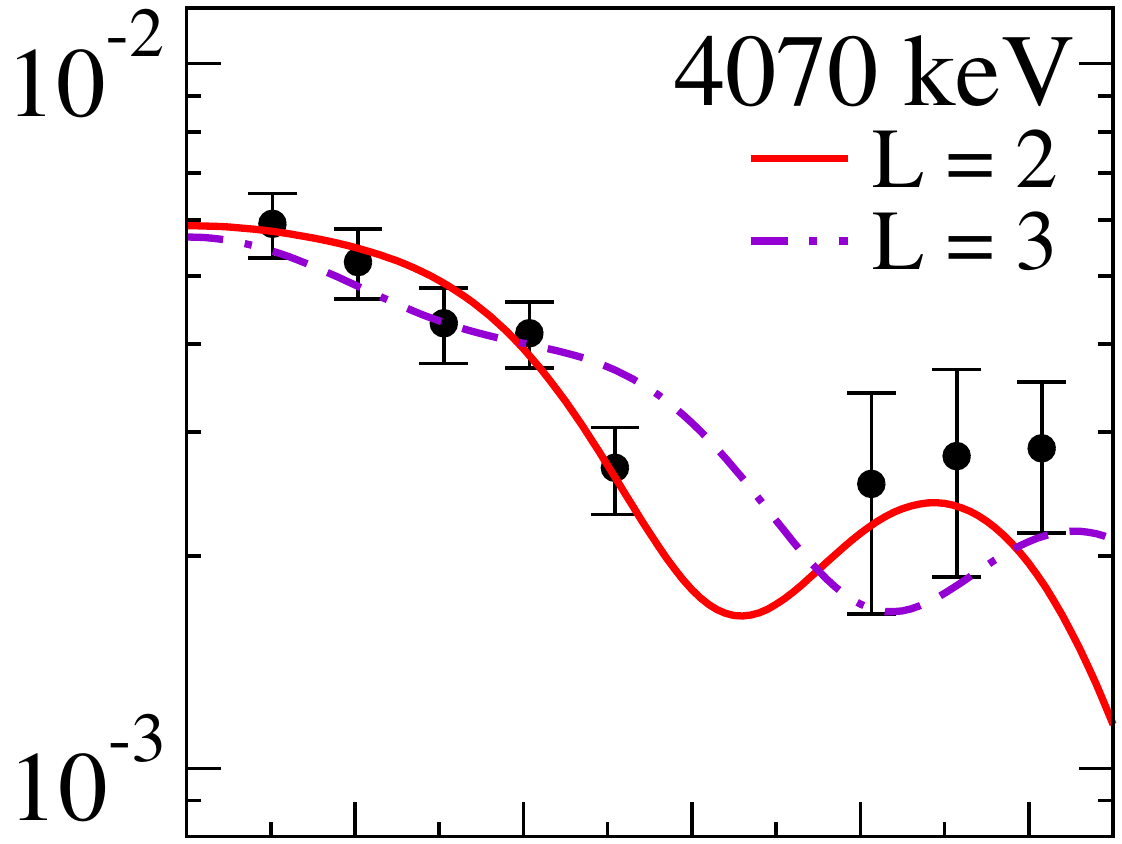}\\
 
 %   \includegraphics[width=0.165\textwidth, height=0.09\textheight]{Images/4268}
 % %   \includegraphics[width=0.153\textwidth, height=0.09\textheight]{Images/4487_fake}
 %   \includegraphics[width=0.153\textwidth, height=0.09\textheight]{Images/4292}\\
 
 %   \includegraphics[width=0.165\textwidth, height=0.12\textheight]{Images/4383} 
 %   \includegraphics[width=0.153\textwidth, height=0.12\textheight]{Images/4451}\\
 
 \includegraphics[width=8.5 cm]{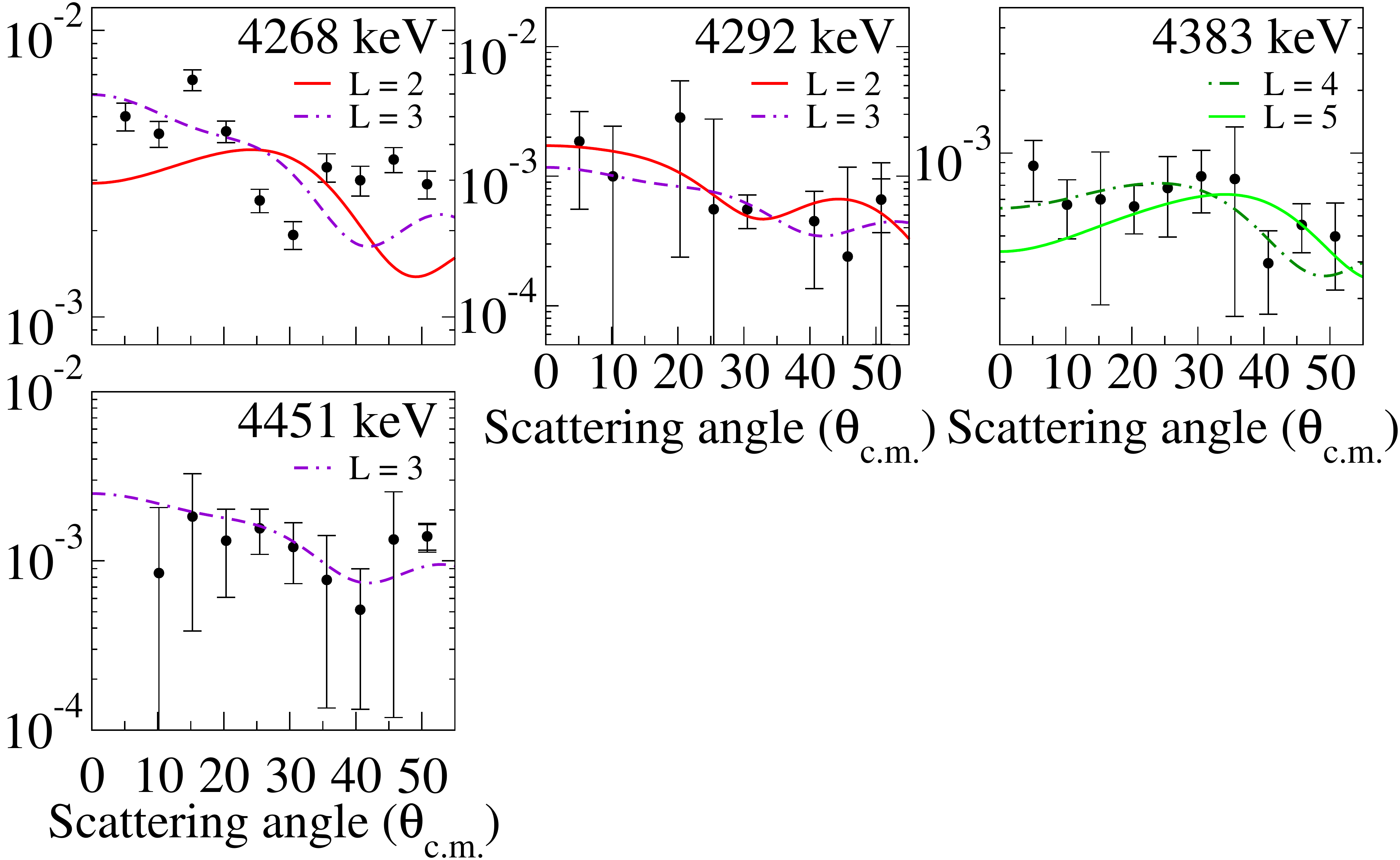}

\caption{\label{Fig:138Ba_pt_multiple_assignment}{Experimental angular distributions for states where multiple DWBA predictions agree reasonably well with the experimental data.}}
\end{figure}

%   \noindent The difference in the shape of the angular distribution for an $L=4$ and $L=5$ transfers is subtle. For a $4^+$ state DWBA predicts a  maximum for the distribution  between $25^\circ- 35^\circ$ while for the $L=5$ transfer, the maximum is between $30^\circ - 40^\circ$. 
%   % Unlike for an $L=0$ transfer, in the case of $L=4~\rm{or}~5$,  for $\theta < 60^\circ$, the  cross section varies  within  an order of magnitude.
%   % Additionally, the experimental DWBA predicts  the first maximum for the $4^+$ state at $\sim 15^\circ$, experimental angular distributions do not always follow this pattern. 
%   Therefore to distinguish a $4^+$ from a $5^-$ we had to compare the experimental angular distributions with DWBA predictions for both $L=4$ and $L=5$ transfers. One such example is shown in Fig.~\ref{Fig:138Ba_pt_4+}. By carefully following this methodology, 
We observe a total of eight $4^+$ states, including the known levels at 1866.611(18)~keV and 2053.892(18)~keV~\cite{Mccutchan2018}. A few of these are discussed below.
%\\
  % 
  \begin{figure}[t]
   \includegraphics[width=8.4 cm]{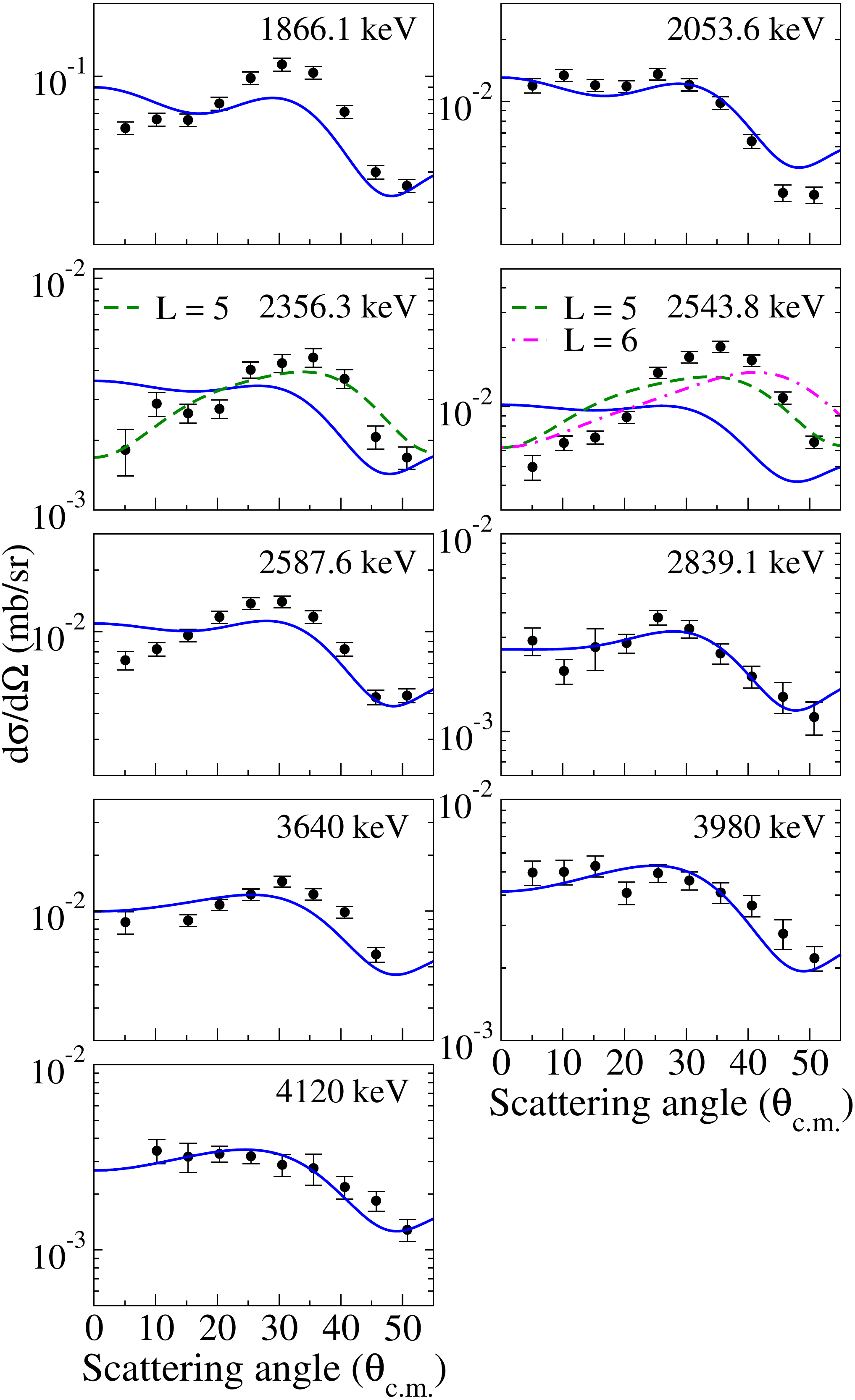} \\
   \caption{\label{Fig:138Ba_pt_4+}{Angular distributions of states for which we could make a definite $J^\pi=4^+$ assignment. The blue lines represent the $L = 4$ normalized DWBA cross sections. The green and magenta dashed lines are included for comparison. }}
  \end{figure}
%\\ 
\\
\\
{\bf $\mathbf{{{\bf E_x}}=2356.3}$~keV:} This is a well-known $4^+$ state~\cite{Mccutchan2018} observed in $(n,\gamma)$, $(n,n'\gamma)$ and $^{136}$Cs $\beta$-decay studies~\cite{Islam1990, Mukhopadhyay2008, Bargholtz1973}. The spin and parity of the state was ascertained from both $\gamma$-ray angular distribution~\cite{Mukhopadhyay2008} and directional correlation~\cite{Bargholtz1973} measurements, which showed strong $E2$ components in the $4^+$~$\to$~$2^+$~$\to$~$0^+$ transitions.   However our analysis shows that the $L=5$ transfer is more consistent with the measured distribution. Therefore the possibility of another closely spaced peak in the region cannot be ruled out.
\\ \\    
{\bf $\mathbf{{{\bf E_x}}=2543.8}$~keV:} The latest NDS~\cite{Mccutchan2018} lists this state at 2544.481(24)~keV, with $J^\pi=4^+$. This level has been observed to make an exclusive $\gamma$-ray transition to the $2_2^+$ state, with an $E2$ multipolarity~\cite{Mukhopadhyay2008}.   
However our angular distribution is in better agreement with both $L=5$ and $L=6$ transfer. Therefore the possibility of another closely spaced peak in the region cannot be ruled out. 
 \\ \\   
   {\bf $\mathbf{{{\bf E_x}}=2587.6}$~keV:} The NDS lists this state with $J^\pi = (5)^+$~\cite{Mccutchan2018}.  $^{136}{\rm Ba}(n,n'\gamma)$ excitation function and angular distribution data suggest spin 5 or $6^+$ for the state~\cite{Mukhopadhyay2008}. In contrast, our angular distribution agrees with an $L=4$ transfer. We therefore assign this state $J^\pi=4^+$. This assignment would be consistent with the $M1+E2$ nature of the $\gamma$ ray transition to the $4_1^+$ state, as reported in Ref.~\cite{Mccutchan2018}. 
%    at larger angles but is better reproduced by an $L=5$ transfer at the smaller angles.  We therefore tentatively assign a $J^\pi=(4^+,5^-)$ to this state.
\\ \\   
    {\bf $\mathbf{{{\bf E_x}}=2839.1}$~keV:} The NDS~\cite{Mccutchan2018} assigns this level a tentative spin-parity of $(4^+)$. Our measured differential cross sections for this level are consistent with $J^\pi=4^+$, which clarifies the ambiguity regarding this state. 
% \\ \\ 
%   {\bf $\mathbf{{{\bf E_x}}=3640}$~keV} No excited states are reported at this excitation energy in the evaluated or unevaluated nuclear datasets~\cite{Mccutchan2018}. Our analysis favors the angular distribution for $J^\pi=4^+$. 
\\ \\ 
   $\mathbf{ E_x}=$ {\bf 3640 and 4120~keV:} No excited states are reported at these energies~\cite{Mccutchan2018}. Our measured differential scattering cross sections are well reproduced assuming that these states have $J^\pi = 4^+$.
   \\ \\
   {\bf $\mathbf{{{\bf E_x}}=3980}$~keV:} The NDS lists a level at 3979.76(20)~keV, with $J=(1)$~\cite{Mccutchan2018}. This is based on prior observations of a 3980 $\to$ 0~keV transition~\cite{Islam1990} and an excitation of the state via the $(\gamma,\gamma')$ reaction~\cite{Metzger1978,Massarczyk2012}. As evident in Fig.~\ref{Fig:138Ba_pt_multiple_assignment}, an $L=4$ transfer reproduces our experimental data reasonably well. In light of the above, we conclude that the corresponding peak might have an unresolved doublet.  
   
% % 
\subsection{\label{Sect:5-_states} $\mathbf{J^\pi=5^-}$ states}
%   \vspace*{-0.6 cm}
  \begin{figure}[h!]
   \includegraphics[width=8.6 cm]{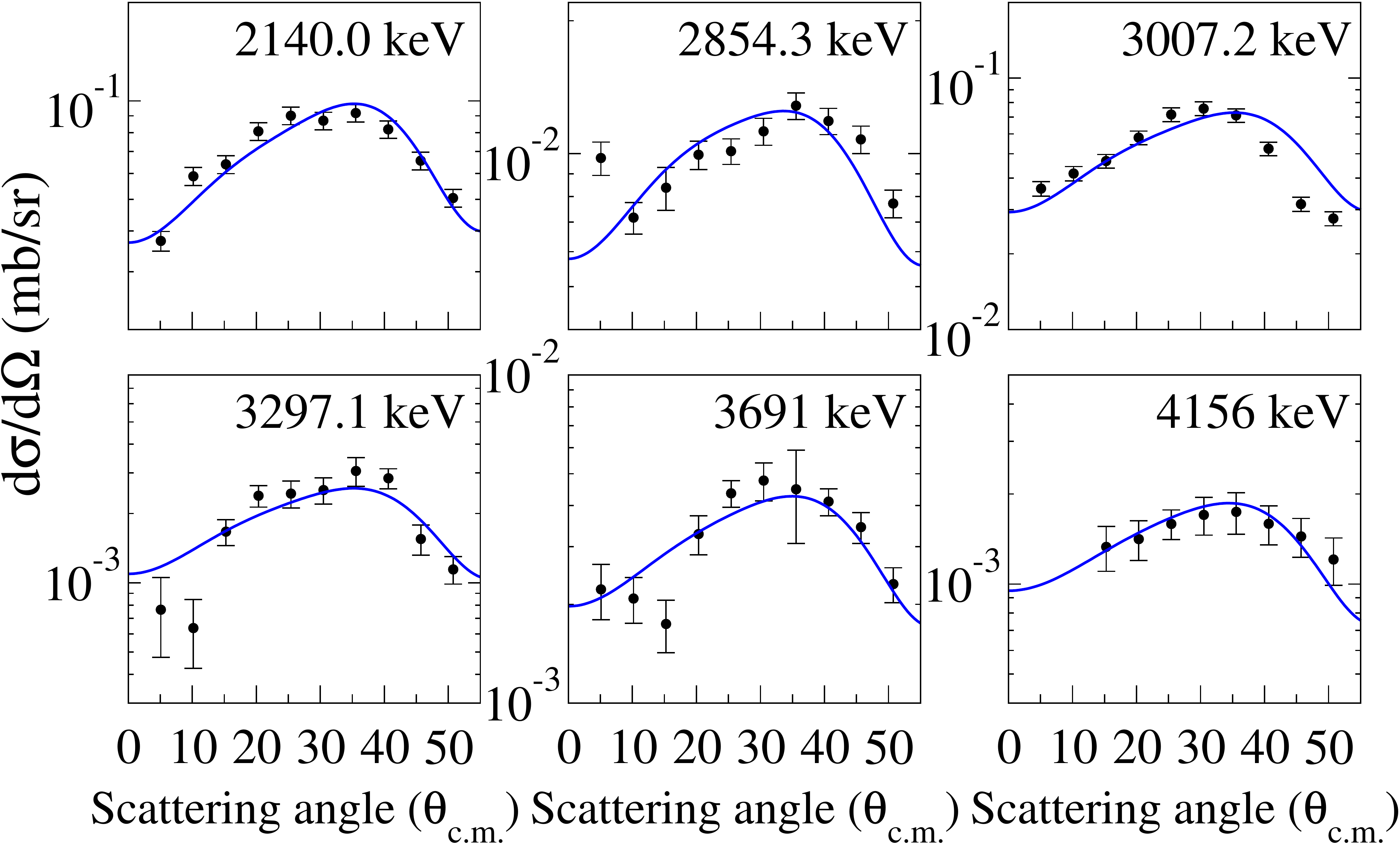}
   \caption{\label{Fig:138Ba_pt_5-}{Experimental angular distributions for all $5^-$ states observed in this work. Normalized DWBA cross sections for the $L=5$ transfer is shown in blue.}}
  \end{figure}
% \\  \\
  \noindent The nuclear data sheets for $^{136}$Ba reports a single $5^-$ state at 2140.237(18)~keV~\cite{Mccutchan2018}. As described below, we identify five more.
\\ \\ 
% 
% \\   
$\mathbf{E_x=}$ {\bf 2854.3, 3007.2, 3297.1 and 4156~keV:} The angular distribution for all these states agree with DWBA predictions for $L=5$ transfer. 
\\ \\    
{\bf $\mathbf{{{\bf E_x}}=3691}$~keV:} This level was assigned a tentative spin assignment of 1 to 3~\cite{Mccutchan2018}, which is consistent with the previous observation~\cite{Islam1990} of a $\gamma$-ray transition to the $2_1^+$ state at 818.5~keV. In comparison, our measured angular distribution for this state is in excellent agreement with DWBA results assuming an $L=5$ transition. This indicates the presence of an unresolved nearly-degenerate $5^-$ state that is strongly populated in the $^{138}{\rm Ba}(p,t)$ reaction. 
% \\ \\
%  {\bf $\mathbf{{{\bf E_x}}=4156}$~keV} No excitation energy is reported in the NDS in the vicinity of  4156~keV~\cite{Mccutchan2018}. In this work, the experimental angular distribution matches well with the DWBA prediction for an $L=5$ transfer.
% 
% 
  \subsection{\label{Sect:6+_states} $\mathbf{J^\pi=6^+}$ states}
  %\vspace*{-0.6 cm}
  \begin{figure}[h!]
   \includegraphics[width=8.4 cm]{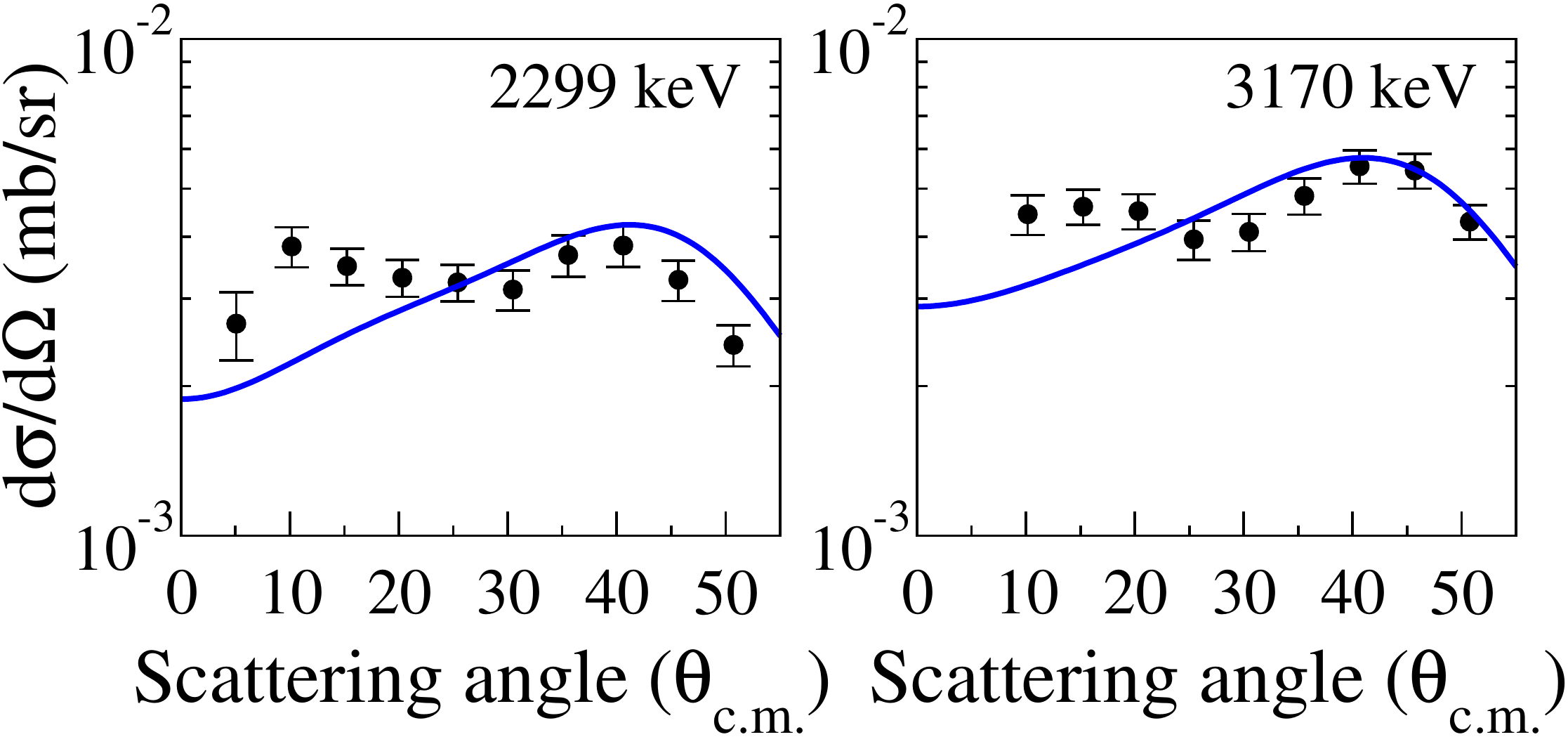}
   \caption{\label{Fig:138Ba_pt_6+}{Experimental angular distribution for the two $6^+$ states observed in this work and normalized DWBA curves for $L=6$ transfer. }} %States with tentative $6^+$ assignments are presented in a different figure. Dashed green lines are DWBA curves for $L=5$ transfer.}}
  \end{figure}
  \noindent The $6_1^+$ state at 2207.147(18)~keV~\cite{Mccutchan2018} is not observed in this work. This is not surprising as the state is fed via a Gamow-Teller transition in $^{136}$Cs $\beta^-$ decay, with a measured $\log$ $ft$ value of around 5.9~\cite{Griffioen1975}. This indicates a $(d_{3/2})_n$ $\to$ $(d_{5/2})_p$ transformation~\cite{Bargholtz1973}, with a dominant $(g_{7/2}, d_{5/2})$ proton configuration for the state. Therefore, the $6_1^+$ state is not expected to be strongly populated in $^{138}{\rm Ba}(p,t)$. This non-observation validates the proton-dominant configuration for the level and is consistent with the small $\log~ft$ value for the decay. We observe two other $6^+$ states that are described below.
  \\
  \\
  $\mathbf{E_x = }$ {\bf 2299.0~keV:} The NDS makes a tentative $(6^-)$ assignment for this state~\cite{Mccutchan2018}. However, as apparent in Fig.~\ref{Fig:138Ba_pt_6+}, we observe a reasonably strong population of this state, with an angular distribution that is consistent with $L = 6$ transfer. Therefore we assign $J^\pi = 6^+$ for this level.
  \\
  \\
  $\mathbf{E_x = }$ {\bf 3170.0~keV:} This level is observed for the first time in our experiment. The measured differential cross sections agree well with an $L = 6$ angular distribution.
% The state at 3858~keV is weakly populated in this work and thus has larger uncertainities. Within the error bars, the angular distributions for this state is well reproduced by both an $L=6$ as well as an $L=5$ transfer. Other states  where we make tentative $6^+$ assignments are discussed later.
%   
%   
  \subsection{\label{Sect:7-_states} $\mathbf{J^\pi=7^-}$ states}
  %\vspace*{-0.6 cm}
  \begin{figure}[h]
   \includegraphics[width=8.4 cm]{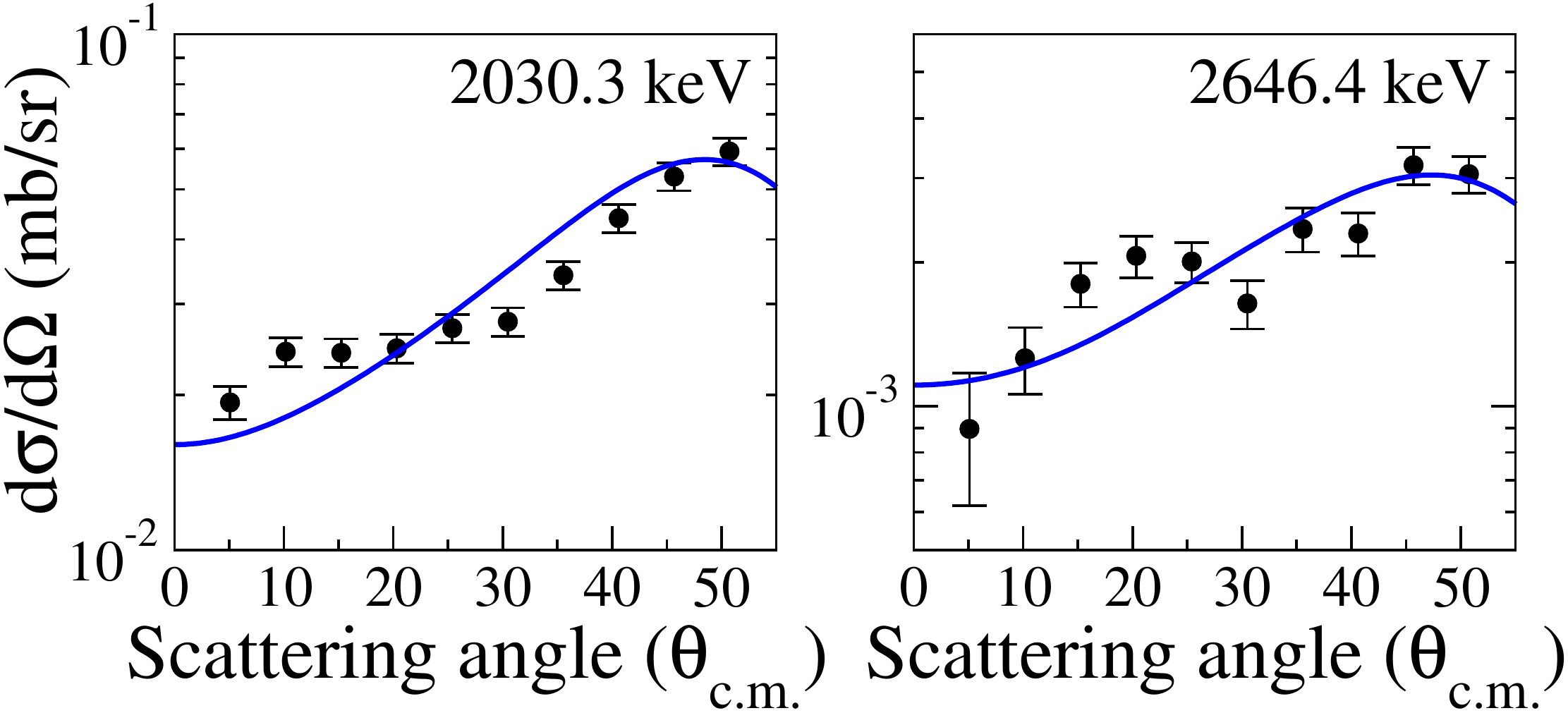}
   \caption{\label{Fig:138Ba_pt_7-}{Experimental angular distribution and normalized $L = 7$ DWBA predictions (solid blue line) for the two $7^-$ states observed in this work. }} 
  \end{figure}
  We observe two $7^-$ states, which includes the known $7_1^-$ state at 2030.535(18)~keV~\cite{Mccutchan2018}.
  \\
  \\
  $\mathbf{E_x = }$ {\bf 2646.4~keV:} This level is not listed  in the NDS~\cite{Mccutchan2018}. As shown in Fig.~\ref{Fig:138Ba_pt_7-}, its angular distribution is consistent with $L = 7$, similar to that of the known $7_1^-$ level. 
% 
% \subsection{\label{Sect:Tentative_assignments}Tentative assignments}
\subsection{\label{Sect:Tentative}Tentative assignments}
In addition to the above, there are several states for whom we could not make conclusive spin-parity assignments. This is because in these cases the measured angular distributions were either consistent with multiple values for $L$ transfer, did not agree with any particular DWBA curve or lacked  statistics. We discuss these observed states by grouping them in two different categories described below.  
\subsubsection{\label{Sect:1-_states}$\mathbf{J^\pi=(1, 2^+)}$ {\bf states}} 
\begin{figure}[h!]
 
 \includegraphics[width=8.6 cm]{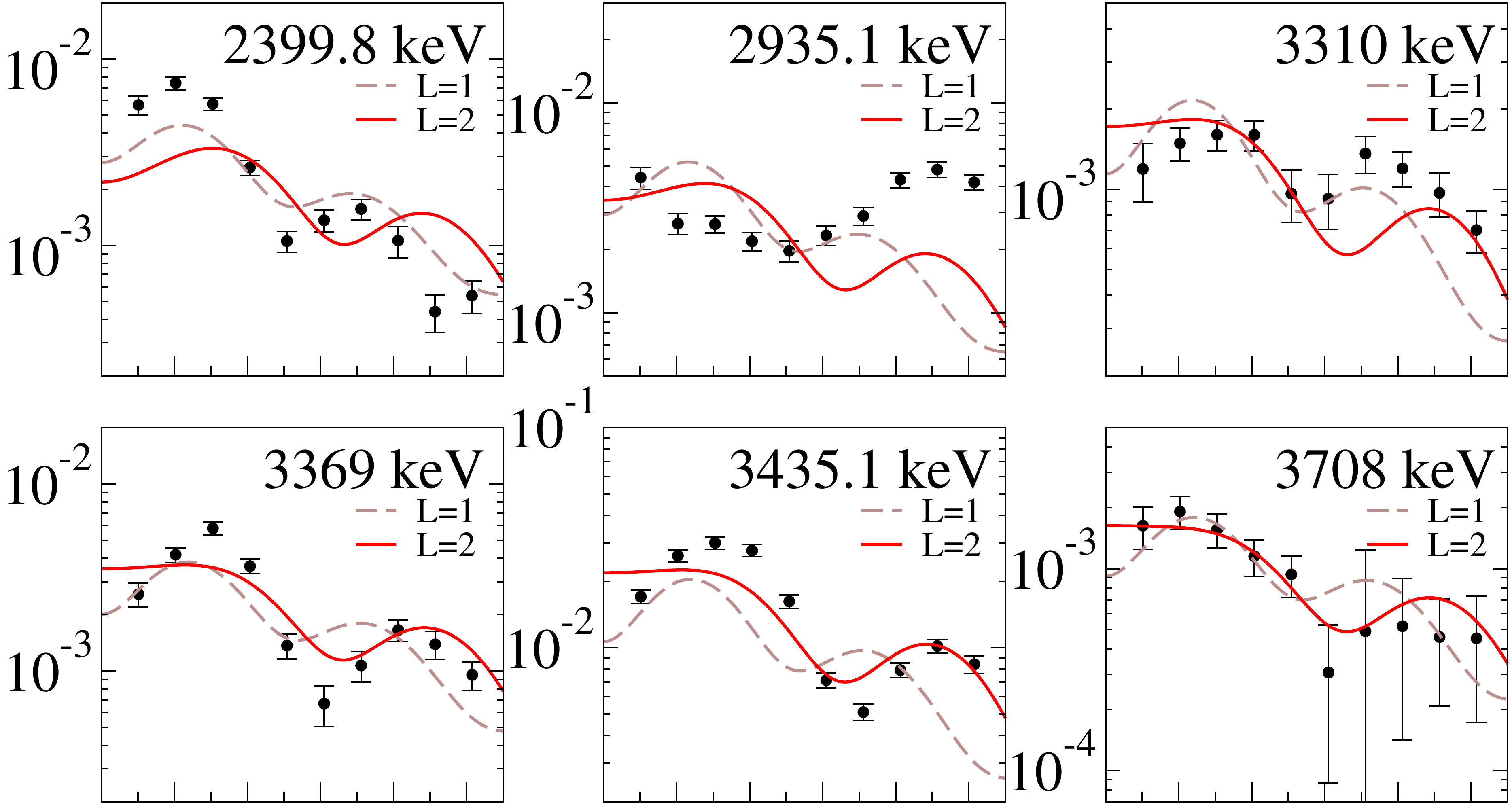} 
 \hspace*{-0.35 cm}
 \includegraphics[width=8.85 cm]{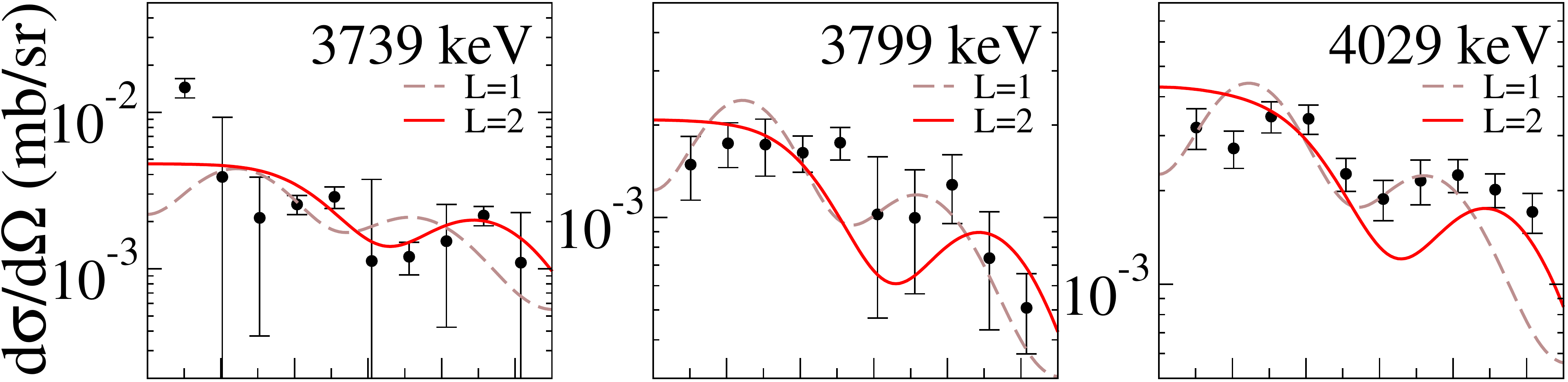}
 \hspace*{-0.035 cm}
 \includegraphics[width=8.85 cm]{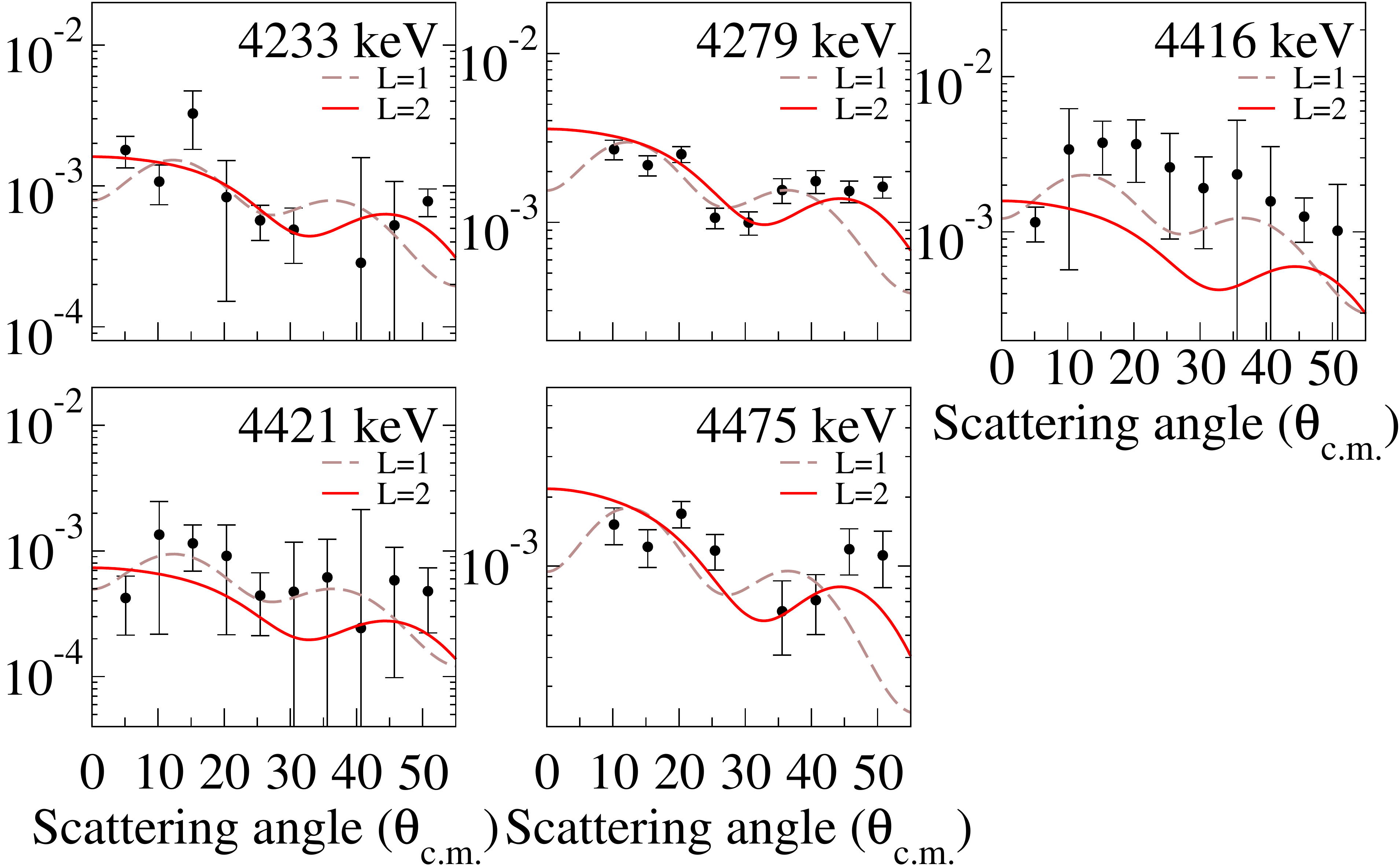}

 %  \includegraphics[width=0.22\textwidth, height=0.21\textwidth]{Images_pt_long_Sept2020/tentatives/2400}
 %  \includegraphics[width=0.22\textwidth, height=0.21\textwidth]{Images_pt_long_Sept2020/tentatives/2935}
 
 %  \includegraphics[width=0.165\textwidth, height=0.09\textheight]{Images/2400} 
 %  \includegraphics[width=0.153\textwidth, height=0.09\textheight]{Images/2935}
 %  \includegraphics[width=0.153\textwidth, height=0.09\textheight]{Images/3310}\\
 %  
 %  \includegraphics[width=0.165\textwidth, height=0.09\textheight]{Images/3369}
 %  \includegraphics[width=0.153\textwidth, height=0.09\textheight]{Images/3435}
 %  \includegraphics[width=0.153\textwidth, height=0.09\textheight]{Images/3708}\\
 %  
 %  \includegraphics[width=0.165\textwidth, height=0.09\textheight]{Images/3739}
 %  \includegraphics[width=0.153\textwidth, height=0.09\textheight]{Images/3799}
 %  \includegraphics[width=0.153\textwidth, height=0.09\textheight]{Images/4029} \\
 
 %  \hspace*{-3.0 cm} 
 %  \includegraphics[width=0.165\textwidth, height=0.09\textheight]{Images/4233}
 %  \includegraphics[width=0.153\textwidth, height=0.09\textheight]{Images/4279}\\
 % % 
 %  \includegraphics[width=0.165\textwidth, height=0.118\textheight]{Images/4416}
 %  \includegraphics[width=0.153\textwidth, height=0.118\textheight]{Images/4421}
 %  \includegraphics[width=0.153\textwidth, height=0.118\textheight]{Images/4475}\\
 
 \caption{\label{Fig:138Ba_pt_1-}{Experimental angular distribution for all the $(1,2^+)$ states observed in this work. The continuous curves are normalized DWBA cross sections. }} %  The dashed brown lines represent $L=1$ transfer and the solid red lines $L=2$ transfer. }} %The solid blue lines are normalized DWBA curves for $L=1$ transfer and the dashed green line is included for comparison. }} % for $L=2$ transfer. }}
\end{figure}

These states had angular distributions that were mainly consistent with both $L=1$ and $L=2$ transitions, as shown in Fig.~\ref{Fig:138Ba_pt_1-}. We assign these states $J^\pi=(1, 2^+)$ because strong excitations of $1^-$ levels are not expected due to the vacant $2p_{3/2}$ neutron shell above $N=82$. 
%and ii.) others that agreed with different values of $L$ transfer for the predicted distributions. For the former 
Furthermore, one cannot conclusively exclude the population of $1^+$ states via mechanisms such as sequential two-step transfer. \\ \\
%A brief discussion of these data and the other observed tentative states are described below.

\noindent{\bf $\mathbf{{{\bf E_x}}=2399.8}$~keV:} The $A=136$ NDS reports this state to have tentative $J^\pi=(1)^+$~\cite{Mccutchan2018}. Two independent $(n,n'\gamma)$ experiments determined different assignments for the spin and the parity of this state. Di\'oszegi {\it et al.}~\cite{DIOSZEGI1985395} measured $\gamma$-ray angular distribution coefficients and assigned it a $J^\pi$ of $(3^{\pm})$, while Mukhopadhyay {\it et al.}~\cite{Mukhopadhyay2008} used $(n,n'\gamma)$ excitation function data for the $\gamma$-ray transition to the $2_1^+$ state. They inferred its spin-parity to be $(1^+)$ and further concluded that the $M1$ strength for the transition was unexpectedly large.  

% We observe a reasonably significant population of this state in our reaction, which rules out a $1^+$ assignment, given the selection rules. 
Our measured angular distribution for this level is shown in Fig.~\ref{Fig:138Ba_pt_1-} and agrees well with DWBA calculations that assume $L=1$ or $L=2$ transfer. We thus assign the state $J^\pi = (1,2^+)$. 
% If this level had $J^\pi = 1^-$ it's  transition to the $2_1^+$ state would be mainly $E1$, with a $B(E1)$ value $\approx9.4\times10^{-4}$ e$^2$~fm$^2$ or about $5.5\times 10^{-4}$~W.u. This is contrary to the assumption of Mukhopadhyay {\it et al.}~\cite{Mukhopadhyay2008}. 
\\
\\
{\bf $\mathbf{{{\bf E_x}}=2935.1}$~keV:} This state is reported in the NDS~\cite{Mccutchan2018} with no spin-parity assignment. Our experimental data agree well with predictions for $L=1$ or $L=2$ transfer. We therefore assign $J^\pi=(1,2^+)$ for this state.
\\ \\ 
{\bf $\mathbf{{{\bf E_x}}=3310}$ keV:} No excited state is reported at this energy~\cite{Mccutchan2018}.  Our measured angular distribution for this state is reproduced well by DWBA curves that assume $L=1$ or $L=2$ transfer. 
%   as well as an $\ell=2$ transfer.   However, chi-squared analysis gives a minimum for the $\ell=2$ transfer.
Hence we assign this state a tentative spin-parity of $(1,2^+)$.
\\ \\ 
{\bf $\mathbf{{{\bf E_x}}=3369}$~keV:} This state is listed as a $J = 1$ level in the NDS~\cite{Mccutchan2018}. Our measured angular distribution agrees well with DWBA calculations for both $L=1$ and $L=2$ transfers. Hence we assign $J^\pi=(1,2^+)$ to this state. 
%    : $1^-, 2^+$. Previous assignment - 1
\\ \\    
{\bf $\mathbf{{{\bf E_x}}=3435.1}$~keV:} This state has known $J^\pi = 1^-$~\cite{Mccutchan2018}. A resonant photon scattering experiment that measured $\gamma$-ray angular distributions and linear polarizations determined the radiative transition to the ground state to be $E1$~\cite{Metzger1978}. However, as apparent in Fig.~\ref{Fig:138Ba_pt_1-}, we are unable to conclusively  rule out $L = 2$ transitions from our data. Because of this one cannot rule out an unresolved doublet at this energy. Therefore we assign $J^\pi=(1,2^+)$ for this state. 
\\ \\
{\bf $\mathbf{{{\bf E_x}}=3708}$~keV:} Two nearly degenerate states are reported in the NDS around this energy, at $E_x = 3706.1$(6) and 3706.4(3)~keV respectively~\cite{Mccutchan2018}. 
The measured angular distribution for our triton peak does not allow us to distinguish between a $J^\pi=1$ or $2^+$ assignment. Thus we tentatively assign our identified 3708~keV level $J^\pi=(1,2^+)$.
\\ \\     
{\bf $\mathbf{{{\bf E_x}}=3739}$~keV:} This state is also reported for the first time in this work. Our measured angular distribution resembles an $L=2$ transfer but since the cross sections have large uncertainties, one cannot neglect an $L=1$ component. We thus make a tentative assignment of $J^\pi=(1, 2^+)$.
\\ \\ 
{\bf $\mathbf{{{\bf E_x}}=3799}$~keV:} The NDS~\cite{Mccutchan2018} lists a state at 3795.34(15)~keV, with an assigned spin-parity of $(1^-,2^+)$, based on $\gamma$-ray angular distribution measurements~\cite{Islam1990}. This is consistent with the results of our analysis. We conservatively assign this state $J^\pi=(1,2^+)$. 
\\ \\    
{\bf $\mathbf{{{\bf E_x}}=4029}$~keV:} There is no reported state at this excitation energy in the $A=136$ NDS~\cite{Mccutchan2018}. This state is  weakly populated in $^{138}{\rm Ba}(p,t)$ and the measured angular distribution is found to be consistent with both $L=1$ and $L=2$ transfer. We thus assign this state $J^\pi = (1,2^+)$.
\\ \\ 
{\bf $\mathbf{{{\bf E_x}}=4233}$~keV:} Recent $^{136}$Ba$(\gamma,\gamma')$ work~\cite{Massarczyk2012} reports a level at 4231.17(20)~keV with spin 1. Here, this state is weakly populated and consequently the uncertainties in the measured differential cross sections are large. The DWBA predictions for $L=1,2,3$ transfers all agree reasonably well with the measured angular distribution. %However, the $\chi^2_{min}$ corresponds to the $L=2$ transfer. 
However, the $3^-$ assignment is unlikely as this state was previously observed in a $^{136}{\rm Ba}(\gamma,\gamma')$ experiment~\cite{Massarczyk2012}, making a transition to the ground state. We therefore assign this state $J^\pi = (1,2^+)$.  
\\ \\ 
{\bf $\mathbf{{{\bf E_x}}=4279}$~keV:} No level is reported at this energy in the NDS~\cite{Mccutchan2018}. Our measured angular distribution indicates tentative values of $J^\pi =(1, 2^+)$. 
\\ \\  
{\bf $\mathbf{{{\bf E_x}}=4416}$~keV:} A level at 4413.28(10)~keV was recently observed in a $^{136}$Ba$(\gamma,\gamma')$ experiment~\cite{Massarczyk2012}, where the authors tentatively assigned it spin 1. In our work, the measured angular distribution is compatible with both $J^\pi=1$ and $2^+$ assignments.
\\ \\   
{\bf $\mathbf{{{\bf E_x}}=4421}$~keV:} This is a previously unreported state that is weakly populated. Its measured angular distribution is consistent with both $J^\pi=1$ and $2^+$. % Due to a lack of the required statistics to make conclusive spin-parity assignments.  
%Large uncertainties on the cross section make definite $J^\pi$ determination difficult. 
\\ \\  
{\bf $\mathbf{{{\bf E_x}}=4475}$~keV:} A level at 4475.18(10)~keV was recently observed in Ref.~\cite{Massarczyk2012} and tentatively assigned $J = 1$. Although our data are consistent with $L = 1$ transfer, our measured angular distributions cannot rule out $L=2$ and $L =3$ transitions for this state. As the state was populated via inelastic photon scattering~\cite{Massarczyk2012}, making a transition to the ground state, this rules out a $3^-$ assignment for the level.

\subsubsection{\label{Sect:other_states} {\bf Other states}} 
These include all levels for which the measured angular distributions agreed with predicted DWBA distributions for different values of $L$ transfer (other than $L = 1$ and 2) or did not have the required statistics/agreement with DWBA to make meaningful comparisons.

% \begin{figure}[h!]
% 
%  \includegraphics[width=0.153\textwidth, height=0.09\textheight]{Images_pt_long_Sept2020/tentatives/4292}
%  \includegraphics[width=0.153\textwidth, height=0.09\textheight]{Images_pt_long_Sept2020/tentatives/4383}\\
%  \includegraphics[width=0.153\textwidth, height=0.09\textheight]{Images_pt_long_Sept2020/tentatives/4451} \\
% 
%  \hspace*{-3.0 cm} 
%  \includegraphics[width=0.153\textwidth, height=0.102\textheight]{Images_pt_long_Sept2020/tentatives/4534}\\
% 
%  %  \includegraphics[width=0.48\textwidth]{Images_pt_long_Sept2020/all_tentative_states_Sep2020_5}
% %  \includegraphics[width=0.48\textwidth]{Images_pt_long_Sept2020/all_tentative_states_Sep2020_6}
%  \caption{\label{Fig:138Ba_pt_multiple_assignment_2} Experimental angular distributions for states where multiple DWBA predictions agree reasonably well with the experimental data. }
% \end{figure}
% 

\noindent {\bf $\mathbf{{{\bf E_x}}=2829.9}$~keV:} No levels are reported at this excitation energy in the NDS~\cite{Mccutchan2018}. Our analysis shows the distribution to be consistent with $L=7$ transfer. However, since the data are limited by statistics, one cannot ignore $L=6$ as well. Hence we tentatively assign this state a $J^\pi$ value of $(6^+,7^-)$.
\\ \\
{\bf $\mathbf{{{\bf E_x}}=2902.0}$~keV:} The NDS report a state about 3~keV higher, at 2905.0(5)~keV~\cite{Mccutchan2018}, with no spin-parity assignment. It is quite likely that we observe the same state.  
%This state was reported as $E_x=2905.8\pm1.5$~keV in the $^{135}$Ba$(n,\gamma)$ reaction in Ref.~\cite{Chrien1974}. No spin and parity assignment is reported for this state. 
Our measured angular distribution for this level agrees well with both $L=4$ and $L=5$ transfer. We thus assign $J^\pi=(4^+,5^-)$ for this state.
\\ \\
{\bf $\mathbf{{{\bf E_x}}=3044.5}$~keV:} This level is reported in the NDS as a $1^{(-)}$ state~\cite{Mccutchan2018}. 
Our measured angular distribution agrees better with a DWBA curve that assumes a combination of $L=0$ and $L=2$ transfer. %However we do not observe the characteristic dip at around $40^\circ$ that one would expect for an $L = 3$ transition. Furthermore, a previous $^{135}{\rm Ba}(n,\gamma)$ study~\cite{Islam1990} observed a strong 3044 $\to$ 0~keV transition, which is not expected for a $3^-$ state. 
Therefore we do not rule out a possible unresolved doublet at this energy and tentatively assign $J^\pi=(0^+,2^+)$ to the observed state.
\\ \\   
{\bf $\mathbf{{{\bf E_x}}=3089}$~keV:} As this state is populated quite weakly, the measured angular distribution lacks the required statistics to make definite conclusions. The shape of the distribution indicates either an $L=4$ or $L=5$ transfer. We thus tentatively assign this state $J^\pi=(4^+,5^-)$.
\\ \\
{\bf $\mathbf{{{\bf E_x}}=3108.7}$~keV:} The NDS lists a $2^+$ state at 3109.59(9)~keV. The measured angular distribution for our observed peak corresponding to this state agrees well with a combination of $L = 0$ and $L = 2$ transitions. This is most likely due to an unresolved doublet. We therefore assign this state $J^\pi = (0^+,2^+)$.
\\ \\
{\bf $\mathbf{{{\bf E_x}}=3210}$~keV:} This is most likely the known 3212.0(5)~keV state that was assigned $J^\pi = 0^{(+)}$,$1,2,3^+$ in the NDS~\cite{Mccutchan2018}. In this work, the state is weakly populated and has large uncertainties in the measured differential cross sections. Based on our DWBA comparisons, we tentatively assign $J^\pi = (2^+,~3^-)$ for this level.
\\ \\  
{\bf $\mathbf{{{\bf E_x}}=3498.7}$~keV:} There are no reported levels at this energy. The nearest states listed in the NDS are at 3505.5(9) and 3508.7(3)~keV, with $J^\pi =0^{(+)},1,2,3^+$ and $J^\pi = (4^+)$ respectively~\cite{Mccutchan2018}. Our measured angular distribution for this level is consistent with both $L = 4$ and $L = 5$ transfer. We therefore make a tentative assignment of $J^\pi = (4^+, 5^-)$ for this state.
\\ \\ 
{\bf $\mathbf{{{\bf E_x}}=3547.9}$~keV:} An unassigned excited state at 3550.70(20)~keV is reported in the NDS~\cite{Mccutchan2018}, which disagrees with our identified level by $\sim~3$~keV. A comparison of the measured angular distribution with DWBA predictions leads us to assign this state $J^\pi=(4)^+$. %({\bf Bernadette should remember to move this text and the figure to the tentative section})
\\ \\ 
{\bf $\mathbf{{{\bf E_x}}=3720}$~keV:} This state is weakly populated and reported for the first time in this work. Our analysis indicates that this state is populated via $L\geq 6$ transfer.
\\ \\    
{\bf $\mathbf{{{\bf E_x}}=3754}$~keV:} This is another newly reported state for which no information is available in the latest NDS~\cite{Mccutchan2018}. Its angular distribution is consistent with DWBA calculations for both $L=4$ and $L=5$ transfers. We thus tentatively assign this state $J^\pi=(4^+,5^-)$. 
\\ \\   
{\bf $\mathbf{{{\bf E_x}}=3808}$~keV:} No level is reported at this excitation energy in the NDS~\cite{Mccutchan2018}. In this work, given the statistics, the angular distribution corresponds to either a $2^+$ or $3^-$ state. 
\\ \\   
{\bf $\mathbf{{{\bf E_x}}=3858}$~keV:} This state is reported for the first time in this work. Its measured angular distribution corresponds to either $L=5$ or $L=6$ transfer. We therefore tentatively assign $J^\pi=(5^-,6^+)$ for this level. 
\\ \\   
% {\bf $\mathbf{{{\bf E_x}}=3868}$~keV:} There are no listed states at this energy in the NDS~\cite{Mccutchan2018}.
% In our work, the angular distribution is only well reproduced only by assuming a combination of $L = 2~(55\%)$ and $L = 6~(45\%)$ transfer. Thus, the identified state could be an unresolved doublet.
% \\ \\   
{\bf $\mathbf{{{\bf E_x}}=3883}$~keV:} A state at 3881.17(10)~keV is reported in the NDS~\cite{Mccutchan2018}, with $J = (1,2^+)$.
%    The authors mention a $\gamma$ transitions from this  level to the $0^+$ ground state, implying the state is either a spin 1 or 2. 
Our angular distribution for this state is in reasonable agreement with DWBA predictions for $J^\pi=7^-$ or $J^\pi = 8^+$. An unresolved doublet cannot be ruled out at this energy. 
\\ \\   
{\bf $\mathbf{{{\bf E_x}}=4064}$~keV:} No information is available in the NDS~\cite{Mccutchan2018} for this excitation energy. The angular distribution is well reproduced by both $L=5$ and $L=6$ transfers. We therefore assign this state $J^\pi=(5^-,6^+)$. 
\\ \\  
{\bf $\mathbf{{{\bf E_x}}=4070}$~keV:} The angular distributions for this weakly populated state is reproduced reasonably well assuming  $J^\pi=2^+$ or $J^\pi=3^-$.  In the absence of any additional information, we tentatively assign this state a value of $J^\pi=(2^+,3^-)$. 
\\ \\   
{\bf $\mathbf{{{\bf E_x}}=4268}$~keV:} This state is reported for the first time. Its angular distribution is not as well reproduced by DWBA calculations. However, we tentatively assign this state a $J^\pi$ value of~$(2^+,3^-)$, based on its measured distribution.
\\ \\  
{\bf $\mathbf{{{\bf E_x}}=4292}$~keV:} This state is weakly populated  and not reported previously. Its angular distribution is reproduced reasonably well by DWBA calculations for both $L=2$ and $L=3$ transfers. Hence we tentatively assign this state $J^\pi = (2^+, 3^-)$.
\\ \\ 
{\bf $\mathbf{{{\bf E_x}}=4383}$~keV:} This state is also reported for the first time. The $J^\pi$ assignments for our measured angular distribution are either $4^+$ or $5^-$. 
\\ \\ 
{\bf $\mathbf{{{\bf E_x}}=4451}$~keV:} This state is reported for the first time here. Its angular distribution is most consistent with $L = 3$ transfer. However, the data lack the required statistics to make a conclusive spin-parity assignment for the state.  
%Large uncertainties on the cross section make definite $J^\pi$ determination difficult. 

\subsection{\label{Sect:Indefinite_assignment} Undetermined assignments}

\begin{figure}[th!]
 \vspace*{0.5 cm}
\includegraphics[width=0.153\textwidth, height=0.09\textheight]{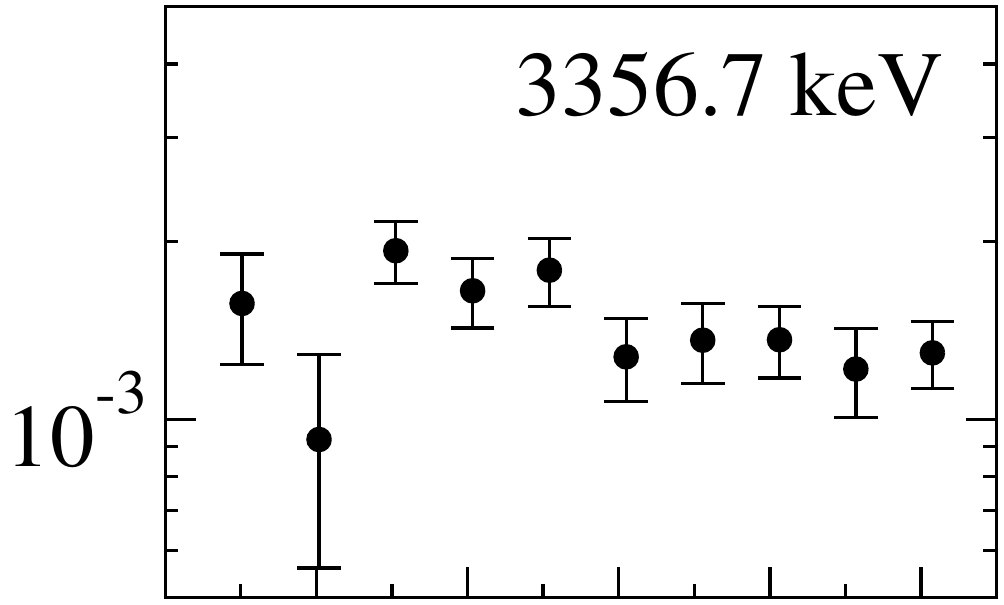}
\includegraphics[width=0.153\textwidth, height=0.09\textheight]{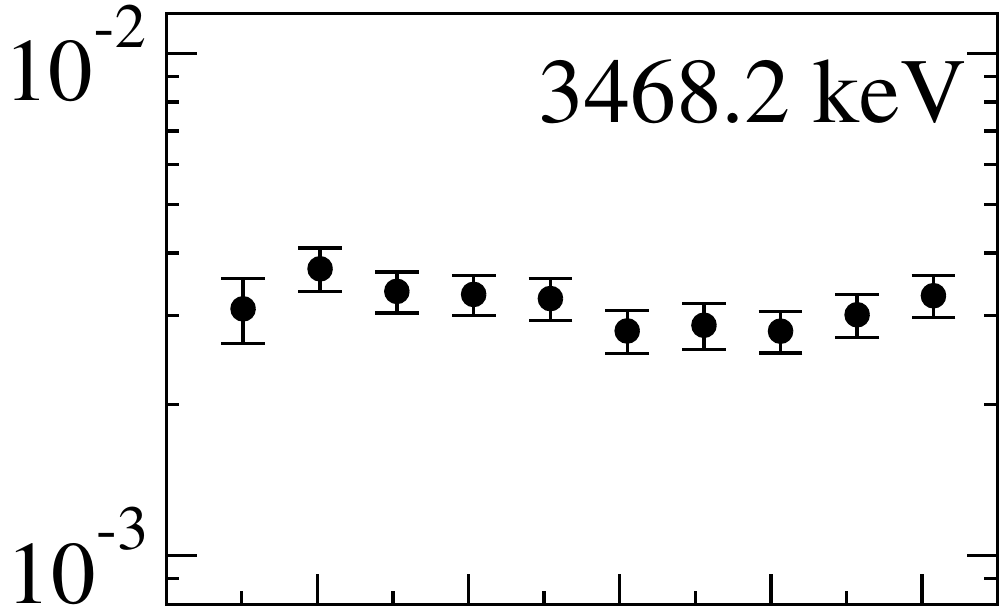}
\includegraphics[width=0.153\textwidth, height=0.09\textheight]{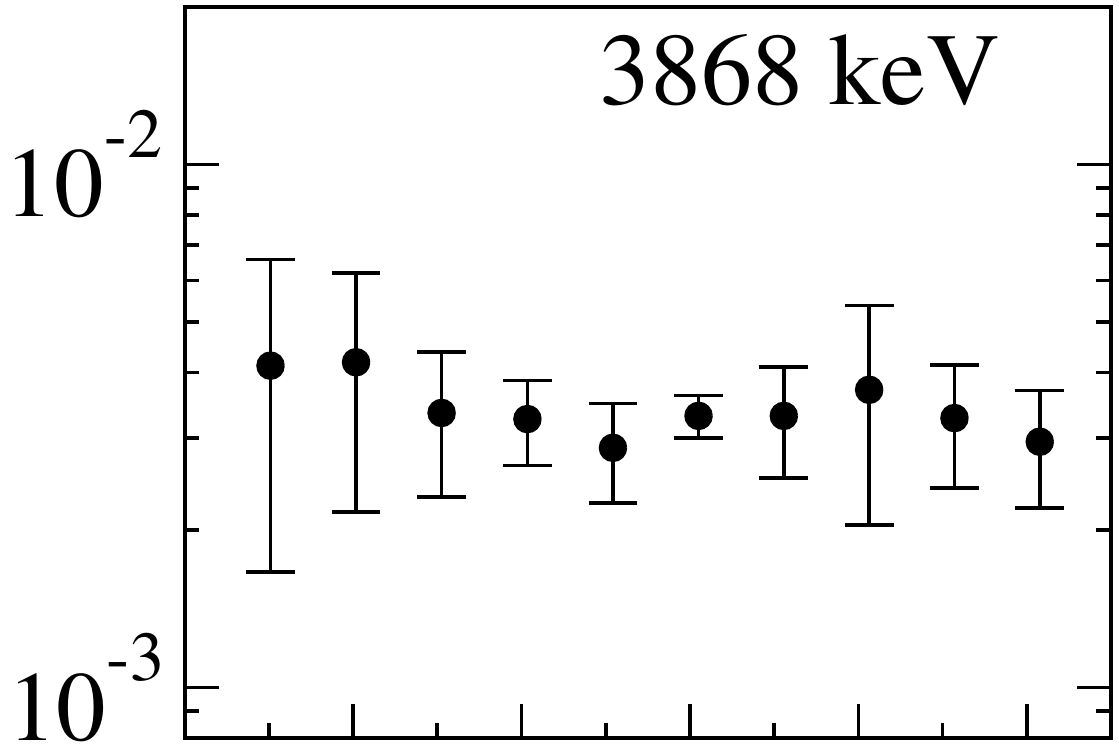}\\
\hspace*{-0.3 cm}
\includegraphics[width=0.165\textwidth, height=0.09\textheight]{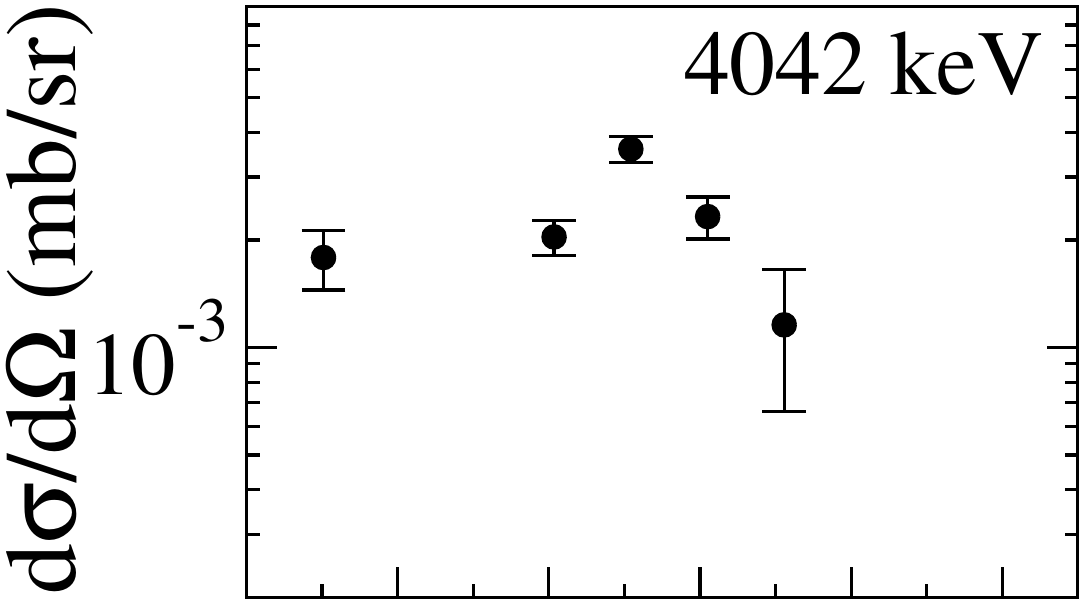}
\includegraphics[width=0.153\textwidth, height=0.09\textheight]{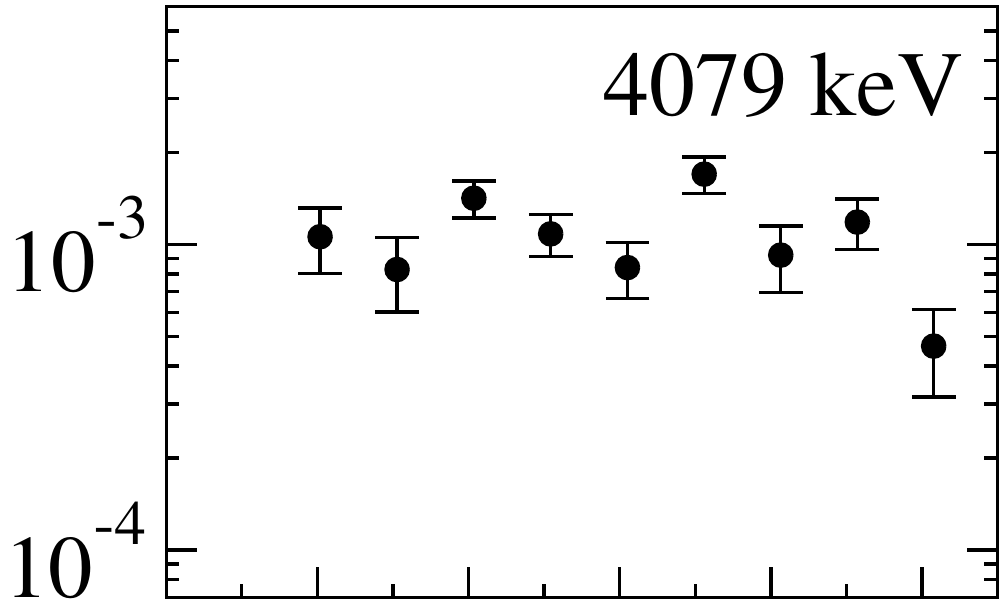}
\includegraphics[width=0.153\textwidth, height=0.09\textheight]{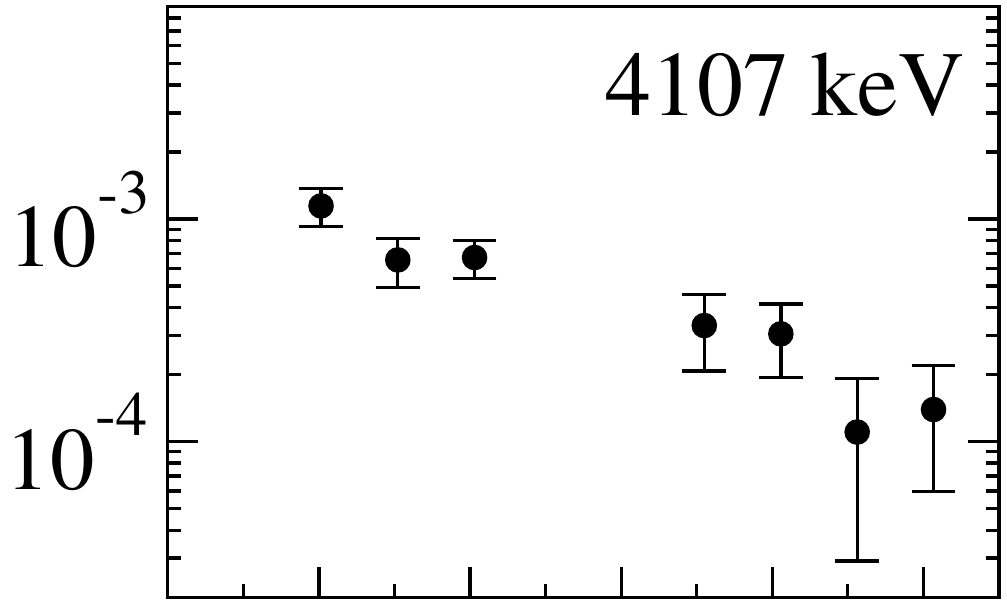}\\
\includegraphics[width=0.153\textwidth, height=0.12\textheight]{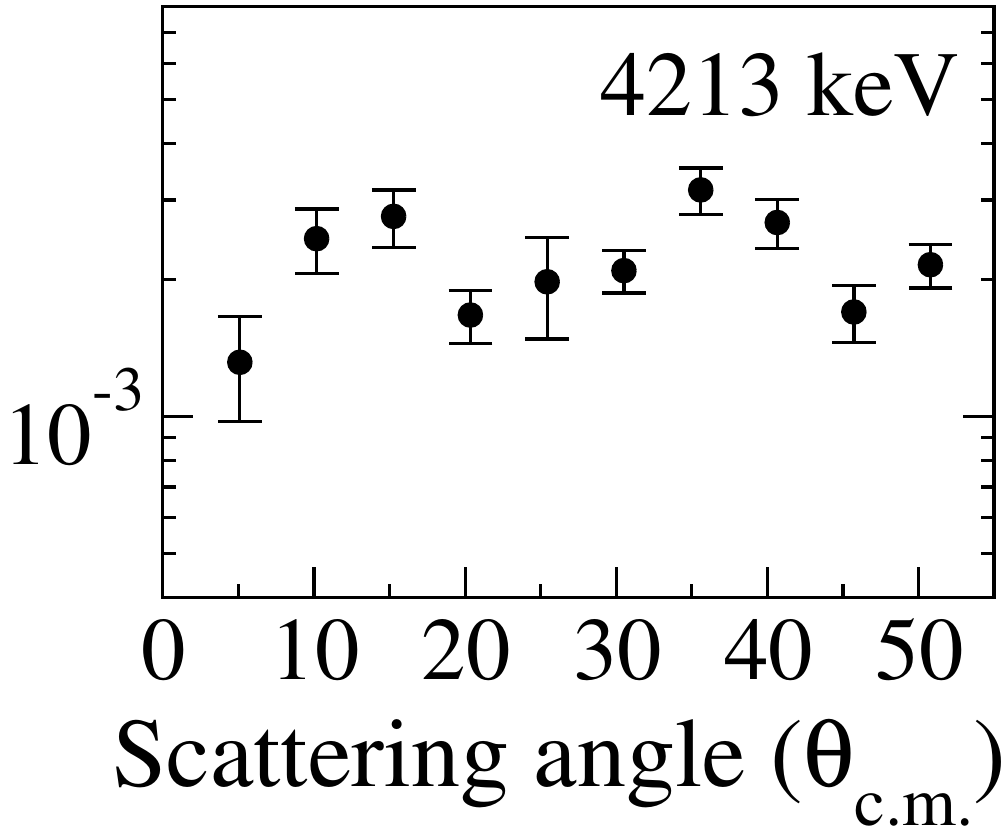}
\includegraphics[width=0.153\textwidth, height=0.12\textheight]{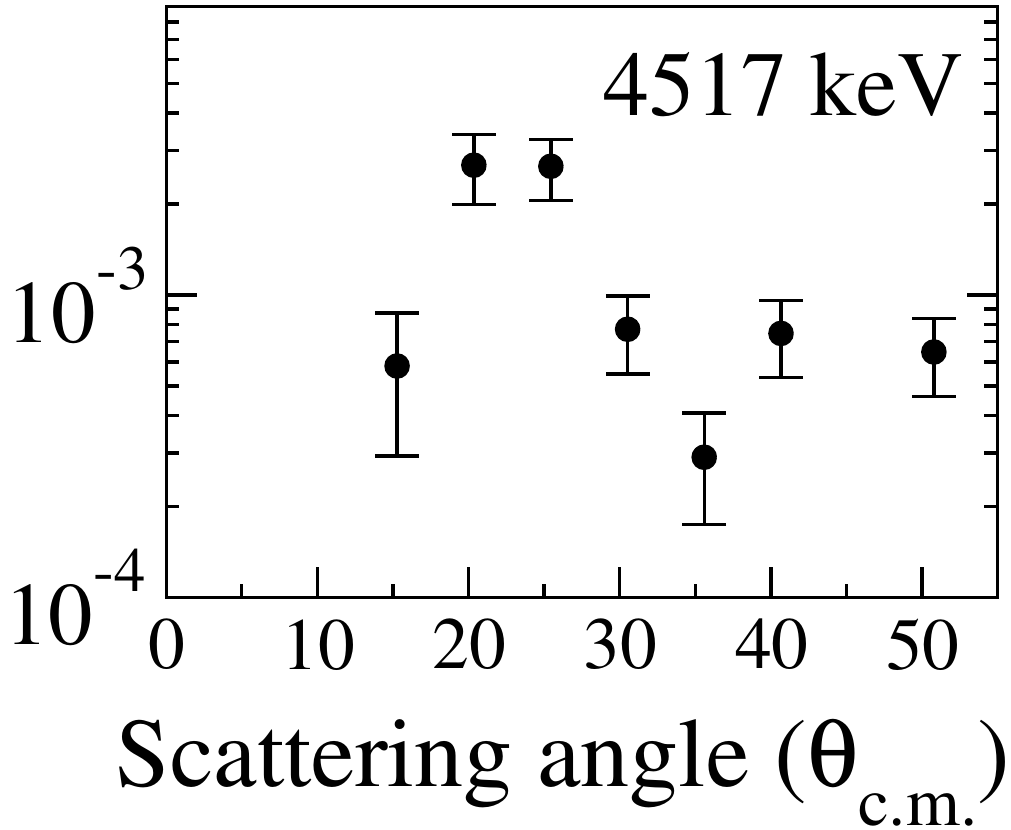}
\includegraphics[width=0.153\textwidth, height=0.12\textheight]{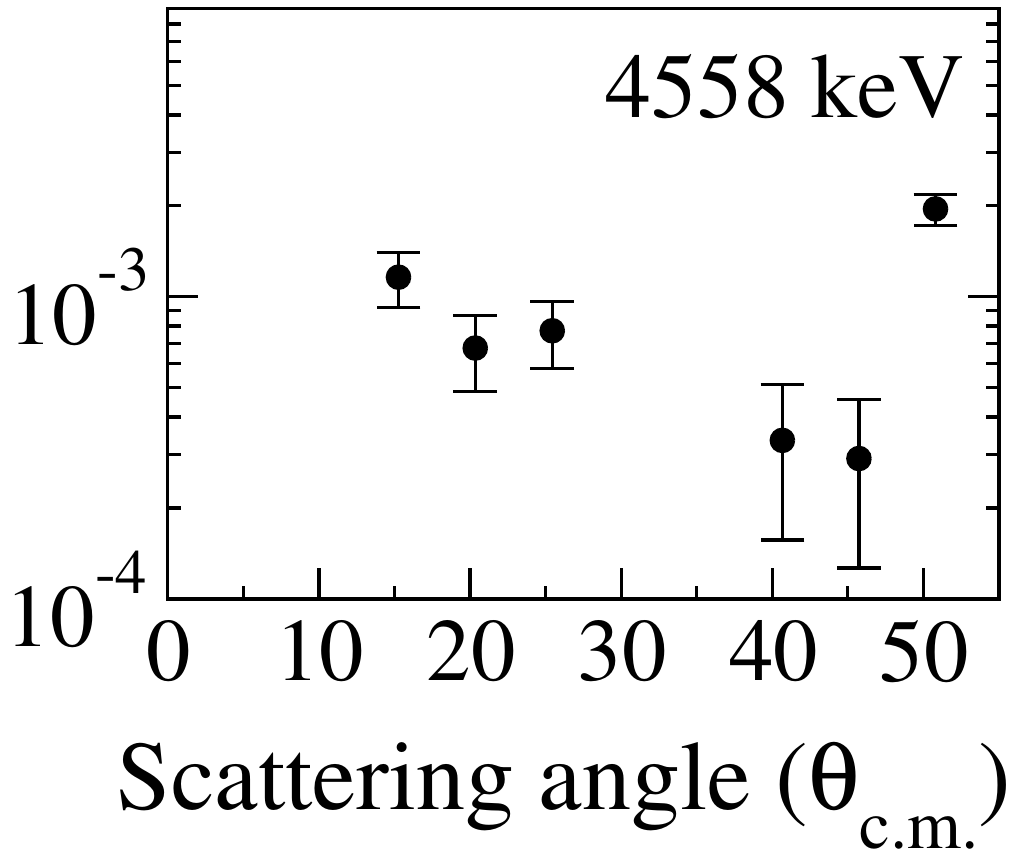} \\
\caption{\label{Fig:138Ba_pt_unassigned}{Excited states in $^{136}$Ba for whom the triton angular distributions are nearly isotropic.}} 
\end{figure}

As shown in Fig~\ref{Fig:138Ba_pt_unassigned}, the angular distributions for nine states are washed out of any diffraction pattern. Consequently we could not make conclusive spin-parity assignments for any of these levels.

% 

% \begin{figure*}[t]
%  %  \includegraphics[scale=0.38]{Feedback_from_colabs/136Ba_calib_spectra_normalized}
%  \centering
%  \includegraphics[scale=0.45]{Feedback_from_colabs/all_strengths_with_tentatives}
%  \caption{\label{fig:strengths_all_Jpi}{$(p,t)$ strengths for each $J=L$ excited state relative to the first observed level. This is similar to what was performed for the $0^+$ states in Ref.~\cite{Rebeiro}, after correcting for kinematic and Q-value effects. }}
% \end{figure*}
% 
% 
% 
\section{{\label{Sect:Conclusion}}Conclusions}    

In summary we used the $^{138}$Ba$(p,t)$ reaction with a high resolution magnetic spectrograph to study excited states in $^{136}$Ba. We identified a total of one hundred and two states, up to an excitation energy of about 4.6 MeV. Comparisons with DWBA calculations were used to make spin-parity assignments for most of these states. We anticipate that the spectroscopic information presented in this paper will further elucidate the low-lying excitations in $^{136}$Ba. This isotope is relevant in the context of $^{136}$Xe $\beta\beta$ decay~\cite{Rebeiro,jespere} as well as for pertinent nuclear structure studies that focus on systematics below the $N = 82$ shell closure~\cite{Mukhopadhyay2008, Massarczyk2012}. As an example, we compare available experimental information for $^{136}$Ba, with predictions from shell model calculations also used for evaluating the nuclear matrix element of $^{136}$Xe $0\nu\beta\beta$ decay~\cite{Rebeiro}. These calculations were carried out in a model space with the five orbitals
$(0g_{7/2}, 1d_{5/2}, 1d_{3/2}, 2s_{1/2}, 0h_{11/2})$ for protons and neutrons. 
We compare two Hamiltonians, whose single-particle energies were adjusted to reproduce experimental spectra relative to the closed shell of $^{132}$Sn, the  
single proton states in $^{133}$Sb and neutron (hole) states in $^{131}$Sn. The first Hamiltonian is from Ref.~\cite{Brown:05} and called sn100pn in the NuShellX interaction library~\cite{nushellx}. It is in proton-neutron ($pn$) formalism, such that that the
neutron-neutron $(nn)$, proton-neutron $(pn)$ and proton-proton $(pp)$ isospin $T=1$
two-body matrix elements (TBME) are all different, with the $pp$ TBME containing the Coulomb interaction. The TBME were obtained from the Brueckner $G$~matrix~\cite{gmatrix} elements of the Paris potential~\cite{paris1,paris2}, and corrected for core-polarization from configuration mixing with orbitals outside of the model space~\cite{gmatrix}. As in Ref.~\cite{Brown:05}, some adjustments were made to improve the spectra for even-even nuclei near $^{132}$Sn.  The second Hamiltonian is GCN50:82~\cite{gcn}, also in the isospin formalism. This was obtained from the $G$-matrix and based upon a realistic CD-Bonn potential~\cite{cdbonn}, with similar adjustments made to the TBME to improve agreement with experimental spectra~\cite{gmatrix}. We used both these Hamiltonians to calculate approximately 300 levels in $^{136}$Ba, up to $J = 6$ and about 4~MeV. 
\begin{figure}[ht]
 \includegraphics[width=8.4 cm]{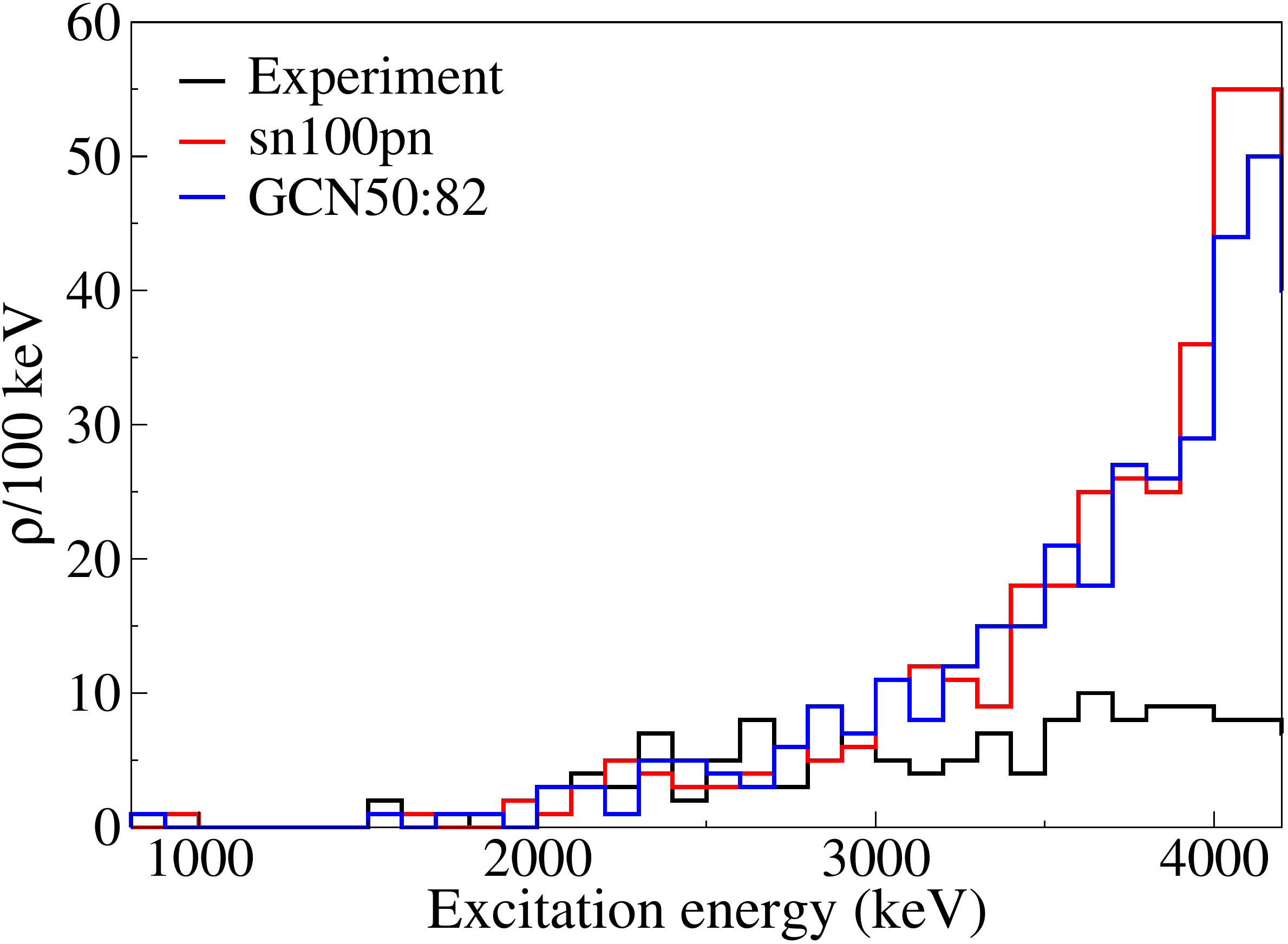}
 \caption{\label{fig:levels_all}{Comparison of measured level densities of states in $^{136}$Ba with shell model predictions.}} 
\end{figure}

\begin{figure}[ht]
 \includegraphics[width=8.4 cm]{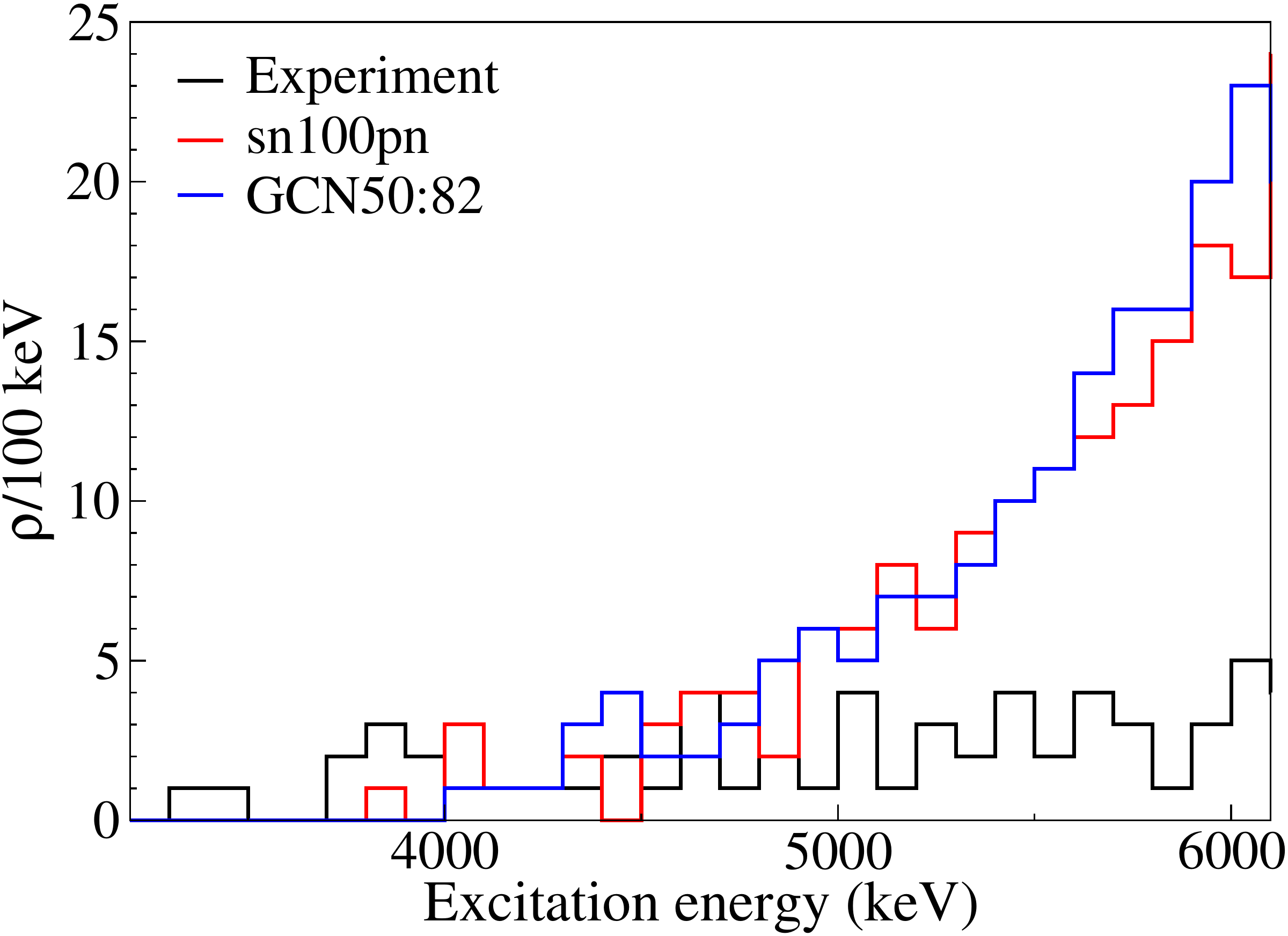}
 \caption{\label{fig:levels_1}{A similar comparison as Fig.~\ref{fig:levels_all}, but for higher-lying $1^-$ states. Here we include all the $J = 1$ states reported in the literature~\cite{ensdf}, as they were mostly identified via inelastic photon scattering which has high selectivity in populating $1^-$ levels.}} 
\end{figure}
Fig.~\ref{fig:levels_all} shows a comparison of the level densities obtained from the calculations, compared with available experimental information. For the latter we include all the known states from the Evaluated Nuclear Structure Data File (ENDSF)~\cite{ensdf} database and the newly observed states reported in this work. We observe that the two Hamiltonians yield very similar results, also showing excellent agreement  with experiment up to around 3~MeV. The calculated values begin to diverge at higher energies. This is not unexpected, as many of the predicted states would have small production cross sections or overlap with other closely spaced states so that they may be difficult to resolve experimentally. Independently, we performed calculations for the $1^-$ states,  with $E_x \gtrsim$~3.5~MeV. As shown in Fig.~\ref{fig:levels_1}, we observe a similar situation on comparing with the large number of $1^-$ states reported in ENSDF~\cite{ensdf}.          
\begin{acknowledgments}
 We thank Marcus Scheck for fruitful discussions. Funding support from the National Research Foundation (NRF), South Africa, under Grant No.~85100, the U.S. Department of Energy Office of Science under grants no. DE-SC0017649 and~DE-FG02-93ER40789 and the National Science Foundation under grant no.~PHY-1811855 are gratefully acknowledged.  
P.A. thanks the Claude Leon Foundation for his postdoctoral fellowship. P.Z.M and J.C.N.O are grateful to the NRF funded MaNuS/MatSci program at UWC for financial support during their M.Sc.  
\end{acknowledgments}

\bibliography{138Ba_pt_ST}

\end{document}